\documentclass[11pt]{article}
\usepackage{amssymb,amsfonts,amsmath,graphicx,hyperref}

\newcommand{\mcm}[3]{\newcommand{#1}[#2]{{\ensuremath{#3}}}}

\usepackage{latexsym}
\usepackage{amssymb}

\mcm{\blank}{0}{(\emptybk)} \mcm{\dashbk}{0}{\mbox{---}}
\mcm{\emptybk}{0}{\:\:} \mcm{\hyph}{0}{\mbox{-}}

\mcm{\diagspace}{0}{\mbox{\hspace{2em}}}

\mcm{\cat}{1}{\mc{#1}} \mcm{\fcat}{1}{\mb{#1}}
\mcm{\mc}{1}{\mathcal{#1}} \mcm{\mr}{1}{\mathrm{#1}}
\mcm{\mi}{1}{\mathit{#1}} \mcm{\mb}{1}{\mathbf{#1}}
\mcm{\scat}{1}{\Bbb{#1}} \mcm{\twid}{1}{\widetilde{#1}}

\mcm{\elt}{0}{\in} \mcm{\sub}{0}{\,\subseteq\,}
\mcm{\such}{0}{\:|\:} \mcm{\without}{0}{\setminus}

\mcm{\atsr}{0}{\Box} \mcm{\eqv}{0}{\,\simeq\,}
\mcm{\iso}{0}{\,\cong\,}
\mcm{\of}{0}{\raisebox{0.2mm}{\ensuremath{\scriptstyle\circ}}}

\mcm{\bdry}{0}{\partial}

\mcm{\Bee}{0}{\cat{B}} \mcm{\Beep}{0}{\cat{B'}}
\mcm{\Eee}{0}{\cat{E}} \mcm{\Eeep}{0}{\cat{E'}}

\mcm{\Ess}{0}{\cat{S}} \mcm{\Tee}{0}{\cat{T}}
\mcm{\Teep}{0}{\cat{T'}} \mcm{\Stee}{0}{\scat{T}}
\mcm{\Steep}{0}{\scat{T'}}

\mcm{\blbk}{0}{\blank^{\blob}}
\mcm{\blob}{0}{\scriptscriptstyle{\bullet}}
\mcm{\stbk}{0}{\blank^{*}} \mcm{\ubl}{0}{{}^{\blob}}
\mcm{\ust}{0}{{}^{*}}

\mcm{\Cartpr}{0}{\pr{\Eee}{T}} \mcm{\Cartprp}{0}{\pr{\Eeep}{T'}}
\mcm{\Mnd}{0}{\triple{T}{\eta}{\mu}}
\mcm{\Zeropr}{0}{\pr{\Set}{\id}}

\mcm{\dopset}{0}{\ftrcat{\Delta^{\op}}{\Set}}
\mcm{\tropset}{0}{\ftrcat{\fcat{TR}^{\op}}{\Set}}

\mcm{\cod}{0}{\mr{cod}} \mcm{\dom}{0}{\mr{dom}}
\mcm{\End}{0}{\mr{End}} \mcm{\Hom}{0}{\mr{Hom}}
\mcm{\ob}{0}{\mr{ob}\,} \mcm{\op}{0}{\mr{op}}

\mcm{\comp}{0}{\mi{comp}} \mcm{\id}{0}{\mi{id}}
\mcm{\ids}{0}{\mi{ids}} \mcm{\mult}{0}{\mi{mult}}
\mcm{\unit}{0}{\mi{unit}}

\mcm{\Ab}{0}{\fcat{Ab}} \mcm{\Alg}{0}{\fcat{Alg}}
\mcm{\Bim}{1}{\fcat{Bim}(#1)} \mcm{\Cat}{0}{\fcat{Cat}}
\mcm{\Cay}{0}{\fcat{Cay}} \mcm{\Cpn}{1}{\pr{\Set/S_{#1}}{T_{#1}}}
\mcm{\fc}{0}{\fcat{fc}} \mcm{\fm}{0}{\fcat{fm}}
\mcm{\Graph}{0}{\fcat{Graph}} \mcm{\Gy}{0}{\fcat{Gy}}
\mcm{\Hpn}{1}{\pr{\Eee_{#1}}{P_{#1}}} \mcm{\Mon}{0}{\mb{Mon}}
\mcm{\Multicat}{0}{\fcat{Multicat}} \mcm{\One}{0}{\fcat{1}}
\mcm{\PD}{1}{\fcat{PD}_{#1}} \mcm{\Prof}{0}{\fcat{Prof}}
\mcm{\Set}{0}{\fcat{Set}} \mcm{\Span}{0}{\fcat{Span}}
\mcm{\Ssq}{0}{\fcat{Ssq}} \mcm{\Struc}{0}{\fcat{Struc}}
\mcm{\Sym}{0}{\fcat{Sym}} \mcm{\TR}{1}{\fcat{TR}(#1)}
\mcm{\Tr}{0}{\fcat{Tr}} \mcm{\Twocat}{0}{\fcat{2\hyph\Cat}}

\mcm{\integers}{0}{\mathbb{Z}}

\mcm{\range}{2}{#1,\,\ldots\,,#2}
\mcm{\bftuple}{2}{\tuplebts{\range{#1}{#2}}}
\mcm{\tuple}{3}{\tuplebts{\range{#1,#2}{#3}}}
\mcm{\rttuple}{1}{\tuplebts{\,\ldots\,,#1}}
\mcm{\abftuple}{2}{\atuplebts{\range{#1}{#2}}}
\mcm{\atuple}{3}{\atuplebts{\range{#1,#2}{#3}}}
\mcm{\arttuple}{1}{\atuplebts{\,\ldots\,,#1}}
\mcm{\sqbftuple}{2}{\obt\range{#1}{#2}\cbt}
\mcm{\pr}{2}{\tuplebts{#1,#2}}
\mcm{\triple}{3}{\tuplebts{#1,#2,#3}}

\mcm{\eend}{2}{#1[#2]} \mcm{\ehom}{3}{#1[#2,#3]}
\mcm{\ftrcat}{2}{[#1,#2]} \mcm{\homset}{3}{#1(#2,#3)}
\mcm{\multihom}{3}{#1(#2;#3)}
\mcm{\relhom}{5}{#1_{#2}(\range{#3}{#4};#5)}

\mcm{\go}{0}{\rTo} \mcm{\goby}{1}{\rTo^{#1}}
\mcm{\goesto}{0}{\,\longmapsto\,} \mcm{\goiso}{0}{\goby{\diso}}
\mcm{\monic}{0}{\rMonic} \mcm{\og}{0}{\lTo}
\mcm{\ogby}{1}{\lTo^{#1}}

\mcm{\gph}{2}{\spn{#1}{T #2}{#2}} \mcm{\graph}{4}{\spaan{#1}{T
#2}{#2}{#3}{#4}} \mcm{\oppair}{2}{\stackrel{\rTo^{#1}}{\lTo_{#2}}}
\mcm{\parpair}{2}{\stackrel{\rTo^{#1}}{\rTo_{#2}}}
\mcm{\spn}{3}{#2 \og #1 \go #3} \mcm{\spaan}{5}{#2 \ogby{#4} #1
\goby{#5} #3}

\mcm{\bktdvslob}{3}
    {\left(
    \begin{diagram}[height=1.5em]
    #1      \\
    \dTo>{\,#2} \\
    #3      \\
    \end{diagram}
    \right)}
\mcm{\slob}{3}{(#1 \goby{#2} #3)} \mcm{\vslob}{3}
    {\left.
    \begin{diagram}[height=1.5em]
    #1      \\
    \dTo>{\,#2} \\
    #3      \\
    \end{diagram}
    \right.}

\newenvironment{tree}
    {\begin{diagram}[height=1em,width=.75em,abut,noPS,tight]}
    {\end{diagram}}

\mcm{\enode}{0}{\circ}

\mcm{\nl}{1}{\stackrel{\textstyle #1}{\node}}
\mcm{\node}{0}{\bullet}

\mcm{\utree}{0}{\node}

\mcm{\diso}{0}{\sim}

\mcm{\vdiso}{0}{\wr}

\mcm{\nat}{0}{\mathbb{N}}

\mcm{\Onepr}{0}{\pr{\Graph}{\fc}}
\newlength{\nllwidth}
\newlength{\nllheight}
\newcommand{\stackbelow}[2]{%
\settowidth{\nllwidth}{\ensuremath{#1}\ensuremath{#2}}%
\settoheight{\nllheight}{\ensuremath{#2}}%
\addtolength{\nllheight}{.3ex}%
\mbox{%
\ensuremath{#1}%
\hspace{-.5\nllwidth}%
\raisebox{-1\nllheight}{\ensuremath{#2}}}}
\mcm{\nlal}{2}{\stackbelow{\nl{#1}}{#2}}
\mcm{\nll}{1}{\stackbelow{\node}{#1}} \mcm{\wun}{0}{\fcat{1}}
\mcm{\atuplebts}{1}{\langle #1 \rangle} \mcm{\tuplebts}{1}{(#1)}
\mcm{\bo}{0}{(} \mcm{\bc}{0}{)}
\mcm{\UBilax}{0}{\fcat{UBicat}_\mr{lax}}
\mcm{\UBiwk}{0}{\fcat{UBicat}_\mr{wk}}
\mcm{\UBistr}{0}{\fcat{UBicat}_\mr{str}}
\mcm{\Bilax}{0}{\fcat{Bicat}_\mr{lax}}
\mcm{\Biwk}{0}{\fcat{Bicat}_\mr{wk}}
\mcm{\Bistr}{0}{\fcat{Bicat}_\mr{str}} \mcm{\rotsub}{0}{\cup
\raisebox{0.1em}{$\scriptstyle{|}$}} \mcm{\pd}{0}{\fcat{pd}}
\mcm{\rep}{1}{\widehat{#1}} \mcm{\ovln}{1}{\overline{#1}}
\mcm{\Gph}{0}{\fcat{Gph}} \mcm{\tr}{0}{\fcat{tr}}

\mcm{\ladj}{0}{\,\dashv\,} \mcm{\zeropd}{0}{\node}
    {\end{diagram}}
\mcm{\END}{0}{\fcat{End}} \mcm{\HOM}{0}{\fcat{Hom}}

\newlength{\gwidth} % the width of a glob
\newlength{\gvert}  % the overall vertical measurement
\newlength{\gdrop}  % the distance a labelled glob protrudes below the
\newlength{\gbaredrop}  % the distance from the textline to the bottom of the
\newlength{\goffset}    % the distance from the centre of the glob to the
\newlength{\gtemp}  % temporary register

\newcommand{\present}[1]{%
\makebox[1\gwidth]{%
\rule[-1\gdrop]{0ex}{1\gvert}%
\raisebox{-1\gbaredrop}{#1}}}

\newcommand{\presentl}[1]{%
\makebox[1\gwidth][l]{%
\rule[-1\gdrop]{0ex}{1\gvert}%
\raisebox{-1\gbaredrop}{#1}}}

\newcommand{\presentr}[1]{%
\makebox[1\gwidth][r]{%
\rule[-1\gdrop]{0ex}{1\gvert}%
\raisebox{-1\gbaredrop}{#1}}}

\newcommand{\ginitdims}[2]{%        % GLOBULAR VERSION
\setlength{\unitlength}{1em}%       % unitlength = 1em
\setlength{\goffset}{.25\unitlength}%   % globular offset = .25em
\setlength{\gwidth}{#1\unitlength}% % width as specified
\setlength{\gvert}{#2\unitlength}%  % vert = #2
\setlength{\gdrop}{.5\gvert}%       %
\addtolength{\gdrop}{-1\goffset}%   %
\setlength{\gbaredrop}{1\gdrop}%    % gdrop = drop = half(vert) - offset
\addtolength{\gvert}{.6\unitlength}%    % total extra clearance of .6em...
\addtolength{\gdrop}{.3\unitlength}}    % ...half of which is at bottom

\newcommand{\cinitdims}[2]{%        % CELLULAR VERSION
\setlength{\unitlength}{1em}%       % unitlength = 1em
\setlength{\goffset}{.35\unitlength}%   % cellular offset = .35em
\setlength{\gwidth}{#1\unitlength}% % width as specified
\setlength{\gvert}{#2\unitlength}%  % vert = #2
\setlength{\gdrop}{.5\gvert}%       %
\addtolength{\gdrop}{-1\goffset}%   %
\setlength{\gbaredrop}{1\gdrop}%    % gdrop = drop = half(vert) - offset
\addtolength{\gvert}{.6\unitlength}%    % total extra clearance of .6em...
\addtolength{\gdrop}{.3\unitlength}}    % ...half of which is at bottom

\newcommand{\gsinitdims}[2]{%       % SMALL GLOBULAR VERSION
\setlength{\unitlength}{0.5em}%     % unitlength = 0.5em
\setlength{\goffset}{.25\unitlength}%   % globular offset = .25em
\setlength{\gwidth}{#1\unitlength}% % width as specified
\setlength{\gvert}{#2\unitlength}%  % vert = #2
\setlength{\gdrop}{.5\gvert}%       %
\addtolength{\gdrop}{-1\goffset}%   %
\setlength{\gbaredrop}{1\gdrop}%    % gdrop = drop = half(vert) - offset
\addtolength{\gvert}{.6\unitlength}%    % total extra clearance of .6em...
\addtolength{\gdrop}{.3\unitlength}}    % ...half of which is at bottom

\newcommand{\sidespic}[1]{%
\settowidth{\gtemp}{\ensuremath{#1}}%
\addtolength{\gwidth}{1\gtemp}}

\newcommand{\abovepic}[1]{%
\settoheight{\gtemp}{\ensuremath{#1}}%
\addtolength{\gvert}{1\gtemp}%
\settodepth{\gtemp}{\ensuremath{#1}}%
\addtolength{\gvert}{1\gtemp}}

\newcommand{\belowpic}[1]{%
\settoheight{\gtemp}{\ensuremath{#1}}%
\addtolength{\gvert}{1\gtemp}%
\addtolength{\gdrop}{1\gtemp}%
\settodepth{\gtemp}{\ensuremath{#1}}%
\addtolength{\gvert}{1\gtemp}%
\addtolength{\gdrop}{1\gtemp}}

\newcommand{\cell}[4]{\put(#1,#2){\makebox(0,0)[#3]{\ensuremath{#4}}}}
\mcm{\zmark}{0}{\scriptstyle{\bullet}}

\newcommand{\pregfst}[1]{%
\begin{picture}(0.5,0.2)(-0.5,-0.2)%
\cell{-0.1}{-0.2}{tr}{#1}%
\cell{0}{0}{c}{\zmark}%
\end{picture}}

\mcm{\gfst}{1}{%
\ginitdims{0.5}{0.4}%
\sidespic{#1}%
\belowpic{#1}%
\presentr{\pregfst{#1}}}

\newcommand{\preglst}[1]{%
\begin{picture}(0.5,0.2)(0,-0.2)%
\cell{0.1}{-0.2}{tl}{#1}%
\cell{0.05}{0}{c}{\zmark}%
\end{picture}}

\mcm{\glst}{1}{%
\ginitdims{.5}{.4}%
\sidespic{#1}%
\belowpic{#1}%
\presentl{\preglst{#1}}}

\newcommand{\preglft}[1]{%
\begin{picture}(0,0.2)(0,-0.2)%
\cell{-0.1}{-0.2}{tr}{#1}%
\cell{0.05}{0}{c}{\zmark}%
\end{picture}}

\mcm{\glft}{1}{%
\ginitdims{0}{.4}%
\belowpic{#1}%
\present{\preglft{#1}}}

\newcommand{\pregrgt}[1]{%
\begin{picture}(0,0.2)(0,-0.2)%
\cell{0.1}{-0.2}{tl}{#1}%
\cell{0.05}{0}{c}{\zmark}%
\end{picture}}

\mcm{\grgt}{1}{%
\ginitdims{0}{.4}%
\belowpic{#1}%
\present{\pregrgt{#1}}}

\newcommand{\pregblw}[1]{%
\begin{picture}(0,0.3)(0,-0.3)
\cell{0}{-0.3}{t}{#1}%
\cell{0.05}{0}{c}{\zmark}%
\end{picture}}

\mcm{\gblw}{1}{%
\ginitdims{0}{.6}%
\belowpic{#1}%
\present{\pregblw{#1}}}

\newcommand{\pregfbw}[1]{%
\begin{picture}(0,0.65)(0,-0.65)
\cell{0}{-0.65}{t}{#1}%
\cell{0.05}{0}{c}{\zmark}%
\end{picture}}

\mcm{\gfbw}{1}{%
\ginitdims{0}{1.3}%
\belowpic{#1}%
\present{\pregfbw{#1}}}

\newcommand{\pregzero}[1]{%
\begin{picture}(0.8,0.4)(-0.4,-0.4)
\cell{0}{-0.4}{t}{#1}%
\cell{0}{0}{c}{\zmark}%
\end{picture}}

\mcm{\gzero}{1}{%
\ginitdims{0.8}{.6}%
\belowpic{#1}%
\sidespic{#1}%
\present{\pregzero{#1}}}

\newcommand{\pregone}[1]{%
\begin{picture}(5,0.4)(0,-0.2)%
\cell{2.5}{0.2}{b}{#1}%
\put(0,0){\vector(1,0){5}}%
\end{picture}}

\mcm{\gone}{1}{%
\ginitdims{5}{0.4}%
\abovepic{#1}%
\present{\pregone{#1}}}

\newcommand{\pregtwo}[3]{%
\begin{picture}(5,3.4)(0,-0.2)%
\cell{2.5}{3.2}{b}{#1}%
\cell{2.5}{-.2}{t}{#2}%
\cell{2.7}{1.5}{l}{#3}%
\qbezier(0,1.5)(2.5,4.5)(5,1.5)%
\qbezier(0,1.5)(2.5,-1.5)(5,1.5)%
\put(5,1.5){\vector(1,-1){0}}%
\put(5,1.5){\vector(1,1){0}}%
\put(2.5,2.5){\vector(0,-1){2}}%
\end{picture}}

\mcm{\gtwo}{3}{%
\ginitdims{5}{3.4}%
\abovepic{#1}%
\belowpic{#2}%
\present{\pregtwo{#1}{#2}{#3}}}

\newcommand{\pregthree}[5]{%
\begin{picture}(5,5.4)(0,-1.2)%
\cell{2.5}{4.2}{b}{#1}%
\cell{1.5}{1.7}{b}{#2}%
\cell{2.5}{-1.2}{t}{#3}%
\cell{2.7}{2.75}{l}{#4}%
\cell{2.7}{0.25}{l}{#5}%
\qbezier(0,1.5)(2.5,6.5)(5,1.5)%
\qbezier(0,1.5)(2.5,-3.5)(5,1.5)%
\put(0,1.5){\vector(1,0){5}}%
\put(2.5,3.5){\vector(0,-1){1.5}}%
\put(2.5,1){\vector(0,-1){1.5}}%
\put(5,1.5){\vector(1,-3){0}}%
\put(5,1.5){\vector(1,3){0}}%
\end{picture}}

\mcm{\gthree}{5}{%
\ginitdims{5}{5.4}%
\abovepic{#1}%
\belowpic{#3}%
\present{\pregthree{#1}{#2}{#3}{#4}{#5}}}

\newcommand{\pregfour}[7]{%
\begin{picture}(5,8.4)(0,-2.7)%
\cell{2.5}{5.7}{b}{#1}%
\cell{1.5}{2.8}{b}{#2}%
\cell{1.5}{0.2}{t}{#3}%
\cell{2.5}{-2.7}{t}{#4}%
\cell{2.7}{4.25}{l}{#5}%
\cell{2.7}{1.5}{l}{#6}%
\cell{2.7}{-1.25}{l}{#7}%
\qbezier(0,1.5)(2.5,9.5)(5,1.5)%
\qbezier(0,1.5)(2.5,4)(5,1.5)%
\qbezier(0,1.5)(2.5,-1)(5,1.5)%
\qbezier(0,1.5)(2.5,-6.5)(5,1.5)%
\put(2.5,5.25){\vector(0,-1){2}}%
\put(2.5,2.5){\vector(0,-1){2}}%
\put(2.5,-0.25){\vector(0,-1){2}}%
\put(5,1.5){\vector(1,-4){0}}%
\put(5,1.5){\vector(4,-3){0}}%
\put(5,1.5){\vector(4,3){0}}%
\put(5,1.5){\vector(1,4){0}}%
\end{picture}}

\mcm{\gfour}{7}{%
\ginitdims{5}{8.4}%
\abovepic{#1}%
\belowpic{#4}%
\present{\pregfour{#1}{#2}{#3}{#4}{#5}{#6}{#7}}}

\newcommand{\pregthreecell}[5]{%
\begin{picture}(8,5)(-4,-2.5)%
\cell{0}{2.5}{b}{#1}%
\cell{0}{-2.5}{t}{#2}%
\cell{-1.7}{0}{r}{#3}%
\cell{1.7}{0}{l}{#4}%
\cell{0}{0.2}{b}{#5}%
\qbezier(-4,0)(0,4.2)(4,0)%
\qbezier(-4,0)(0,-4.2)(4,0)%
\qbezier(-0.5,1.8)(-2.5,0)(-0.5,-1.8)%
\qbezier(0.5,1.8)(2.5,0)(0.5,-1.8)%
\put(-1,0){\vector(1,0){2}}%
\put(4,0){\vector(1,-1){0}}%
\put(4,0){\vector(1,1){0}}%
\put(-0.5,-1.8){\vector(1,-1){0}}%
\put(0.5,-1.8){\vector(-1,-1){0}}%
\end{picture}}

\mcm{\gthreecell}{5}{%
\ginitdims{8}{5}%
\abovepic{#1}%
\belowpic{#2}%
\present{\pregthreecell{#1}{#2}{#3}{#4}{#5}}}

\newcommand{\pregthreecellu}{%
\begin{picture}(5,3.4)(-0.5,-0.2)%
\qbezier(-.5,1.5)(2,4.5)(4.5,1.5)%
\qbezier(-.5,1.5)(2,-1.5)(4.5,1.5)%
\qbezier(1.5,2.7)(0.5,1.5)(1.5,0.3)%
\qbezier(2.5,2.7)(3.5,1.5)(2.5,0.3)%
\put(1.3,1.5){\vector(1,0){1.4}}%
\put(4.5,1.5){\vector(1,-1){0}}%
\put(4.5,1.5){\vector(1,1){0}}%
\put(1.5,0.3){\vector(2,-3){0}}%
\put(2.5,0.3){\vector(-2,-3){0}}%
\end{picture}}

\mcm{\gthreecellu}{0}{%
\ginitdims{5}{3.4}%
\present{\pregthreecellu}}

\newcommand{\pregtwocentre}[3]{%
\begin{picture}(5,3.4)(0,-0.2)%
\cell{2.5}{3.2}{b}{#1}%
\cell{2.5}{-.2}{t}{#2}%
\cell{2.5}{1.5}{c}{#3}%
\qbezier(0,1.5)(2.5,4.5)(5,1.5)%
\qbezier(0,1.5)(2.5,-1.5)(5,1.5)%
\put(5,1.5){\vector(1,-1){0}}%
\put(5,1.5){\vector(1,1){0}}%
\put(2.5,2.5){\vector(0,-1){2}}%
\end{picture}}

\mcm{\gtwocentre}{3}{%
\ginitdims{5}{3.4}%
\abovepic{#1}%
\belowpic{#2}%
\present{\pregtwocentre{#1}{#2}{#3}}}

\newcommand{\pregspecialone}[9]{%
\begin{picture}(8,8)(-4,-4)%
\cell{0}{3.9}{b}{#1}%
\cell{-2}{-0.2}{t}{#2}%
\cell{0}{-3.9}{t}{#3}%
\cell{-1.5}{1.1}{r}{#4}%
\cell{0.2}{1.5}{l}{#5}%
\cell{1.5}{1.1}{l}{#6}%
\cell{0.2}{-2}{l}{#7}%
\cell{-0.9}{2.3}{b}{#8}%
\cell{0.9}{2.3}{b}{#9}%
\qbezier(-4,0)(0,8)(4,0)%
\qbezier(-4,0)(0,-8)(4,0)%
\qbezier(-0.5,3.4)(-3.5,2)(-0.5,0.6)%
\qbezier(0.5,3.4)(3.5,2)(0.5,0.6)%
\put(-4,0){\vector(1,0){8}}%
\put(0,3.4){\vector(0,-1){2.8}}%
\put(0,-0.8){\vector(0,-1){2.4}}%
\put(-1.5,2.2){\vector(1,0){1.2}}%
\put(0.3,2.2){\vector(1,0){1.2}}%
\put(4,0){\vector(1,-2){0}}%
\put(4,0){\vector(1,2){0}}%
\put(-0.5,0.6){\vector(2,-1){0}}%
\put(0.5,0.6){\vector(-2,-1){0}}%
\end{picture}}

\mcm{\gspecialone}{9}{%
\ginitdims{8}{8}%
\abovepic{#1}%
\belowpic{#3}%
\present{\pregspecialone{#1}{#2}{#3}{#4}{#5}{#6}{#7}{#8}{#9}}}

\newcommand{\pregspecialtwo}{%
\begin{picture}(5,3.4)(0,-0.2)%
\qbezier(0,1.5)(2.5,4.5)(5,1.5)%
\qbezier(0,1.5)(2.5,-1.5)(5,1.5)%
\qbezier(1.7,2.5)(0,1.5)(1.7,0.5)%
\qbezier(3.3,2.5)(5,1.5)(3.3,0.5)%
\put(5,1.5){\vector(1,-1){0}}%
\put(5,1.5){\vector(1,1){0}}%
\put(1.7,0.5){\vector(3,-2){0}}%
\put(3.3,0.5){\vector(-3,-2){0}}%
\put(2.5,2.5){\vector(0,-1){2}}%
\put(1.2,1.5){\vector(1,0){1}}%
\put(2.8,1.5){\vector(1,0){1}}%
\end{picture}}

\mcm{\gspecialtwo}{0}{%
\ginitdims{5}{3.4}%
\present{\pregspecialtwo}}

\newcommand{\pregspecialthree}{%
\begin{picture}(5,5.4)(0,-1.2)%
\qbezier(0,1.5)(2.5,6.5)(5,1.5)%
\qbezier(0,1.5)(2.5,-3.5)(5,1.5)%
\qbezier(2,3.5)(1,2.75)(2,2)%
\qbezier(3,3.5)(4,2.75)(3,2)%
\qbezier(2,1)(1,0.25)(2,-0.5)%
\qbezier(3,1)(4,0.25)(3,-0.5)%
\put(0,1.5){\vector(1,0){5}}%
\put(1.5,2.75){\vector(1,0){2}}%
\put(1.5,0.25){\vector(1,0){2}}%
\put(5,1.5){\vector(1,-3){0}}%
\put(5,1.5){\vector(1,3){0}}%
\put(2,2){\vector(1,-1){0}}%
\put(3,2){\vector(-1,-1){0}}%
\put(2,-0.5){\vector(1,-1){0}}%
\put(3,-0.5){\vector(-1,-1){0}}%
\end{picture}}

\mcm{\gspecialthree}{0}{%
\ginitdims{5}{5.4}%
\present{\pregspecialthree}}

\newcommand{\pregonew}[1]{%
\begin{picture}(8,0.4)(0,-0.2)%
\cell{4}{0.2}{b}{#1}%
\put(0,0){\vector(1,0){8}}%
\end{picture}}

\mcm{\gonew}{1}{%
\ginitdims{8}{0.4}%
\abovepic{#1}%
\present{\pregonew{#1}}}

\mcm{\gzersu}{0}{%
\gsinitdims{0}{.6}%
\present{\pregblw{}}}

\mcm{\gonesu}{0}{%
\gsinitdims{5}{0.4}%
\present{\pregone{}}}

\mcm{\gtwosu}{0}{%
\gsinitdims{5}{3.4}%
\present{\pregtwo{}{}{}}}

\mcm{\gthreesu}{0}{%
\gsinitdims{5}{5.4}%
\present{\pregthree{}{}{}{}{}}}

\mcm{\gfoursu}{0}{%
\gsinitdims{5}{8.4}%
\present{\pregfour{}{}{}{}{}{}{}}}

\newcommand{\precone}[1]{%
\begin{picture}(4.2,0.4)(-0.3,-0.2)%
\cell{1.8}{0.2}{b}{#1}%
\put(0,0){\vector(1,0){3.6}}%
\end{picture}}

\mcm{\cone}{1}{%
\cinitdims{4.2}{0.4}%
\abovepic{#1}%
\present{\precone{#1}}}

\mcm{\gfstsu}{0}{%
\gsinitdims{0.5}{0.4}%
\presentr{\pregfst{}}}

\mcm{\glstsu}{0}{%
\gsinitdims{0.5}{0.4}%
\presentl{\preglst{}}}

\newcommand{\prectwodbl}[3]%
{\begin{picture}(4.2,3.4)(-0.1,-0.2)%
\cell{2}{3.2}{b}{#1}%
\cell{2}{-0.2}{t}{#2}%
\cell{2.3}{1.5}{l}{#3}%
\qbezier(0,2)(2,4)(4,2)%
\qbezier(0,1)(2,-1)(4,1)%
\put(4,2){\vector(1,-1){0}}%
\put(4,1){\vector(1,1){0}}%
\put(1.9,2.5){\line(0,-1){1.8}}%
\put(2.1,2.5){\line(0,-1){1.8}}%
\cell{2.01}{0.4}{b}{\vee}%
\end{picture}}

\mcm{\ctwodbl}{3}{%
\cinitdims{4.2}{3.4}%
\abovepic{#1}%
\belowpic{#2}%
\present{\prectwodbl{#1}{#2}{#3}}}

\newcommand{\precthreedbl}[5]{%
\begin{picture}(4.2,5.4)(-0.1,-0.2)%
\cell{2}{5.2}{b}{#1}%
\cell{1}{2.7}{b}{#2}%
\cell{2}{-.2}{t}{#3}%
\cell{2.3}{3.75}{l}{#4}%
\cell{2.3}{1.25}{l}{#5}%
\qbezier(0,3)(2,7)(4,3)%
\qbezier(0,2)(2,-2)(4,2)%
\put(0,2.5){\vector(1,0){4}}%
\put(1.9,4.5){\line(0,-1){1.3}}%
\put(2.1,4.5){\line(0,-1){1.3}}%
\cell{2.01}{2.9}{b}{\vee}%
\put(1.9,2){\line(0,-1){1.3}}%
\put(2.1,2){\line(0,-1){1.3}}%
\cell{2.01}{0.4}{b}{\vee}%
\put(4,3){\vector(1,-3){0}}%
\put(4,2){\vector(1,3){0}}%
\end{picture}}

\mcm{\cthreedbl}{5}{%
\cinitdims{4.2}{5.4}%
\abovepic{#1}%
\belowpic{#3}%
\present{\precthreedbl{#1}{#2}{#3}{#4}{#5}}}

\newcommand{\precthreecelltrp}[5]{%
\begin{picture}(8.2,5)(-4.1,-2.5)%
\cell{0}{2.5}{b}{#1}%
\cell{0}{-2.5}{t}{#2}%
\cell{-1.8}{0}{r}{#3}%
\cell{1.8}{0}{l}{#4}%
\cell{0}{0.3}{b}{#5}%
\qbezier(-4,0.5)(0,4)(4,0.5)%
\qbezier(-4,-0.5)(0,-4)(4,-0.5)%
\qbezier(-0.6,2)(-2.6,0)(-0.6,-2)%
\qbezier(-0.4,2)(-2.4,0)(-0.5,-1.9)%
\cell{-0.6}{-2}{b}{\lrcorner}%
\qbezier(0.4,2)(2.4,0)(0.5,-1.9)%
\qbezier(0.6,2)(2.6,0)(0.6,-2)%
\cell{0.65}{-2}{b}{\llcorner}%
\put(-1,0.15){\line(1,0){1.7}}%
\put(-1,0){\line(1,0){2}}%
\put(-1,-0.15){\line(1,0){1.7}}%
\cell{1.15}{0}{r}{>}%
\put(4,0.5){\vector(1,-1){0}}%
\put(4,-0.5){\vector(1,1){0}}%
\end{picture}}

\mcm{\cthreecelltrp}{5}{%
\cinitdims{8.2}{5}%
\abovepic{#1}%
\belowpic{#2}%
\present{\precthreecelltrp{#1}{#2}{#3}{#4}{#5}}}

\newcommand{\prectwo}[3]%
{\begin{picture}(4.2,3.4)(-0.1,-0.2)%
\cell{2}{3.2}{b}{#1}%
\cell{2}{-0.2}{t}{#2}%
\cell{2.2}{1.5}{l}{#3}%
\qbezier(0,2)(2,4)(4,2)%
\qbezier(0,1)(2,-1)(4,1)%
\put(4,2){\vector(1,-1){0}}%
\put(4,1){\vector(1,1){0}}%
\put(2,2.5){\vector(0,-1){2}}%
\end{picture}}

\mcm{\ctwo}{3}{%
\cinitdims{4.2}{3.4}%
\abovepic{#1}%
\belowpic{#2}%
\present{\prectwo{#1}{#2}{#3}}}

\newcommand{\precthree}[5]{%
\begin{picture}(4.2,5.4)(-0.1,-0.2)%
\cell{2}{5.2}{b}{#1}%
\cell{1}{2.7}{b}{#2}%
\cell{2}{-.2}{t}{#3}%
\cell{2.2}{3.75}{l}{#4}%
\cell{2.2}{1.25}{l}{#5}%
\qbezier(0,3)(2,7)(4,3)%
\qbezier(0,2)(2,-2)(4,2)%
\put(0,2.5){\vector(1,0){4}}%
\put(2,4.5){\vector(0,-1){1.5}}%
\put(2,2){\vector(0,-1){1.5}}%
\put(4,3){\vector(1,-3){0}}%
\put(4,2){\vector(1,3){0}}%
\end{picture}}

\mcm{\cthree}{5}{%
\cinitdims{4.2}{5.4}%
\abovepic{#1}%
\belowpic{#3}%
\present{\precthree{#1}{#2}{#3}{#4}{#5}}}

\newcommand{\prectwoop}[3]%
{\begin{picture}(4.2,3.4)(-0.1,-0.2)%
\cell{2}{3.2}{b}{#1}%
\cell{2}{-0.2}{t}{#2}%
\cell{2.2}{1.5}{l}{#3}%
\qbezier(0,2)(2,4)(4,2)%
\qbezier(0,1)(2,-1)(4,1)%
\put(0,2){\vector(-1,-1){0}}%
\put(0,1){\vector(-1,1){0}}%
\put(2,2.5){\vector(0,-1){2}}%
\end{picture}}

\mcm{\ctwoop}{3}{%
\cinitdims{4.2}{3.4}%
\abovepic{#1}%
\belowpic{#2}%
\present{\prectwoop{#1}{#2}{#3}}}

\newcommand{\prectwopar}[4]{%
\begin{picture}(4.2,3.4)(-0.1,-0.2)%
\cell{2}{3.2}{b}{#1}%
\cell{2}{-0.2}{t}{#2}%
\cell{1.6}{1.5}{r}{#3}%
\cell{2.4}{1.5}{l}{#4}%
\qbezier(0,2)(2,4)(4,2)%
\qbezier(0,1)(2,-1)(4,1)%
\put(4,2){\vector(1,-1){0}}%
\put(4,1){\vector(1,1){0}}%
\put(1.8,2.5){\vector(0,-1){2}}%
\put(2.2,2.5){\vector(0,-1){2}}%
\end{picture}}

\mcm{\ctwopar}{4}{%
\cinitdims{4.2}{3.4}%
\abovepic{#1}%
\belowpic{#2}%
\present{\prectwopar{#1}{#2}{#3}{#4}}}

\newcommand{\precthreein}[5]{%
\begin{picture}(4.2,5.4)(-0.1,-0.2)%
\cell{2}{5.2}{b}{#1}%
\cell{1}{2.7}{b}{#2}%
\cell{2}{-.2}{t}{#3}%
\cell{2.2}{3.75}{l}{#4}%
\cell{2.2}{1.25}{l}{#5}%
\qbezier(0,3)(2,7)(4,3)%
\qbezier(0,2)(2,-2)(4,2)%
\put(0,2.5){\vector(1,0){4}}%
\put(2,4.5){\vector(0,-1){1.5}}%
\put(2,0.5){\vector(0,1){1.5}}%
\put(4,3){\vector(1,-3){0}}%
\put(4,2){\vector(1,3){0}}%
\end{picture}}

\mcm{\cthreein}{5}{%
\cinitdims{4.2}{5.4}%
\abovepic{#1}%
\belowpic{#3}%
\present{\precthreein{#1}{#2}{#3}{#4}{#5}}}

\newcommand{\precthreecell}[5]{%
\begin{picture}(8.2,5)(-4.1,-2.5)%
\cell{0}{2.5}{b}{#1}%
\cell{0}{-2.5}{t}{#2}%
\cell{-1.7}{0}{r}{#3}%
\cell{1.7}{0}{l}{#4}%
\cell{0}{0.2}{b}{#5}%
\qbezier(-4,0.5)(0,4)(4,0.5)%
\qbezier(-4,-0.5)(0,-4)(4,-0.5)%
\qbezier(-0.5,2)(-2.5,0)(-0.5,-2)%
\qbezier(0.5,2)(2.5,0)(0.5,-2)%
\put(-1,0){\vector(1,0){2}}%
\put(4,0.5){\vector(1,-1){0}}%
\put(4,-0.5){\vector(1,1){0}}%
\put(-0.5,-2){\vector(1,-1){0}}%
\put(0.5,-2){\vector(-1,-1){0}}%
\end{picture}}

\mcm{\cthreecell}{5}{%
\cinitdims{8.2}{5}%
\abovepic{#1}%
\belowpic{#2}%
\present{\precthreecell{#1}{#2}{#3}{#4}{#5}}}

\newcommand{\precthreecellpar}[6]{%
\begin{picture}(8.2,5)(-4.1,-2.5)%
\cell{0}{2.5}{b}{#1}%
\cell{0}{-2.5}{t}{#2}%
\cell{-1.7}{0}{r}{#3}%
\cell{1.7}{0}{l}{#4}%
\cell{0}{0.4}{b}{#5}%
\cell{0}{-0.4}{t}{#6}%
\qbezier(-4,0.5)(0,4)(4,0.5)%
\qbezier(-4,-0.5)(0,-4)(4,-0.5)%
\qbezier(-0.5,2)(-2.5,0)(-0.5,-2)%
\qbezier(0.5,2)(2.5,0)(0.5,-2)%
\put(-1,0.2){\vector(1,0){2}}%
\put(-1,-0.2){\vector(1,0){2}}%
\put(4,0.5){\vector(1,-1){0}}%
\put(4,-0.5){\vector(1,1){0}}%
\put(-0.5,-2){\vector(1,-1){0}}%
\put(0.5,-2){\vector(-1,-1){0}}%
\end{picture}}

\mcm{\cthreecellpar}{6}{%
\cinitdims{8.2}{5}%
\abovepic{#1}%
\belowpic{#2}%
\present{\precthreecellpar{#1}{#2}{#3}{#4}{#5}{#6}}}

\newcommand{\prectwov}[5]{%
\begin{picture}(3.4,4.2)(0.8,0.9)%
\cell{2.5}{5.1}{b}{#1}%
\cell{2.5}{0.9}{t}{#2}%
\cell{0.8}{3}{r}{#3}%
\cell{4.2}{3}{l}{#4}%
\cell{2.5}{3.2}{b}{#5}%
\qbezier(2,5)(0,3)(2,1)%
\qbezier(3,5)(5,3)(3,1)%
\put(2,1){\vector(1,-1){0}}%
\put(3,1){\vector(-1,-1){0}}%
\put(1.5,3){\vector(1,0){2}}%
\end{picture}}

\mcm{\ctwov}{5}{%
\cinitdims{3.4}{4.2}%
\abovepic{#1}%
\belowpic{#2}%
\sidespic{#3}%
\sidespic{#4}%
\present{\prectwov{#1}{#2}{#3}{#4}{#5}}}

\newcommand{\precthreecellv}[7]{%
\begin{picture}(5,8.2)(0.5,-1.6)%
\cell{3}{6.6}{b}{#1}%
\cell{3}{-1.6}{t}{#2}%
\cell{0.5}{2.5}{r}{#3}%
\cell{5.5}{2.5}{l}{#4}%
\cell{3}{4.2}{b}{#5}%
\cell{3}{0.8}{t}{#6}%
\cell{3.2}{2.5}{l}{#7}%
\qbezier(3.5,6.5)(7,2.5)(3.5,-1.5)%
\qbezier(2.5,6.5)(-1,2.5)(2.5,-1.5)%
\put(2.5,-1.5){\vector(1,-1){0}}%
\put(3.5,-1.5){\vector(-1,-1){0}}%
\qbezier(1,3)(3,5)(5,3)%
\qbezier(1,2)(3,0)(5,2)%
\put(5,3){\vector(1,-1){0}}%
\put(5,2){\vector(1,1){0}}%
\put(3,3.5){\vector(0,-1){2}}%
\end{picture}}

\mcm{\cthreecellv}{7}{%
\cinitdims{5}{8.2}%
\abovepic{#1}%
\belowpic{#2}%
\sidespic{#3}%
\sidespic{#4}%
\present{\precthreecellv{#1}{#2}{#3}{#4}{#5}{#6}{#7}}}

\newcommand{\pretopez}[2]{%
\begin{picture}(2.6,2.3)(-1.3,-2.2)%
\cell{0}{-2.2}{t}{#1}%
\cell{0}{-1.2}{c}{#2}%
\qbezier(0,0)(-2,-2)(0,-2)%
\qbezier(0,0)(2,-2)(0,-2)%
\put(0,0){\vector(-1,1){0}}%
\end{picture}}

\mcm{\topez}{2}{%
\ginitdims{2.6}{2.3}%
\belowpic{#1}%
\present{\pretopez{#1}{#2}}}

\newcommand{\pretopea}[3]{%
\begin{picture}(4,1.9)(-2,-0,2)%
\cell{0}{1.7}{b}{#1}%
\cell{0}{-0.2}{t}{#2}%
\cell{0}{0.7}{c}{#3}%
\qbezier(-2,0)(0,3)(2,0)%
\put(-2,0){\vector(1,0){4}}%
\put(2,0){\vector(2,-3){0}}%
\end{picture}}

\mcm{\topea}{3}{%
\ginitdims{4}{1.9}%
\abovepic{#1}%
\belowpic{#2}%
\present{\pretopea{#1}{#2}{#3}}}

\newcommand{\pretopeb}[4]{%
\begin{picture}(4,2.2)(-2,-0.2)%
\cell{-1.1}{1}{br}{#1}%
\cell{1.1}{1}{bl}{#2}%
\cell{0}{-0.2}{t}{#3}%
\cell{0}{0.8}{c}{#4}%
\put(-2,0){\vector(1,1){2}}%
\put(0,2){\vector(1,-1){2}}%
\put(-2,0){\vector(1,0){4}}%
\end{picture}}

\mcm{\topeb}{4}{%
\ginitdims{4}{2.2}%
\belowpic{#3}%
\present{\pretopeb{#1}{#2}{#3}{#4}}}

\newcommand{\pretopec}[5]{%
\begin{picture}(4,2.2)(-2,-0.2)%
\cell{-1.8}{1}{br}{#1}%
\cell{0}{2.2}{b}{#2}%
\cell{1.8}{1}{bl}{#3}%
\cell{0}{-0.2}{t}{#4}%
\cell{0}{0.8}{c}{#5}%
\put(-2,0){\vector(1,2){1}}%
\put(-1,2){\vector(1,0){2}}%
\put(1,2){\vector(1,-2){1}}%
\put(-2,0){\vector(1,0){4}}%
\end{picture}}

\mcm{\topec}{5}{%
\ginitdims{4}{2.2}%
\sidespic{#1}%
\abovepic{#2}%
\sidespic{#3}%
\belowpic{#4}%
\present{\pretopec{#1}{#2}{#3}{#4}{#5}}}

\newcommand{\pretoped}[6]{%
\begin{picture}(4,2.5)(-2,-0.2)%
\cell{-2}{0.6}{br}{#1}%
\cell{-0.7}{2.2}{br}{#2}%
\cell{0.7}{2.2}{bl}{#3}%
\cell{2}{0.6}{bl}{#4}%
\cell{0}{-0.2}{t}{#5}%
\cell{0}{0.8}{c}{#6}%
\put(-2,0){\vector(1,3){0.5}}%
\put(-1.5,1.5){\vector(3,2){1.5}}%
\put(0,2.5){\vector(3,-2){1.5}}%
\put(1.5,1.5){\vector(1,-3){0.5}}%
\put(-2,0){\vector(1,0){4}}%
\end{picture}}

\mcm{\toped}{6}{%
\ginitdims{4}{2.5}%
\sidespic{#1}%
\abovepic{#2}%
\abovepic{#3}%
\sidespic{#4}%
\belowpic{#5}%
\present{\pretoped{#1}{#2}{#3}{#4}{#5}{#6}}}

\newcommand{\pretopeq}[5]{%
\begin{picture}(4,2.5)(-2,-0.2)%
\cell{-2}{0.6}{br}{#1}%
\cell{-1}{2.2}{br}{#2}%
\cell{2}{0.6}{bl}{#3}%
\cell{0}{-0.2}{t}{#4}%
\cell{0}{0.8}{c}{#5}%
\put(-2,0){\vector(1,3){0.5}}%
\put(-1.5,1.5){\vector(1,1){1}}%
\cell{0.9}{2.3}{c}{\ddots}
\put(1.5,1.5){\vector(1,-3){0.5}}%
\put(-2,0){\vector(1,0){4}}%
\end{picture}}

\mcm{\topeq}{5}{%
\ginitdims{4}{2.5}%
\sidespic{#1}%
\abovepic{#2}%
\sidespic{#3}%
\belowpic{#4}%
\present{\pretopeq{#1}{#2}{#3}{#4}{#5}}}

\newcommand{\pretopebase}[1]{%
\begin{picture}(4,0.4)(0,-0.2)%
\cell{2}{0.2}{b}{#1}%
\put(0,0){\vector(1,0){4}}%
\end{picture}}

\mcm{\topebase}{1}{%
\ginitdims{4}{0.4}%
\abovepic{#1}%
\present{\pretopebase{#1}}}

\newcommand{\pretopezs}[2]{%
\begin{picture}(2.6,2.3)(-1.3,-2.2)%
\cell{0}{-2.2}{t}{#1}%
\cell{0}{-1.2}{c}{#2}%
\qbezier(0,0)(-2,-2)(0,-2)%
\qbezier(0,0)(2,-2)(0,-2)%
\end{picture}}

\mcm{\topezs}{2}{%
\ginitdims{2.6}{2.3}%
\belowpic{#1}%
\present{\pretopezs{#1}{#2}}}

\newcommand{\pretopeas}[3]{%
\begin{picture}(4,1.9)(-2,-0,2)%
\cell{0}{1.7}{b}{#1}%
\cell{0}{-0.2}{t}{#2}%
\cell{0}{0.7}{c}{#3}%
\qbezier(-2,0)(0,3)(2,0)%
\put(-2,0){\line(1,0){4}}%
\end{picture}}

\mcm{\topeas}{3}{%
\ginitdims{4}{1.9}%
\abovepic{#1}%
\belowpic{#2}%
\present{\pretopeas{#1}{#2}{#3}}}

\newcommand{\pretopebs}[4]{%
\begin{picture}(4,2.2)(-2,-0.2)%
\cell{-1.1}{1}{br}{#1}%
\cell{1.1}{1}{bl}{#2}%
\cell{0}{-0.2}{t}{#3}%
\cell{0}{0.8}{c}{#4}%
\put(-2,0){\line(1,1){2}}%
\put(0,2){\line(1,-1){2}}%
\put(-2,0){\line(1,0){4}}%
\end{picture}}

\mcm{\topebs}{4}{%
\ginitdims{4}{2.2}%
\belowpic{#3}%
\present{\pretopebs{#1}{#2}{#3}{#4}}}

\newcommand{\pretopecs}[5]{%
\begin{picture}(4,2.2)(-2,-0.2)%
\cell{-1.8}{1}{br}{#1}%
\cell{0}{2.2}{b}{#2}%
\cell{1.8}{1}{bl}{#3}%
\cell{0}{-0.2}{t}{#4}%
\cell{0}{0.8}{c}{#5}%
\put(-2,0){\line(1,2){1}}%
\put(-1,2){\line(1,0){2}}%
\put(1,2){\line(1,-2){1}}%
\put(-2,0){\line(1,0){4}}%
\end{picture}}

\mcm{\topecs}{5}{%
\ginitdims{4}{2.2}%
\sidespic{#1}%
\abovepic{#2}%
\sidespic{#3}%
\belowpic{#4}%
\present{\pretopecs{#1}{#2}{#3}{#4}{#5}}}

\newcommand{\pretopeds}[6]{%
\begin{picture}(4,2.5)(-2,-0.2)%
\cell{-2}{0.6}{br}{#1}%
\cell{-0.7}{2.2}{br}{#2}%
\cell{0.7}{2.2}{bl}{#3}%
\cell{2}{0.6}{bl}{#4}%
\cell{0}{-0.2}{t}{#5}%
\cell{0}{0.8}{c}{#6}%
\put(-2,0){\line(1,3){0.5}}%
\put(-1.5,1.5){\line(3,2){1.5}}%
\put(0,2.5){\line(3,-2){1.5}}%
\put(1.5,1.5){\line(1,-3){0.5}}%
\put(-2,0){\line(1,0){4}}%
\end{picture}}

\mcm{\topeds}{6}{%
\ginitdims{4}{2.5}%
\sidespic{#1}%
\abovepic{#2}%
\abovepic{#3}%
\sidespic{#4}%
\belowpic{#5}%
\present{\pretopeds{#1}{#2}{#3}{#4}{#5}{#6}}}

\newcommand{\pretopeqs}[5]{%
\begin{picture}(4,2.5)(-2,-0.2)%
\cell{-2}{0.6}{br}{#1}%
\cell{-1}{2.2}{br}{#2}%
\cell{2}{0.6}{bl}{#3}%
\cell{0}{-0.2}{t}{#4}%
\cell{0}{0.8}{c}{#5}%
\put(-2,0){\line(1,3){0.5}}%
\put(-1.5,1.5){\line(1,1){1}}%
\cell{0.9}{2.3}{c}{\ddots}
\put(1.5,1.5){\line(1,-3){0.5}}%
\put(-2,0){\line(1,0){4}}%
\end{picture}}

\mcm{\topeqs}{5}{%
\ginitdims{4}{2.5}%
\sidespic{#1}%
\abovepic{#2}%
\sidespic{#3}%
\belowpic{#4}%
\present{\pretopeqs{#1}{#2}{#3}{#4}{#5}}}

\newcommand{\pretopebases}[1]{%
\begin{picture}(4,0.4)(0,-0.2)%
\cell{2}{0.2}{b}{#1}%
\put(0,0){\line(1,0){4}}%
\end{picture}}

\mcm{\topebases}{1}{%
\ginitdims{4}{0.4}%
\abovepic{#1}%
\present{\pretopebases{#1}}}

\newcommand{\pregdots}[6]{%
\begin{picture}(5,8.4)(0,-2.7)%
\cell{2.5}{5.7}{b}{#1}%
\cell{1.5}{2.8}{b}{#2}%
\cell{1.5}{0.2}{t}{#3}%
\cell{2.5}{-2.7}{t}{#4}%
\cell{2.7}{4.25}{l}{#5}%
\cell{2.7}{-1.25}{l}{#6}%
\qbezier(0,1.5)(2.5,9.5)(5,1.5)%
\qbezier(0,1.5)(2.5,4)(5,1.5)%
\qbezier(0,1.5)(2.5,-1)(5,1.5)%
\qbezier(0,1.5)(2.5,-6.5)(5,1.5)%
\put(2.5,5.25){\vector(0,-1){2}}%
\put(2.5,-0.25){\vector(0,-1){2}}%
\cell{2.5}{1.7}{c}{\vdots}%
\put(5,1.5){\vector(1,-4){0}}%
\put(5,1.5){\vector(4,-3){0}}%
\put(5,1.5){\vector(4,3){0}}%
\put(5,1.5){\vector(1,4){0}}%
\end{picture}}

\mcm{\gdots}{6}{%
\ginitdims{5}{8.4}%
\abovepic{#1}%
\belowpic{#4}%
\present{\pregdots{#1}{#2}{#3}{#4}{#5}{#6}}}

\newlength{\volt}
\makeatletter

\def\diagram{\m@th\leftwidth=\z@ \rightwidth=\z@ \topheight=\z@
\botheight=\z@ \setbox\@picbox\hbox\bgroup}

\def\enddiagram{\egroup\wd\@picbox\rightwidth\unitlength
\ht\@picbox\topheight\unitlength \dp\@picbox\botheight\unitlength
\hskip\leftwidth\unitlength\box\@picbox}

\def\bfig{\begin{diagram}}
\def\efig{\end{diagram}}
\newcount\wideness \newcount\leftwidth \newcount\rightwidth
\newcount\highness \newcount\topheight \newcount\botheight

\def\ratchet#1#2{\ifnum#1<#2 \global #1=#2 \fi}

\def\putbox(#1,#2)#3{%
\horsize{\wideness}{#3} \divide\wideness by 2 {\advance\wideness
by #1 \ratchet{\rightwidth}{\wideness}} {\advance\wideness by -#1
\ratchet{\leftwidth}{\wideness}} \vertsize{\highness}{#3}
\divide\highness by 2 {\advance\highness by #2
\ratchet{\topheight}{\highness}} {\advance\highness by -#2
\ratchet{\botheight}{\highness}} \put(#1,#2){\makebox(0,0){$#3$}}}

\def\putlbox(#1,#2)#3{%
\horsize{\wideness}{#3} {\advance\wideness by #1
\ratchet{\rightwidth}{\wideness}} {\ratchet{\leftwidth}{-#1}}
\vertsize{\highness}{#3} \divide\highness by 2 {\advance\highness
by #2 \ratchet{\topheight}{\highness}} {\advance\highness by -#2
\ratchet{\botheight}{\highness}}
\put(#1,#2){\makebox(0,0)[l]{$#3$}}}

\def\putrbox(#1,#2)#3{%
\horsize{\wideness}{#3} {\ratchet{\rightwidth}{#1}}
{\advance\wideness by -#1 \ratchet{\leftwidth}{\wideness}}
\vertsize{\highness}{#3} \divide\highness by 2 {\advance\highness
by #2 \ratchet{\topheight}{\highness}} {\advance\highness by -#2
\ratchet{\botheight}{\highness}}
\put(#1,#2){\makebox(0,0)[r]{$#3$}}}

\def\adjust[#1]{} % For compatibility

\newcount \coefa
\newcount \coefb
\newcount \coefc
\newcount\tempcounta
\newcount\tempcountb
\newcount\tempcountc
\newcount\tempcountd
\newcount\xext
\newcount\yext
\newcount\xoff
\newcount\yoff
\newcount\gap%
\newcount\arrowtypea
\newcount\arrowtypeb
\newcount\arrowtypec
\newcount\arrowtyped
\newcount\arrowtypee
\newcount\height
\newcount\width
\newcount\xpos
\newcount\ypos
\newcount\run
\newcount\rise
\newcount\arrowlength
\newcount\halflength
\newcount\arrowtype
\newdimen\tempdimen
\newdimen\xlen
\newdimen\ylen
\newsavebox{\tempboxa}%
\newsavebox{\tempboxb}%
\newsavebox{\tempboxc}%

\newdimen\w@dth

\def\setw@dth#1#2{\setbox\z@\hbox{\m@th$#1$}\w@dth=\wd\z@
\setbox\@ne\hbox{\m@th$#2$}\ifnum\w@dth<\wd\@ne \w@dth=\wd\@ne \fi
\advance\w@dth by 1.2em}

\def\t@^#1_#2{\allowbreak\def\n@one{#1}\def\n@two{#2}\mathrel
{\setw@dth{#1}{#2} \mathop{\hbox to
\w@dth{\rightarrowfill}}\limits \ifx\n@one\empty\else
^{\box\z@}\fi \ifx\n@two\empty\else _{\box\@ne}\fi}}
\def\t@@^#1{\@ifnextchar_{\t@^{#1}}{\t@^{#1}_{}}}
\def\to{\@ifnextchar^{\t@@}{\t@@^{}}}

\def\t@left^#1_#2{\def\n@one{#1}\def\n@two{#2}\mathrel{\setw@dth{#1}{#2}
\mathop{\hbox to \w@dth{\leftarrowfill}}\limits
\ifx\n@one\empty\else ^{\box\z@}\fi \ifx\n@two\empty\else
_{\box\@ne}\fi}}
\def\t@@left^#1{\@ifnextchar_{\t@left^{#1}}{\t@left^{#1}_{}}}
\def\toleft{\@ifnextchar^{\t@@left}{\t@@left^{}}}

\def\two@^#1_#2{\allowbreak
\def\n@one{#1}\def\n@two{#2}\mathrel{\setw@dth{#1}{#2}
\mathop{\vcenter{\lineskip\z@\baselineskip\z@
                 \hbox to \w@dth{\rightarrowfill}%
                 \hbox to \w@dth{\rightarrowfill}}%
       }\limits
\ifx\n@one\empty\else ^{\box\z@}\fi \ifx\n@two\empty\else
_{\box\@ne}\fi}}
\def\tw@@^#1{\@ifnextchar _{\two@^{#1}}{\two@^{#1}_{}}}
\def\two{\@ifnextchar ^{\tw@@}{\tw@@^{}}}

\def\tofr@^#1_#2{\def\n@one{#1}\def\n@two{#2}\mathrel{\setw@dth{#1}{#2}
\mathop{\vcenter{\hbox to \w@dth{\rightarrowfill}\kern-1.7ex
                 \hbox to \w@dth{\leftarrowfill}}%
       }\limits
\ifx\n@one\empty\else ^{\box\z@}\fi \ifx\n@two\empty\else
_{\box\@ne}\fi}}
\def\t@fr@^#1{\@ifnextchar_ {\tofr@^{#1}}{\tofr@^{#1}_{}}}
\def\tofro{\@ifnextchar^ {\t@fr@}{\t@fr@^{}}}

\def\mon{\mathop{\m@th\hbox to
      14.6\P@{\lasyb\char'51\hskip-2.1\P@$\arrext$\hss
$\mathord\rightarrow$}}\limits} % width of \epi
\def\leftmono{\mathrel{\m@th\hbox to
14.6\P@{$\mathord\leftarrow$\hss$\arrext$\hskip-2.1\P@\lasyb\char'50%
}}\limits} % width of \epi
\mathchardef\arrext="0200       % amr minus for arrow extension (see \into)

\def\settypes(#1,#2,#3){\arrowtypea#1 \arrowtypeb#2 \arrowtypec#3}
\def\settoheight#1#2{\setbox\@tempboxa\hbox{#2}#1\ht\@tempboxa\relax}%
\def\settodepth#1#2{\setbox\@tempboxa\hbox{#2}#1\dp\@tempboxa\relax}%
\def\settokens`#1`#2`#3`#4`{%
     \def\tokena{#1}\def\tokenb{#2}\def\tokenc{#3}\def\tokend{#4}}
\def\setsqparms[#1`#2`#3`#4;#5`#6]{%
\arrowtypea #1 \arrowtypeb #2 \arrowtypec #3 \arrowtyped #4
\width #5 \height #6 }
\def\setpos(#1,#2){\xpos=#1 \ypos#2}

\def\settriparms[#1`#2`#3;#4]{\settripairparms[#1`#2`#3`1`1;#4]}%

\def\settripairparms[#1`#2`#3`#4`#5;#6]{%
\arrowtypea #1 \arrowtypeb #2 \arrowtypec #3 \arrowtyped #4
\arrowtypee #5 \width #6 \height #6 }

\def\resetparms{\settripairparms[1`1`1`1`1;500]\width 500}%default values%

\resetparms

\def\mvector(#1,#2)#3{%%
\put(0,0){\vector(#1,#2){#3}}%
\put(0,0){\vector(#1,#2){26}}%
}
\def\evector(#1,#2)#3{{%%
\arrowlength #3
\put(0,0){\vector(#1,#2){\arrowlength}}%
\advance \arrowlength by-30
\put(0,0){\vector(#1,#2){\arrowlength}}%
}}

\def\horsize#1#2{%
\settowidth{\tempdimen}{$#2$}%
#1=\tempdimen \divide #1 by\unitlength }

\def\vertsize#1#2{%
\settoheight{\tempdimen}{$#2$}%
#1=\tempdimen
\settodepth{\tempdimen}{$#2$}%
\advance #1 by\tempdimen \divide #1 by\unitlength }

\def\putvector(#1,#2)(#3,#4)#5#6{{%
\ifnum3<\arrowtype \putdashvector(#1,#2)(#3,#4)#5\arrowtype \else
\ifnum\arrowtype<-3 \putdashvector(#1,#2)(#3,#4)#5\arrowtype \else
\xpos=#1 \ypos=#2 \run=#3 \rise=#4 \arrowlength=#5 \ifnum
\arrowtype<0
    \ifnum \run=0
        \advance \ypos by-\arrowlength
    \else
        \tempcounta \arrowlength
        \multiply \tempcounta by\rise
        \divide \tempcounta by\run
        \ifnum\run>0
            \advance \xpos by\arrowlength
            \advance \ypos by\tempcounta
        \else
            \advance \xpos by-\arrowlength
            \advance \ypos by-\tempcounta
        \fi
    \fi
    \multiply \arrowtype by-1
    \multiply \rise by-1
    \multiply \run by-1
\fi \ifcase \arrowtype
\or \put(\xpos,\ypos){\vector(\run,\rise){\arrowlength}}%
\or \put(\xpos,\ypos){\mvector(\run,\rise)\arrowlength}%
\or \put(\xpos,\ypos){\evector(\run,\rise){\arrowlength}}%
\fi\fi\fi }}

\def\putsplitvector(#1,#2)#3#4{%%
\xpos #1 \ypos #2 \arrowtype #4 \halflength #3 \arrowlength #3
\gap 140 \advance \halflength by-\gap \divide \halflength by2
\ifnum\arrowtype>0
   \ifcase \arrowtype
   \or \put(\xpos,\ypos){\line(0,-1){\halflength}}%
       \advance\ypos by-\halflength
       \advance\ypos by-\gap
       \put(\xpos,\ypos){\vector(0,-1){\halflength}}%
   \or \put(\xpos,\ypos){\line(0,-1)\halflength}%
       \put(\xpos,\ypos){\vector(0,-1)3}%
       \advance\ypos by-\halflength
       \advance\ypos by-\gap
       \put(\xpos,\ypos){\vector(0,-1){\halflength}}%
   \or \put(\xpos,\ypos){\line(0,-1)\halflength}%
       \advance\ypos by-\halflength
       \advance\ypos by-\gap
       \put(\xpos,\ypos){\evector(0,-1){\halflength}}%
   \fi
\else \arrowtype=-\arrowtype
   \ifcase\arrowtype
   \or \advance \ypos by-\arrowlength
       \put(\xpos,\ypos){\line(0,1){\halflength}}%
       \advance\ypos by\halflength
       \advance\ypos by\gap
       \put(\xpos,\ypos){\vector(0,1){\halflength}}%
   \or \advance \ypos by-\arrowlength
       \put(\xpos,\ypos){\line(0,1)\halflength}%
       \put(\xpos,\ypos){\vector(0,1)3}%
       \advance\ypos by\halflength
       \advance\ypos by\gap
       \put(\xpos,\ypos){\vector(0,1){\halflength}}%
   \or \advance \ypos by-\arrowlength
       \put(\xpos,\ypos){\line(0,1)\halflength}%
       \advance\ypos by\halflength
       \advance\ypos by\gap
       \put(\xpos,\ypos){\evector(0,1){\halflength}}%
   \fi
\fi }

\def\putmorphism(#1)(#2,#3)[#4`#5`#6]#7#8#9{{%
\run #2 \rise #3 \ifnum\rise=0
  \puthmorphism(#1)[#4`#5`#6]{#7}{#8}#9%
\else\ifnum\run=0
  \putvmorphism(#1)[#4`#5`#6]{#7}{#8}#9%
\else
\setpos(#1)%
\arrowlength #7 \arrowtype #8 \ifnum\run=0 \else\ifnum\rise=0
\else \ifnum\run>0
    \coefa=1
\else
   \coefa=-1
\fi \ifnum\arrowtype>0
   \coefb=0
   \coefc=-1
\else
   \coefb=\coefa
   \coefc=1
   \arrowtype=-\arrowtype
\fi \width=2 \multiply \width by\run \divide \width by\rise
\ifnum \width<0  \width=-\width\fi \advance\width by60 \if l#9
\width=-\width\fi
\putbox(\xpos,\ypos){#4}%            %node 1
{\multiply \coefa by\arrowlength%      %node 2
\advance\xpos by\coefa \multiply \coefa by\rise \divide \coefa
by\run \advance \ypos by\coefa
\putbox(\xpos,\ypos){#5} }%
{\multiply \coefa by\arrowlength%      %label
\divide \coefa by2 \advance \xpos by\coefa \advance \xpos by\width
\multiply \coefa by\rise \divide \coefa by\run \advance \ypos
by\coefa
\if l#9%
   \putrbox(\xpos,\ypos){#6}%
\else\if r#9%
   \putlbox(\xpos,\ypos){#6}%
\fi\fi }%
{\multiply \rise by-\coefc%             %arrow
\multiply \run by-\coefc \multiply \coefb by\arrowlength \advance
\xpos by\coefb \multiply \coefb by\rise \divide \coefb by\run
\advance \ypos by\coefb \multiply \coefc by70 \advance \ypos
by\coefc \multiply \coefc by\run \divide \coefc by\rise \advance
\xpos by\coefc \multiply \coefa by140 \multiply \coefa by\run
\divide \coefa by\rise \advance \arrowlength by\coefa
\ifcase\arrowtype
\or \put(\xpos,\ypos){\vector(\run,\rise){\arrowlength}}%
\or \put(\xpos,\ypos){\mvector(\run,\rise){\arrowlength}}%
\or \put(\xpos,\ypos){\evector(\run,\rise){\arrowlength}}%
\fi}\fi\fi\fi\fi}}

\newcount\numbdashes \newcount\lengthdash \newcount\increment

\def\howmanydashes{% Actually returns both number and length
\numbdashes=\arrowlength \lengthdash=40 \divide\numbdashes by
\lengthdash \lengthdash=\arrowlength \divide\lengthdash by
\numbdashes
\increment=\lengthdash \multiply\lengthdash by 3
\divide\lengthdash by 5 }

\def\putdashvector(#1)(#2,#3)#4#5{%
\ifnum#3=0 \putdashhvector(#1){#4}#5 \else \ifnum#2=0
\putdashvvector(#1){#4}#5\fi\fi}

\def\putdashhvector(#1,#2)#3#4{{%
\arrowlength=#3 \howmanydashes
\multiput(#1,#2)(\increment,0){\numbdashes}%
{\vrule height .4pt width \lengthdash\unitlength} \arrowtype=#4
\xpos=#1 \ifnum\arrowtype<0 \advance\arrowtype by 7 \fi
\ifcase\arrowtype \or \advance\xpos by 10
    \put(\xpos,#2){\vector(-1,0){\lengthdash}}
    \advance\xpos by 40
    \put(\xpos,#2){\vector(-1,0){\lengthdash}}
\or \advance \xpos by 10
    \put(\xpos,#2){\vector(-1,0){\lengthdash}}
    \advance\xpos by  \arrowlength
    \advance\xpos by  -50
    \put(\xpos,#2){\vector(-1,0){\lengthdash}}
\or \advance\xpos by 10
    \put(\xpos,#2){\vector(-1,0){\lengthdash}}
\or \advance\xpos by \arrowlength
    \advance\xpos by -\lengthdash
    \put(\xpos,#2){\vector(1,0){\lengthdash}}
\or {\advance\xpos by 10
    \put(\xpos,#2){\vector(1,0){\lengthdash}}}
    \advance\xpos by \arrowlength
    \advance\xpos by -\lengthdash
    \put(\xpos,#2){\vector(1,0){\lengthdash}}
\or \advance\xpos by \arrowlength
    \advance\xpos by -\lengthdash
    \put(\xpos,#2){\vector(1,0){\lengthdash}}
    \advance\xpos by -40
    \put(\xpos,#2){\vector(1,0){\lengthdash}}
   \fi
}}

\def\putdashvvector(#1,#2)#3#4{{%
\arrowlength=#3 \howmanydashes \ypos=#2 \advance\ypos by
-\arrowlength
\multiput(#1,#2)(0,\increment){\numbdashes}%
    {\vrule width .4pt height \lengthdash\unitlength}
\arrowtype=#4 \ypos=#2 \ifnum\arrowtype<0 \advance\arrowtype by 7
\fi \ifcase\arrowtype \or \advance\ypos by \arrowlength
\advance\ypos by -40
    \put(#1,\ypos){\vector(0,1){\lengthdash}}
    \advance\ypos by -40
    \put(#1,\ypos){\vector(0,1){\lengthdash}}
\or \advance\ypos by 10
    \put(#1,\ypos){\vector(0,1){\lengthdash}}
    \advance\ypos by \arrowlength \advance\ypos by -40
    \put(#1,\ypos){\vector(0,1){\lengthdash}}
\or \advance\ypos by \arrowlength \advance\ypos by -40
    \put(#1,\ypos){\vector(0,1){\lengthdash}}
\or \advance\ypos by 10
    \put(#1,\ypos){\vector(0,-1){\lengthdash}}
\or \advance\ypos by 10
    \put(#1,\ypos){\vector(0,-1){\lengthdash}}
    \advance\ypos by \arrowlength \advance\ypos by -40
    \put(#1,\ypos){\vector(0,-1){\lengthdash}}
\or \advance\ypos by 10
    \put(#1,\ypos){\vector(0,-1){\lengthdash}}
    \advance\ypos by 40
    \put(#1,\ypos){\vector(0,-1){\lengthdash}}
\fi }}

\def\puthmorphism(#1,#2)[#3`#4`#5]#6#7#8{{%
\xpos #1 \ypos #2 \width #6 \arrowlength #6 \arrowtype=#7
\putbox(\xpos,\ypos){#3\vphantom{#4}}%
{\advance \xpos by\arrowlength
\putbox(\xpos,\ypos){\vphantom{#3}#4}}%
\horsize{\tempcounta}{#3}%
\horsize{\tempcountb}{#4}%
\divide \tempcounta by2 \divide \tempcountb by2 \advance
\tempcounta by30 \advance \tempcountb by30 \advance \xpos
by\tempcounta \advance \arrowlength by-\tempcounta \advance
\arrowlength by-\tempcountb
\putvector(\xpos,\ypos)(1,0)\arrowlength\arrowtype \divide
\arrowlength by2 \advance \xpos by\arrowlength
\vertsize{\tempcounta}{#5}%
\divide\tempcounta by2 \advance \tempcounta by20
\if a#8 %
   \advance \ypos by\tempcounta
   \putbox(\xpos,\ypos){#5}%
\else
   \advance \ypos by-\tempcounta
   \putbox(\xpos,\ypos){#5}%
\fi}}

\def\putvmorphism(#1,#2)[#3`#4`#5]#6#7#8{{%
\xpos #1 \ypos #2 \arrowlength #6 \arrowtype #7
\settowidth{\xlen}{$#5$}%
\putbox(\xpos,\ypos){#3}%
{\advance \ypos by-\arrowlength
\putbox(\xpos,\ypos){#4}}%
{\advance\arrowlength by-140 \advance \ypos by-70 \ifdim\xlen>0pt
   \if m#8%
      \putsplitvector(\xpos,\ypos)\arrowlength\arrowtype
   \else
   \putvector(\xpos,\ypos)(0,-1)\arrowlength\arrowtype
   \fi
\else
   \putvector(\xpos,\ypos)(0,-1)\arrowlength\arrowtype
\fi}%
\ifdim\xlen>0pt
   \divide \arrowlength by2
   \advance\ypos by-\arrowlength
   \if l#8%
      \advance \xpos by-40
      \putrbox(\xpos,\ypos){#5}%
   \else\if r#8%
      \advance \xpos by40
      \putlbox(\xpos,\ypos){#5}%
   \else
      \putbox(\xpos,\ypos){#5}%
   \fi\fi
\fi }}

\def\putsquarep<#1>(#2)[#3;#4`#5`#6`#7]{{%
\setsqparms[#1]%
\setpos(#2)%
\settokens`#3`%
\puthmorphism(\xpos,\ypos)[\tokenc`\tokend`{#7}]{\width}{\arrowtyped}b%
\advance\ypos by \height
\puthmorphism(\xpos,\ypos)[\tokena`\tokenb`{#4}]{\width}{\arrowtypea}a%
\putvmorphism(\xpos,\ypos)[``{#5}]{\height}{\arrowtypeb}l%
\advance\xpos by \width
\putvmorphism(\xpos,\ypos)[``{#6}]{\height}{\arrowtypec}r%
}}

\def\putsquare{\@ifnextchar <{\putsquarep}{\putsquarep%
   <\arrowtypea`\arrowtypeb`\arrowtypec`\arrowtyped;\width`\height>}}
\def\square{\@ifnextchar< {\squarep}{\squarep
   <\arrowtypea`\arrowtypeb`\arrowtypec`\arrowtyped;\width`\height>}}
\def\squarep<#1>[#2`#3`#4`#5;#6`#7`#8`#9]{{%       %     #2------>#3
\setsqparms[#1]%                                   %      |       |
\diagram%                                          %      |       |
\putsquarep<\arrowtypea`\arrowtypeb`\arrowtypec`%  %    #7|       |#8
\arrowtyped;\width`\height>%                       %      |       |
(0,0)[#2`#3`#4`{#5};#6`#7`#8`{#9}]%                %      |       |
\enddiagram%                                       %      v       v
}}                                                 %     #4------>#5
\def\putptrianglep<#1>(#2,#3)[#4`#5`#6;#7`#8`#9]{{%
\settriparms[#1]%
\xpos=#2 \ypos=#3 \advance\ypos by \height
\puthmorphism(\xpos,\ypos)[#4`#5`{#7}]{\height}{\arrowtypea}a%
\putvmorphism(\xpos,\ypos)[`#6`{#8}]{\height}{\arrowtypeb}l%
\advance\xpos by\height
\putmorphism(\xpos,\ypos)(-1,-1)[``{#9}]{\height}{\arrowtypec}r%
}}

\def\putptriangle{\@ifnextchar <{\putptrianglep}{\putptrianglep
   <\arrowtypea`\arrowtypeb`\arrowtypec;\height>}}
\def\ptriangle{\@ifnextchar <{\ptrianglep}{\ptrianglep
   <\arrowtypea`\arrowtypeb`\arrowtypec;\height>}}
\def\ptrianglep<#1>[#2`#3`#4;#5`#6`#7]{{%%    %      #2----->#3
\settriparms[#1]%                             %      |      /
\diagram%                                     %      |     /
\putptrianglep<\arrowtypea`\arrowtypeb`%      %    #6|    /#7
\arrowtypec;\height>%                         %      |   /
(0,0)[#2`#3`#4;#5`#6`{#7}]%                   %      |  /
\enddiagram%%                                 %      v v
}}                                            %      #4

\def\putqtrianglep<#1>(#2,#3)[#4`#5`#6;#7`#8`#9]{{%
\settriparms[#1]%
\xpos=#2 \ypos=#3 \advance\ypos by\height
\puthmorphism(\xpos,\ypos)[#4`#5`{#7}]{\height}{\arrowtypea}a%
\putmorphism(\xpos,\ypos)(1,-1)[``{#8}]{\height}{\arrowtypeb}l%
\advance\xpos by\height
\putvmorphism(\xpos,\ypos)[`#6`{#9}]{\height}{\arrowtypec}r%
}}

\def\putqtriangle{\@ifnextchar <{\putqtrianglep}{\putqtrianglep
   <\arrowtypea`\arrowtypeb`\arrowtypec;\height>}}
\def\qtriangle{\@ifnextchar <{\qtrianglep}{\qtrianglep
   <\arrowtypea`\arrowtypeb`\arrowtypec;\height>}}
\def\qtrianglep<#1>[#2`#3`#4;#5`#6`#7]{{%%    %        #2----->#3
\settriparms[#1]%                             %         \      |
\width=\height                                %          \     |
\diagram%                                     %         #6\    |#7
\putqtrianglep<\arrowtypea`\arrowtypeb`%      %            \   |
\arrowtypec;\height>%                         %             \  |
(0,0)[#2`#3`#4;#5`#6`{#7}]%                   %              v v
\enddiagram%%                                 %               #4
}}

\def\putdtrianglep<#1>(#2,#3)[#4`#5`#6;#7`#8`#9]{{%
\settriparms[#1]%
\xpos=#2 \ypos=#3
\puthmorphism(\xpos,\ypos)[#5`#6`{#9}]{\height}{\arrowtypec}b%
\advance\xpos by \height \advance\ypos by\height
\putmorphism(\xpos,\ypos)(-1,-1)[``{#7}]{\height}{\arrowtypea}l%
\putvmorphism(\xpos,\ypos)[#4``{#8}]{\height}{\arrowtypeb}r%
}}

\def\putdtriangle{\@ifnextchar <{\putdtrianglep}{\putdtrianglep
   <\arrowtypea`\arrowtypeb`\arrowtypec;\height>}}
\def\dtriangle{\@ifnextchar <{\dtrianglep}{\dtrianglep
   <\arrowtypea`\arrowtypeb`\arrowtypec;\height>}}
\def\dtrianglep<#1>[#2`#3`#4;#5`#6`#7]{{%%    %                  / |
\settriparms[#1]%                             %                 /  |
\width=\height                                %              #5/   |#6
\diagram%                                     %               /    |
\putdtrianglep<\arrowtypea`\arrowtypeb`%      %              /     |
\arrowtypec;\height>%                         %             v      v
(0,0)[#2`#3`#4;#5`#6`{#7}]%                   %            #3----->#4
\enddiagram%%                                 %                #7
}}

\def\putbtrianglep<#1>(#2,#3)[#4`#5`#6;#7`#8`#9]{{%
\settriparms[#1]%
\xpos=#2 \ypos=#3
\puthmorphism(\xpos,\ypos)[#5`#6`{#9}]{\height}{\arrowtypec}b%
\advance\ypos by\height
\putmorphism(\xpos,\ypos)(1,-1)[``{#8}]{\height}{\arrowtypeb}r%
\putvmorphism(\xpos,\ypos)[#4``{#7}]{\height}{\arrowtypea}l%
}}

\def\putbtriangle{\@ifnextchar <{\putbtrianglep}{\putbtrianglep
   <\arrowtypea`\arrowtypeb`\arrowtypec;\height>}}
\def\btriangle{\@ifnextchar <{\btrianglep}{\btrianglep
   <\arrowtypea`\arrowtypeb`\arrowtypec;\height>}}
\def\btrianglep<#1>[#2`#3`#4;#5`#6`#7]{{%%   %              | \
\settriparms[#1]%                            %              |  \
\width=\height                               %            #5|   \#6
\diagram%                                    %              |    \
\putbtrianglep<\arrowtypea`\arrowtypeb`%     %              |     \
\arrowtypec;\height>%                        %              v      v
(0,0)[#2`#3`#4;#5`#6`{#7}]%                  %              #3----->#4
\enddiagram%%                                %                 #7
}}

\def\putAtrianglep<#1>(#2,#3)[#4`#5`#6;#7`#8`#9]{{%
\settriparms[#1]%
\xpos=#2 \ypos=#3 {\multiply \height by2
\puthmorphism(\xpos,\ypos)[#5`#6`{#9}]{\height}{\arrowtypec}b}%
\advance\xpos by\height \advance\ypos by\height
\putmorphism(\xpos,\ypos)(-1,-1)[#4``{#7}]{\height}{\arrowtypea}l%
\putmorphism(\xpos,\ypos)(1,-1)[``{#8}]{\height}{\arrowtypeb}r%
}}

\def\putAtriangle{\@ifnextchar <{\putAtrianglep}{\putAtrianglep
   <\arrowtypea`\arrowtypeb`\arrowtypec;\height>}}
\def\Atriangle{\@ifnextchar <{\Atrianglep}{\Atrianglep
   <\arrowtypea`\arrowtypeb`\arrowtypec;\height>}}
\def\Atrianglep<#1>[#2`#3`#4;#5`#6`#7]{{%%         %         /   \
\settriparms[#1]%                                  %        /     \
\width=\height                                     %     #5/       \#6
\diagram%                                          %      /         \
\putAtrianglep<\arrowtypea`\arrowtypeb`%           %     /           \
\arrowtypec;\height>%                              %    v             v
(0,0)[#2`#3`#4;#5`#6`{#7}]%                        %   #3------------>#4
\enddiagram%%                                      %          #7
}}

\def\putAtrianglepairp<#1>(#2)[#3;#4`#5`#6`#7`#8]{{%
\settripairparms[#1]%
\setpos(#2)%
\settokens`#3`%
\puthmorphism(\xpos,\ypos)[\tokenb`\tokenc`{#7}]{\height}{\arrowtyped}b%
\advance\xpos by\height
\puthmorphism(\xpos,\ypos)[\phantom{\tokenc}`\tokend`{#8}]%
{\height}{\arrowtypee}b%
\advance\ypos by\height
\putmorphism(\xpos,\ypos)(-1,-1)[\tokena``{#4}]{\height}{\arrowtypea}l%
\putvmorphism(\xpos,\ypos)[``{#5}]{\height}{\arrowtypeb}m%
\putmorphism(\xpos,\ypos)(1,-1)[``{#6}]{\height}{\arrowtypec}r%
}}

\def\putAtrianglepair{\@ifnextchar <{\putAtrianglepairp}{\putAtrianglepairp%
   <\arrowtypea`\arrowtypeb`\arrowtypec`\arrowtyped`\arrowtypee;\height>}}
\def\Atrianglepair{\@ifnextchar <{\Atrianglepairp}{\Atrianglepairp%
   <\arrowtypea`\arrowtypeb`\arrowtypec`\arrowtyped`\arrowtypee;\height>}}

\def\Atrianglepairp<#1>[#2;#3`#4`#5`#6`#7]{{%           %  #2a
\settripairparms[#1]%                         %           / | \
\settokens`#2`%                               %          /  |  \
\width=\height                                %       #3/  #4   \#5
\diagram%                                     %        /    |    \
\putAtrianglepairp                            %       /     |     \
<\arrowtypea`\arrowtypeb`\arrowtypec`%        %      v      v      v
\arrowtyped`\arrowtypee;\height>%             %     #2b---->#2c---->#2d
(0,0)[{#2};#3`#4`#5`#6`{#7}]%                 %         #6     #7
\enddiagram%%
}}

\def\putVtrianglep<#1>(#2,#3)[#4`#5`#6;#7`#8`#9]{{%
\settriparms[#1]%
\xpos=#2 \ypos=#3 \advance\ypos by\height {\multiply\height by2
\puthmorphism(\xpos,\ypos)[#4`#5`{#7}]{\height}{\arrowtypea}a}%
\putmorphism(\xpos,\ypos)(1,-1)[`#6`{#8}]{\height}{\arrowtypeb}l%
\advance\xpos by\height \advance\xpos by\height
\putmorphism(\xpos,\ypos)(-1,-1)[``{#9}]{\height}{\arrowtypec}r%
}}

\def\putVtriangle{\@ifnextchar <{\putVtrianglep}{\putVtrianglep
   <\arrowtypea`\arrowtypeb`\arrowtypec;\height>}}
\def\Vtriangle{\@ifnextchar <{\Vtrianglep}{\Vtrianglep
   <\arrowtypea`\arrowtypeb`\arrowtypec;\height>}}
\def\Vtrianglep<#1>[#2`#3`#4;#5`#6`#7]{{%%     %        #2------------->#3
\settriparms[#1]%                              %         \             /
\width=\height                                 %          \           /
\diagram%                                      %         #6\         /#7
\putVtrianglep<\arrowtypea`\arrowtypeb`%       %            \       /
\arrowtypec;\height>%                          %             \     /
(0,0)[#2`#3`#4;#5`#6`{#7}]%                    %              v   v
\enddiagram%%                                  %               #4
}}

\def\putVtrianglepairp<#1>(#2)[#3;#4`#5`#6`#7`#8]{{
\settripairparms[#1]%
\setpos(#2)%
\settokens`#3`%
\advance\ypos by\height
\putmorphism(\xpos,\ypos)(1,-1)[`\tokend`{#6}]{\height}{\arrowtypec}l%
\puthmorphism(\xpos,\ypos)[\tokena`\tokenb`{#4}]{\height}{\arrowtypea}a%
\advance\xpos by\height
\puthmorphism(\xpos,\ypos)[\phantom{\tokenb}`\tokenc`{#5}]%
{\height}{\arrowtypeb}a%
\putvmorphism(\xpos,\ypos)[``{#7}]{\height}{\arrowtyped}m%
\advance\xpos by\height
\putmorphism(\xpos,\ypos)(-1,-1)[``{#8}]{\height}{\arrowtypee}r%
}}

\def\putVtrianglepair{\@ifnextchar <{\putVtrianglepairp}{\putVtrianglepairp%
    <\arrowtypea`\arrowtypeb`\arrowtypec`\arrowtyped`\arrowtypee;\height>}}
\def\Vtrianglepair{\@ifnextchar <{\Vtrianglepairp}{\Vtrianglepairp%
    <\arrowtypea`\arrowtypeb`\arrowtypec`\arrowtyped`\arrowtypee;\height>}}
\def\Vtrianglepairp<#1>[#2;#3`#4`#5`#6`#7]{{%  %  #2a---->#2b---->#2c
\settripairparms[#1]%                          %   \      |      /
\settokens`#2`%                                %    \     |     /
\diagram%                                      %   #5\   #6    /#7
\putVtrianglepairp                             %      \   |   /
<\arrowtypea`\arrowtypeb`\arrowtypec`%         %       \  |  /
\arrowtyped`\arrowtypee;\height>%              %        v v v
(0,0)[{#2};#3`#4`#5`#6`{#7}]%                  %         #2d
\enddiagram%%
}}

\def\putCtrianglep<#1>(#2,#3)[#4`#5`#6;#7`#8`#9]{{%
\settriparms[#1]%
\xpos=#2 \ypos=#3 \advance\ypos by\height
\putmorphism(\xpos,\ypos)(1,-1)[``{#9}]{\height}{\arrowtypec}l%
\advance\xpos by\height \advance\ypos by\height
\putmorphism(\xpos,\ypos)(-1,-1)[#4`#5`{#7}]{\height}{\arrowtypea}l%
{\multiply\height by 2
\putvmorphism(\xpos,\ypos)[`#6`{#8}]{\height}{\arrowtypeb}r}%
}}

\def\putCtriangle{\@ifnextchar <{\putCtrianglep}{\putCtrianglep
    <\arrowtypea`\arrowtypeb`\arrowtypec;\height>}}
\def\Ctriangle{\@ifnextchar <{\Ctrianglep}{\Ctrianglep
    <\arrowtypea`\arrowtypeb`\arrowtypec;\height>}}
\def\Ctrianglep<#1>[#2`#3`#4;#5`#6`#7]{{%%   %                / |
\settriparms[#1]%                            %             #5/  |
\width=\height                               %              /   |
\diagram%                                    %             v    |
\putCtrianglep<\arrowtypea`\arrowtypeb`%     %           #3     |#6
\arrowtypec;\height>%                        %             \    |
(0,0)[#2`#3`#4;#5`#6`{#7}]%                  %            #7\   |
\enddiagram%%                                %               \  |
}}                                           %                v v
\def\putDtrianglep<#1>(#2,#3)[#4`#5`#6;#7`#8`#9]{{%
\settriparms[#1]%
\xpos=#2 \ypos=#3 \advance\xpos by\height \advance\ypos by\height
\putmorphism(\xpos,\ypos)(-1,-1)[``{#9}]{\height}{\arrowtypec}r%
\advance\xpos by-\height \advance\ypos by\height
\putmorphism(\xpos,\ypos)(1,-1)[`#5`{#8}]{\height}{\arrowtypeb}r%
{\multiply\height by 2
\putvmorphism(\xpos,\ypos)[#4`#6`{#7}]{\height}{\arrowtypea}l}%
}}

\def\putDtriangle{\@ifnextchar <{\putDtrianglep}{\putDtrianglep
    <\arrowtypea`\arrowtypeb`\arrowtypec;\height>}}
\def\Dtriangle{\@ifnextchar <{\Dtrianglep}{\Dtrianglep
   <\arrowtypea`\arrowtypeb`\arrowtypec;\height>}}
\def\Dtrianglep<#1>[#2`#3`#4;#5`#6`#7]{{%%  %          | \
\settriparms[#1]%                           %          |  \#6
\width=\height                              %          |   \
\diagram%                                   %          |    v
\putDtrianglep<\arrowtypea`\arrowtypeb`%    %        #5|    #3
\arrowtypec;\height>%                       %          |    /
(0,0)[#2`#3`#4;#5`#6`{#7}]%                 %          |   /#7
\enddiagram%%                               %          |  /
}}                                          %          v v
\def\setrecparms[#1`#2]{\width=#1 \height=#2}%
\def\recursep<#1`#2>[#3;#4`#5`#6`#7`#8]{{\m@th
\width=#1 \height=#2 \settokens`#3`
\settowidth{\tempdimen}{$\tokena$} \ifdim\tempdimen=0pt
  \savebox{\tempboxa}{\hbox{$\tokenb$}}%
  \savebox{\tempboxb}{\hbox{$\tokend$}}%
  \savebox{\tempboxc}{\hbox{$#6$}}%
\else
  \savebox{\tempboxa}{\hbox{$\hbox{$\tokena$}\times\hbox{$\tokenb$}$}}%
  \savebox{\tempboxb}{\hbox{$\hbox{$\tokena$}\times\hbox{$\tokend$}$}}%
  \savebox{\tempboxc}{\hbox{$\hbox{$\tokena$}\times\hbox{$#6$}$}}%
\fi \ypos=\height \divide\ypos by 2 \xpos=\ypos \advance\xpos by
\width \bfig
\putCtrianglep<-1`1`1;\ypos>(0,0)[`\tokenc`;#5`#6`{#7}]%
\puthmorphism(\ypos,0)[\tokend`\usebox{\tempboxb}`{#8}]{\width}{-1}b%
\puthmorphism(\ypos,\height)[\tokenb`\usebox{\tempboxa}`{#4}]{\width}{-1}a%
\advance\ypos by \width
\putvmorphism(\ypos,\height)[``\usebox{\tempboxc}]{\height}1r%
\efig }}

\def\recurse{\@ifnextchar <{\recursep}{\recursep<\width`\height>}}

\def\puttwohmorphisms(#1,#2)[#3`#4;#5`#6]#7#8#9{{%
\puthmorphism(#1,#2)[#3`#4`]{#7}0a \ypos=#2 \advance\ypos by 20
\puthmorphism(#1,\ypos)[\phantom{#3}`\phantom{#4}`#5]{#7}{#8}a
\advance\ypos by -40
\puthmorphism(#1,\ypos)[\phantom{#3}`\phantom{#4}`#6]{#7}{#9}b }}

\def\puttwovmorphisms(#1,#2)[#3`#4;#5`#6]#7#8#9{{%
\putvmorphism(#1,#2)[#3`#4`]{#7}0a \xpos=#1 \advance\xpos by -20
\putvmorphism(\xpos,#2)[\phantom{#3}`\phantom{#4}`#5]{#7}{#8}l
\advance\xpos by 40
\putvmorphism(\xpos,#2)[\phantom{#3}`\phantom{#4}`#6]{#7}{#9}r }}

\def\puthcoequalizer(#1)[#2`#3`#4;#5`#6`#7]#8#9{{%
\setpos(#1)%
\puttwohmorphisms(\xpos,\ypos)[#2`#3;#5`#6]{#8}11%
\advance\xpos by #8
\puthmorphism(\xpos,\ypos)[\phantom{#3}`#4`#7]{#8}1{#9} }}

\def\putvcoequalizer(#1)[#2`#3`#4;#5`#6`#7]#8#9{{%
\setpos(#1)%
\puttwovmorphisms(\xpos,\ypos)[#2`#3;#5`#6]{#8}11%
\advance\ypos by -#8
\putvmorphism(\xpos,\ypos)[\phantom{#3}`#4`#7]{#8}1{#9} }}

\def\putthreehmorphisms(#1)[#2`#3;#4`#5`#6]#7(#8)#9{{%
\setpos(#1) \settypes(#8)
\if a#9 %
     \vertsize{\tempcounta}{#5}%
     \vertsize{\tempcountb}{#6}%
     \ifnum \tempcounta<\tempcountb \tempcounta=\tempcountb \fi
\else
     \vertsize{\tempcounta}{#4}%
     \vertsize{\tempcountb}{#5}%
     \ifnum \tempcounta<\tempcountb \tempcounta=\tempcountb \fi
\fi \advance \tempcounta by 60
\puthmorphism(\xpos,\ypos)[#2`#3`#5]{#7}{\arrowtypeb}{#9}
\advance\ypos by \tempcounta
\puthmorphism(\xpos,\ypos)[\phantom{#2}`\phantom{#3}`#4]{#7}{\arrowtypea}{#9}
\advance\ypos by -\tempcounta \advance\ypos by -\tempcounta
\puthmorphism(\xpos,\ypos)[\phantom{#2}`\phantom{#3}`#6]{#7}{\arrowtypec}{#9}
}}

\def\setarrowtoks[#1`#2`#3`#4`#5`#6]{%
\def\toka{#1}
\def\tokb{#2}
\def\tokc{#3}
\def\tokd{#4}
\def\toke{#5}
\def\tokf{#6}
}
\def\hex{\@ifnextchar <{\hexp}{\hexp<1000`400>}}
\def\hexp<#1`#2>[#3`#4`#5`#6`#7`#8;#9]{%
\setarrowtoks[#9] \yext=#2 \advance \yext by #2 \xext=#1
\advance\xext by \yext \bfig
\putCtriangle<-1`0`1;#2>(0,0)[`#5`;\tokb``\tokd] \xext=#1
\yext=#2 \advance \yext by #2
\putsquare<1`0`0`1;\xext`\yext>(#2,0)[#3`#4`#7`#8;\toka```\tokf]
\advance \xext by #2
\putDtriangle<0`1`-1;#2>(\xext,0)[`#6`;`\tokc`\toke] \efig }

\makeatother

\begin{document}

\title{Entropic Geometry of Crowd Dynamics}
\author{Vladimir G. Ivancevic and Darryn J. Reid \\
Land Operations Division, Defence Science \& Technology Organisation,
Australia}\date{}\maketitle

\abstract{We propose an entropic geometrical model of psycho--physical crowd dynamics (with dissipative crowd kinematics), using Feynman action--amplitude formalism that operates on three synergetic levels: macro, meso and micro. The intent is to explain the dynamics of crowds simultaneously and consistently across these three levels, in order to characterize their geometrical properties particularly with respect to behavior regimes and the state changes between them. Its most natural statistical descriptor is crowd entropy $S$ that satisfies the Prigogine's extended second law of thermodynamics, $\partial_tS\geq 0$ (for any nonisolated multi-component system). Qualitative similarities and superpositions between individual and crowd configuration manifolds motivate our claim that goal-directed crowd movement operates under entropy conservation, $\partial_tS = 0$, while natural crowd dynamics operates under (monotonically) increasing entropy function, $\partial_tS > 0$. Between these two distinct topological phases lies a phase transition with a chaotic inter-phase. Both inertial crowd dynamics and its dissipative kinematics represent diffusion processes on the crowd manifold governed by the Ricci flow, with the associated Perelman entropy--action.

 }

\newpage
\tableofcontents
\newpage
%%%%%%%%%%%%%%%%%%%%%%%%%%%%%%%%%%%%%%%%%%%%%%%%%%%%%
\section{Introduction}

Recall that the term \emph{cognition}\footnote{Latin: ``cognoscere = to know"} is used in several loosely related ways to refer to a faculty for the human--like
processing of information, applying knowledge and changing preferences.
In psychology, cognition refers
to an information processing view of an individual psychological functions
(see \cite{Cog1,Cog2,Cog3,Cog4,Ashcraft}). More generally, cognitive processes
can be natural and artificial, conscious and not
conscious; therefore, they are analyzed from different perspectives and in
different contexts, e.g., anesthesia, neurology, psychology, philosophy,
logic (both Aristotelian and mathematical), systemics, computer science,
artificial intelligence (AI) and computational intelligence (CI). Both in
psychology and in AI/CI, cognition refers to the mental functions, mental
processes and states of intelligent entities (humans, human organizations,
highly autonomous robots), with a particular focus toward the study of
comprehension, inferencing, decision--making, planning and learning (see,
e.g. \cite{Busemeyer2}). The recently developed Scholarpedia, the free peer
reviewed web encyclopedia of computational neuroscience is largely based on
cognitive neuroscience (see, e.g. \cite{Pessoa}). The concept of cognition
is closely related to such abstract concepts as mind, reasoning, perception,
intelligence, learning, and many others that describe numerous capabilities
of the human mind and expected properties of AI/CI (see \cite{NeuFuz,CompMind} and references therein).

Yet disembodied cognition is a myth, albeit one that has had profound influence in Western science since Rene Descartes and others gave it credence during the Scientific Revolution. In fact, the mind-body separation had much more to do with explanation of method than with explanation of the mind and cognition, yet it is with respect to the latter that its impact is most widely felt. We find it to be an unsustainable assumption in the realm of crowd behavior. Mental intention is
(almost immediately) followed by a physical action, that is, a human or
animal movement \cite{Schoner}. In animals, this physical action would be
jumping, running, flying, swimming, biting or grabbing. In humans, it can be
talking, walking, driving, or shooting, etc. Mathematical description of
human/animal movement in terms of the corresponding neuro-musculo-skeletal
equations of motion, for the purpose of prediction and control, is
formulated within the realm of biodynamics (see \cite%
{GaneshIEEE,IJMMS1,SIAM,VladNick,LieLagr,GaneshSprSml,GaneshWSc,GaneshSprBig}).

The crowd (or, collective) psycho--physical behavior is clearly formed by
some kind of \textit{superposition, contagion, emergence,} or \textit{%
convergence} from the individual agents' behavior. Le Bon's 1895 contagion
theory, presented in ``The Crowd: A Study of the Popular Mind" influenced
many 20th century figures. Sigmund Freud criticized Le Bon's concept of
\textquotedblleft collective soul," asserting that crowds do not have a soul
of their own. The main idea of Freudian crowd behavior theory was that
people who were in a crowd acted differently towards people than those who
were thinking individually: the minds of the group would merge together to
form a collective way of thinking. This idea was further developed in
Jungian famous \textquotedblleft collective unconscious" \cite{Jung}. The
term \textquotedblleft collective behavior" \cite{Blumer} refers to social
processes and events which do not reflect existing social structure (laws,
conventions, and institutions), but which emerge in a \textquotedblleft
spontaneous" way. Collective behavior might also be defined as action which
is neither conforming (in which actors follow prevailing norms) nor deviant
(in which actors violate those norms). According to the emergence theory
\cite{Turner}, crowds begin as collectivities composed of people with mixed
interests and motives; especially in the case of less stable crowds
(expressive, acting and protest crowds) norms may be vague and changing;
people in crowds make their own rules as they go along. According to
currently popular convergence theory, crowd behavior is not a product of the
crowd itself, but is carried into the crowd by particular individuals, thus
crowds amount to a convergence of like--minded individuals.

We propose that the contagion and convergence theories may be unified by acknowledging that both factors may coexist, even within a single scenario: we propose to refer to this third approach as \emph{behavioral composition}. It represents a substantial philosophical shift from traditional analytical approaches, which have assumed either reduction of a whole into parts or the emergence of the whole from the parts. In particular, both contagion and convergence are related to social entropy, which is the natural decay of structure (such as law, organization, and convention) in a social system \cite{Downarowicz}. Thus, social entropy provides an entry point into realizing a behavioral--compositional theory of crowd dynamics.

Thus, while all mentioned psycho-social theories of crowd behavior are explanatory only, in this paper we attempt to formulate a geometrically predictive model--theory of crowd psycho-physical behavior.

We propose the entropy formulation of crowd dynamics as a three--step process involving individual psycho-physical dynamics and collective psycho-physical dynamics.

\section{Generic three--step crowd behavioral dynamics}

In this section we propose a generic crowd psycho--physical dynamics as a three--step process based on a general partition function formalism. Note that the number of variables $X_{i}$ in the standard partition function from statistical mechanics (see equation (\ref{partFun1}) in Appendix) need not be countable, in which case
the set of coordinates $\{x^{i}\}$ becomes a field\ $\phi =\phi (x),$ so\
the sum is to be replaced by the \emph{Euclidean path integral} (that is a
Wick--rotated Feynman transition amplitude in imaginary time, see subsection \ref{aggreg}), as
\begin{equation*}
Z(\phi )=\int \mathcal{D}[\phi ]\exp \left[ -H(\phi )\right] .
\end{equation*}
More generally, in quantum field theory, instead of the field Hamiltonian $%
H(\phi )$ we have the action $S(\phi )$ of the theory. Both Euclidean path
integral,
\begin{equation}
Z(\phi )=\int \mathcal{D}[\phi ]\exp \left[ -S(\phi )\right] ,\qquad \text{%
real path integral in imaginary time}  \label{Eucl}
\end{equation}%
and Lorentzian one,
\begin{equation}
Z(\phi )=\int \mathcal{D}[\phi ]\exp \left[ iS(\phi )\right] ,\qquad \text{%
complex path integral in real time}  \label{Lor}
\end{equation}%
--r epresent quantum field theory (QFT) partition functions. We will give formal definitions of the above path integrals (i.e., general partition functions) in section 3. For the moment, we only remark that the Lorentzian path integral
(\ref{Lor}) represents a QFT generalization of the (nonlinear) Schr\"{o}dinger equation, while the Euclidean path integral (\ref{Eucl}) in the (rectified) real time represents a
statistical field theory (SFT) generalization of the Fokker--Planck equation.

Now, following the framework of the Extended Second Law of Thermodynamics
(see Appendix), $\partial _{t}S\geq 0,$ for entropy $S$ in any complex
system described by its partition function, we formulate a generic crowd psycho--physical dynamics, based on above partition functions, as the
following three--step process:

1. Individual psycho--physical dynamics ($\mathcal{ID}$) is a transition
process from an entropy--growing \textquotedblleft loading" phase of mental
preparation, to the entropy--conserving \textquotedblleft hitting/shooting"
phase of physical action. Formally, $\mathcal{ID}$ is given by the
phase--transition map:
\begin{equation}
\mathcal{ID}:~\overset{``\mathrm{LOADING}":\,\partial _{t}S>0}{\overbrace{%
\mathrm{MENTAL~PREPARATION}}}~\Longrightarrow ~\overset{``\mathrm{HITTING}%
":\,\partial _{t}S=0}{\overbrace{\mathrm{PHYSICAL~ACTION}}}  \label{id}
\end{equation}%
defined by the individual (chaotic) phase--transition amplitude
\begin{equation*}
\left\langle \overset{\partial _{t}S=0}{\mathrm{PHYS.~ACTION}}\right\vert
CHAOS\left\vert \overset{\partial _{t}S>0}{\mathrm{MENTAL~PREP.}}%
\right\rangle _{\mathrm{ID}}:=\int \mathcal{D}[\Phi ]\,\mathrm{e}^{iS_{%
\mathrm{ID}}[\Phi ]},
\end{equation*}%
where the right-hand-side is the Lorentzian path-integral (or complex
path-integral in real time, see Appendix), with the individual
psycho--physical action
\begin{equation*}
S_{\mathrm{ID}}[\Phi ]=\int_{t_{ini}}^{t_{fin}}{L}_{\mathrm{ID}}[\Phi ]\,dt,
\end{equation*}%
where ${L}_{\mathrm{ID}}[\Phi ]$ is the psycho--physical Lagrangian,
consisting of mental cognitive potential and physical kinetic energy.

2. Aggregate psycho--physical dynamics ($\mathcal{AD}$) represents the
behavioral composition--transition map:
\begin{equation}
\mathcal{AD}:~\overset{``\mathrm{LOADING}":\,\partial _{t}S>0}{\sum_{i\in
\mathrm{AD}}\overbrace{\mathrm{MENTAL~PREPARATION}}}~\Longrightarrow
\sum_{i\in \mathrm{AD}}\overset{``\mathrm{HITTING}":\,\partial _{t}S=0}{%
\overbrace{\mathrm{PHYSICAL~ACTION}}}_{i}  \label{ad}
\end{equation}%
where the (weighted) aggregate sum is taken over all individual agents,
assuming equipartition of the total psycho--physical energy. It is defined
by the aggregate (chaotic) phase--transition amplitude
\begin{equation*}
\left\langle \overset{\partial _{t}S=0}{\mathrm{PHYS.~ACTION}}\right\vert
CHAOS\left\vert \overset{\partial _{t}S>0}{\mathrm{MENTAL~PREP.}}%
\right\rangle _{\mathrm{AD}}:=\int \mathcal{D}[\Phi ]\,\mathrm{e}^{-S_{%
\mathrm{AD}}[\Phi ]},
\end{equation*}%
with the Euclidean path-integral in real time,
that is the SFT--partition function, based on the
aggregate psycho--physical action
\begin{equation*}
S_{\mathrm{AD}}[\Phi ]=\int_{t_{ini}}^{t_{fin}}{L}_{\mathrm{AD}}[\Phi
]\,dt,\qquad \mathrm{with}\qquad {L}_{\mathrm{AD}}[\Phi ]=\sum_{i\in \mathrm{%
AD}}{L}_{\mathrm{ID}}^{i}[\Phi ].
\end{equation*}

3. Crowd psycho--physical dynamics ($\mathcal{CD}$) represents the
cumulative transition map:
\begin{equation}
\mathcal{CD}:~\overset{``\mathrm{LOADING}":\,\partial _{t}S>0}{\sum_{i\in
\mathrm{CD}}\overbrace{\mathrm{MENTAL~PREPARATION}}}~\Longrightarrow
\sum_{i\in \mathrm{CD}}\overset{``\mathrm{HITTING}":\,\partial _{t}S=0}{%
\overbrace{\mathrm{PHYSICAL~ACTION}}}_{i}  \label{cd}
\end{equation}%
where the (weighted) cumulative sum is taken over all individual agents,
assuming equipartition of the total psycho--physical energy. It is defined
by the crowd (chaotic) phase--transition amplitude
\begin{equation*}
\left\langle \overset{\partial _{t}S=0}{\mathrm{PHYS.~ACTION}}\right\vert
CHAOS\left\vert \overset{\partial _{t}S>0}{\mathrm{MENTAL~PREP.}}%
\right\rangle _{\mathrm{CD}}:=\int \mathcal{D}[\Phi ]\,\mathrm{e}^{iS_{%
\mathrm{CD}}[\Phi ]},
\end{equation*}%
with the general Lorentzian path-integral, that is, the QFT--partition function), based on the crowd psycho--physical action
\begin{equation*}
S_{\mathrm{CD}}[\Phi ]=\int_{t_{ini}}^{t_{fin}}{L}_{\mathrm{CD}}[\Phi
]\,dt,\qquad \mathrm{with}\qquad {L}_{\mathrm{CD}}[\Phi ]=\sum_{i\in \mathrm{%
CD}}{L}_{\mathrm{ID}}^{i}[\Phi ]=\sum_{k=\,\text{\# of }\mathrm{ADs}%
\text{ in CD}}{L}_{\mathrm{AD}}^{k}[\Phi ].
\end{equation*}

All three entropic phase--transition maps, $\mathcal{ID}$, $\mathcal{AD}$ and $\mathcal{CD}$, are
spatio--temporal biodynamic cognition systems, evolving within their
respective configuration manifolds (i.e., sets of their respective degrees-of-freedom with equipartition of energy), according to biphasic
action--functional formalisms with psycho--physical Lagrangian functions $L_{%
\mathrm{ID}}$, $L_{\mathrm{AD}}$ and $L_{\mathrm{CD}}$, each consisting of:
\begin{enumerate}
\item Cognitive mental potential (which is a mental preparation for the
physical action), and

\item Physical kinetic energy (which describes the physical action itself).
\end{enumerate}

To develop $\mathcal{ID}$, $\mathcal{AD}$ and $\mathcal{CD}$ formalisms, we extend into a
physical (or, more precisely, biodynamic) crowd domain a purely--mental individual
Life--Space Foam (LSF) framework for motivational cognition \cite{IA}, based
on the quantum--probability concept.\footnote{The quantum probability concept is based on the following physical facts
\cite{Complexity,QuLeap}
\begin{enumerate}
\item {The time--dependent Schr\"{o}dinger equation} represents a {%
complex--valued generalization} of the real--valued {Fokker--Planck equation}
for describing the spatio--temporal {probability density function} for the
system exhibiting {continuous--time Markov stochastic process}.

\item The {Feynman path integral }(including integration over continuous
spectrum and summation over discrete spectrum) is a generalization of the
time--dependent Schr\"{o}dinger equation, including both continuous--time
and discrete--time Markov stochastic processes.

\item Both Schr\"{o}dinger equation and path integral give `physical
description' of any system they are modelling in terms of its physical
energy, instead of an abstract probabilistic description of the
Fokker--Planck equation.
\end{enumerate}
Therefore, the Feynman path integral, as a generalization of the (nonlinear)
time--dependent Schr\"{o}dinger equation, gives a unique physical
description for the general Markov stochastic process, in terms of the
physically based generalized probability density functions, valid both for
continuous--time and discrete--time Markov systems. Its basic consequence is this: a different way for calculating probabilities. The
difference is rooted in the fact that \textsl{sum of squares is different
from the square of sums}, as is explained in the following text. Namely, in
Dirac--Feynman quantum formalism, each possible route from the initial
system state $A$ to the final system state $B$ is called a {history}. This
history comprises any kind of a route, ranging from continuous and smooth
deterministic (mechanical--like) paths to completely discontinues and random
Markov chains (see, e.g., \cite{Gardiner}). Each history (labelled by index $%
i$) is quantitatively described by a {complex number}.

In this way, the overall probability of the system's transition from some
initial state $A$ to some final state $B$ is given {not} by adding up the
probabilities for each history--route, but by `head--to--tail' adding up the
sequence of amplitudes making--up each route first (i.e., performing the
sum--over--histories) -- to get the total amplitude as a `resultant vector',
and then squaring the total amplitude to get the overall transition
probability.

Here we emphasize that the domain of validity of the `quantum' is not
restricted to the microscopic world \cite{Ume93}. There are macroscopic
features of classically behaving systems, which cannot be explained without
recourse to the quantum dynamics. This field theoretic model leads to the
view of the phase transition as a condensation that is comparable to the
formation of fog and rain drops from water vapor, and that might serve to
model both the gamma and beta phase transitions. According to such a model,
the production of activity with long--range correlation in the brain takes
place through the mechanism of spontaneous breakdown of symmetry (SBS),
which has for decades been shown to describe long-range correlation in
condensed matter physics. The adoption of such a field theoretic approach
enables modelling of the whole cerebral hemisphere and its hierarchy of
components down to the atomic level as a fully integrated macroscopic
quantum system, namely as a macroscopic system which is a quantum system not
in the trivial sense that it is made, like all existing matter, by quantum
components such as atoms and molecules, but in the sense that some of its
macroscopic properties can best be described with recourse to quantum
dynamics (see \cite{FreVit06} and references therein). Also, according to
Freeman and Vitielo, \textit{many--body quantum field theory} appears to be
the only existing theoretical tool capable to explain the dynamic origin of
long--range correlations, their rapid and efficient formation and
dissolution, their interim stability in ground states, the multiplicity of
coexisting and possibly non--interfering ground states, their degree of
ordering, and their rich textures relating to sensory and motor facets of
behaviors. It is historical fact that many--body quantum field theory has
been devised and constructed in past decades exactly to understand features
like ordered pattern formation and phase transitions in condensed matter
physics that could not be understood in classical physics, similar to those
in the brain.}

The psycho--physical approach to $\mathcal{ID}$, $\mathcal{AD}$ and $\mathcal{CD}$ is based
on \textit{entropic motor control} \cite{Hong1,Hong2}, which deals with
neuro-physiological feedback information and environmental uncertainty. The
probabilistic nature of human motor action can be characterized by entropies
at the level of the organism, task, and environment. Systematic changes in
motor adaptation are characterized as task--organism and
environment--organism tradeoffs in entropy. Such compensatory adaptations
lead to a view of goal--directed motor control as the product of an
underlying conservation of entropy across the task--organism--environment
system. In particular, an experiment conducted in \cite{Hong2} examined the
changes in entropy of the coordination of isometric force output under
different levels of task demands and feedback from the environment. The goal
of the study was to examine the hypothesis that human motor adaptation can
be characterized as a process of entropy conservation that is reflected in
the compensation of entropy between the task, organism motor output, and
environment. Information entropy of the coordination dynamics relative phase
of the motor output was made conditional on the idealized situation of human
movement, for which the goal was always achieved. Conditional entropy of the
motor output decreased as the error tolerance and feedback frequency were
decreased. Thus, as the likelihood of meeting the task demands was decreased
increased task entropy and/or the amount of information from the environment
is reduced increased environmental entropy, the subjects of this experiment
employed fewer coordination patterns in the force output to achieve the
goal. The conservation of entropy supports the view that context dependent
adaptations in human goal--directed action are guided fundamentally by
natural law and provides a novel means of examining human motor behavior.
This is fundamentally related to the \textit{Heisenberg uncertainty principle%
} \cite{QuLeap} and further supports the argument for the primacy of a
probabilistic approach toward the study of biodynamic cognition systems.\footnote{Our entropic action--amplitude formalism represents a kind of a generalization of the
Haken-Kelso-Bunz (HKB) model of self-organization in the individual's motor
system \cite{HKB,Kelso95}, including: multi-stability, phase
transitions and hysteresis effects, presenting a contrary view to
the purely feedback driven systems. HKB uses the concepts of
synergetics (order parameters, control parameters, instability,
etc) and the mathematical tools of nonlinearly coupled (nonlinear)
dynamical systems to account for self-organized behavior both at
the cooperative, coordinative level and at the level of the
individual coordinating elements. The HKB model stands as a
building block upon which numerous extensions and elaborations
have been constructed. In particular, it has been possible to
derive it from a realistic model of the cortical sheet in which
neural areas undergo a reorganization that is mediated by intra-
and inter-cortical connections. Also, the HKB model describes
phase transitions (`switches') in coordinated human movement as
follows: (i) when the agent begins in the anti-phase mode and
speed of movement is increased, a spontaneous switch to
symmetrical, in-phase movement occurs; (ii) this transition
happens swiftly at a certain critical frequency; (iii) after the
switch has occurred and the movement rate is now decreased the
subject remains in the symmetrical mode, i.e. she does not switch
back; and (iv) no such transitions occur if the subject begins
with symmetrical, in-phase movements. The HKB dynamics of the
order parameter relative phase as is given by a nonlinear
first-order ODE:
$$\dot{\phi} = (\alpha + 2 \beta r^2) \sin\phi - \beta r^2
\sin2\phi,
$$
where $\phi$ is the phase relation (that characterizes the
observed patterns of behavior, changes abruptly at the transition
and is only weakly dependent on parameters outside the phase
transition), $r$ is the oscillator amplitude, while $\alpha,\beta$
are coupling parameters (from which the critical frequency where
the phase transition occurs can be calculated).}

On the other hand, it is well--known that humans possess more
degrees of freedom than are needed to perform any defined motor
task, but are required to co-ordinate them in order to reliably
accomplish high-level goals, while faced with intense motor
variability. In an attempt to explain how this takes place,
Todorov and Jordan have formulated an alternative
theory of human motor co-ordination based on the concept of
stochastic optimal feedback control \cite{Todorov}. They were able to conciliate
the requirement of goal achievement (e.g., grasping an object)
with that of motor variability (biomechanical degrees of freedom).
Moreover, their theory accommodates the idea that the human motor
control mechanism uses internal `functional synergies' to regulate
task--irrelevant (redundant) movement.

Also, a developing field in coordination dynamics involves the theory of
social coordination, which attempts to relate the DC to normal
human development of complex social cues following certain
patterns of interaction. This work is aimed at understanding how
human social interaction is mediated by meta-stability of neural
networks. fMRI and EEG are particularly useful in mapping
thalamocortical response to social cues in experimental studies.
In particular, a new theory called the \emph{Phi complex} has been
developed by S. Kelso and collaborators, to provide experimental
results for the theory of social coordination dynamics (see the
recent nonlinear dynamics paper discussing social coordination and
EEG dynamics \cite{Tognoli}). According to this theory, a pair of
phi rhythms, likely generated in the mirror neuron system, is the
hallmark of human social coordination. Using a dual--EEG recording
system, the authors monitored the interactions of eight pairs of
subjects as they moved their fingers with and without a view of
the other individual in the pair.

Finally, the chaotic psycho--physical phase transitions embedded in $\mathcal{CD}$ may give a formal description for a phenomenon called \emph{crowd turbulence} by D. Helbing, depicting crowd disasters caused by the panic stampede that can occur at high pedestrian densities and which is a serious concern during mass events like soccer championship games or annual pilgrimage in Makkah (see \cite{HelbingNature,HelbingPRL,HelbingPRE,HelbingACS}).

\section{Formal crowd dynamics}

In this section we formally develop a three--step crowd psycho--physical dynamics, conceptualized by transition maps (\ref{id})--(\ref{ad})--(\ref{cd}), in agreement with Haken's synergetics \cite{Haken1,Haken2}.
We first develop a macro--level individual psycho--physical dynamics $\cal ID$. Then we generalize $\cal ID$ into an `orchestrated' behavioral--compositional crowd dynamics $\cal CD$, using a quantum--like micro--level formalism with individual agents representing `crowd quanta'. Finally we develop a meso--level aggregate statistical--field dynamics $\cal AD$, such that composition of the aggregates $\cal AD$ makes--up the crowd.

\subsection{Individual behavioral dynamics ($\cal ID$)}

$\cal ID$ transition map (\ref{id}) is developed using the following action--amplitude formalism (see \cite{IA,IAY}):
\begin{enumerate}
\item Macroscopically, as a smooth Riemannian $n-$manifold $M_{\rm ID}$ (see Appendix) with steady
force--fields and behavioral paths, modelled by a real--valued classical
{action functional} $S_{\rm ID}[\Phi ]$, of the form
\begin{equation*}
S_{\rm ID}[\Phi ]=\int_{t_{ini}}^{t_{fin}}{L}_{\rm ID}[\Phi ]\,dt,
\end{equation*}%
(where macroscopic paths, fields and geometries are commonly denoted by an
abstract field symbol $\Phi ^{i}$) with the potential--energy based {Lagrangian $L$} given by
\begin{equation*}
{L}_{\rm ID}[\Phi ]=\int d^{n}x\,\mathcal{L}_{\rm ID}(\Phi _{i},\partial _{x^{j}}\Phi
^{i}),
\end{equation*}%
where $\mathcal{L}$ is Lagrangian density, the integral is taken over all $n$ local coordinates $x^{j}=x^{j}(t)$ of the
ID, and $\partial _{x^{j}}\Phi ^{i}$ are time and space partial derivatives
of the $\Phi ^{i}-$variables over coordinates. The standard {least action
principle}
\begin{equation*}
\delta S_{\rm ID}[\Phi ]=0,
\end{equation*}%
gives, in the form of the Euler--Lagrangian equations, a shortest
path, an extreme force--field, with a geometry of minimal
curvature and topology without holes. We will see below that high Riemannian curvature generates chaotic behavior, while holes in the manifold produce topologically induced phase transitions.

\item Microscopically, as a collection of wildly fluctuating and jumping
paths (histories), force--fields and geometries/topologies, modelled by a
complex--valued adaptive path integral, formulated by defining a
multi--phase and multi--path (multi--field and multi--geometry) {transition
amplitude} from the entropy--growing state of Mental~Preparation to the
entropy--conserving state of Physical~Action,
\begin{equation}
\langle \mathrm{Physical~Action\,|\,Mental~Preparation}\rangle_{\mathrm{ID}} :=\int_{%
\mathrm{ID}}\mathcal{D}[\Phi ]\,\mathrm{e}^{iS_{\mathrm{ID}}[\Phi ]}\,
\label{pathInt2}
\end{equation}%
where the functional ID--measure $\mathcal{D}[w\Phi]$ is defined as a weighted product
\begin{equation}
\mathcal{D}[w\Phi]=\lim_{N\to\infty}\prod_{s=1}^{N}w_sd\Phi
_{s}^i, \qquad ({i=1,...,n=con+dis}),\label{prod1}
\end{equation} representing an $\infty-$dimensional neural network
\cite{IA}, with weights $w_s$ updating by the general rule
\begin{equation*}
new\;value(t+1)\;=\;old\;value(t)\;+\;innovation(t).
\end{equation*}
More precisely, the weights $w_s=w_s(t)$ in (\ref{prod1}) are updated according to one of the two
standard neural learning schemes, in which the micro--time level
is traversed in discrete steps, i.e., if $t=t_0,t_1,...,t_s$ then
$t+1=t_1,t_2,...,t_{s+1}$:\footnote{The traditional neural networks approaches are
known for their classes of functions they can
represent. Here we are talking about functions in an
\emph{extensional} rather than merely \emph{intensional} sense;
that is, function can be read as input/output behavior
\cite{Barendregt,Benthem,Forster,Hankin}. This limitation has
been attributed to their low-dimensionality (the largest neural
networks are limited to the order of $10^5$ dimensions
\cite{Izh}). The proposed path integral approach represents a new
family of function-representation methods, which potentially
offers a basis for a fundamentally more expansive solution.}
\begin{enumerate}
    \item A \textit{self--organized}, \textit{unsupervised}
    (e.g., Hebbian--like \cite{Hebb}) learning rule:
\begin{equation}
w_s(t+1)=w_s(t)+ \frac{\sigma}{\eta}(w_s^{d}(t)-w_s^{a}(t)),
\label{Hebb}
\end{equation}
where $\sigma=\sigma(t),\,\eta=\eta(t)$ denote \textit{signal} and
\textit{noise}, respectively, while superscripts $d$ and $a$
denote \textit{desired} and \textit{achieved} micro--states,
respectively; or
    \item A certain form of a \textit{supervised gradient descent
    learning}:
\begin{equation}
w_s(t+1)\,=\,w_s(t)-\eta \nabla J(t), \label{gradient}
\end{equation}
where $\eta $ is a small constant, called the \textit{step size},
or the \textit{learning rate,} and $\nabla J(n)$ denotes the
gradient of the `performance hyper--surface' at the $t-$th
iteration.
\end{enumerate}(Note that we could also use a reward--based, {reinforcement
learning} rule \cite{SB}, in which system learns its {optimal
policy}:
~$
innovation(t)=|reward(t)-penalty(t)|.\,
$)
\end{enumerate}
In this way, we effectively derive a unique and
globally smooth, causal and entropic phase--transition map (\ref{id}),
performed at a macroscopic (global) time--level from some initial time $%
t_{ini}$ to the final time $t_{fin}$. Thus, we have obtained
macro--objects in the ID: a single path described by Newtonian--like
equation of motion, a single force--field described by Maxwellian--like
field equations, and a single obstacle--free Riemannian geometry (with
global topology without holes).

\begin{figure}[htb]
\centering \includegraphics[width=12cm]{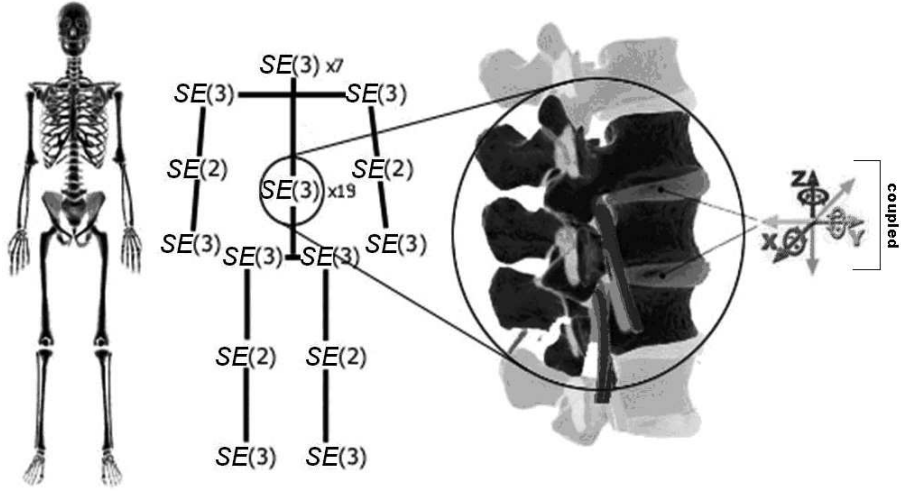}
\caption{Riemannian configuration manifold $M_{\rm ID}$ of human biodynamics is defined
as a topological product $M=\prod_{i}SE(3)^{i}$ of constrained Euclidean $SE(3)$--groups of rigid body motion in 3D Euclidean space (see \cite{GaneshSprBig,GaneshADG}), acting in all major (synovial) human joints. The manifold $M$ is a dynamical structure activated/controlled by potential covariant forces (\ref{cforc1}) produced by a synergetic action of about 640 skeletal muscles \cite{GaneshSprSml}.}
\label{SpineSE(3)}
\end{figure}

In particular, on the macro--level, we have the ID--paths, that is biodynamical trajectories generated by the Hamilton action principle
$$
\delta S_{\rm ID}[x]=0,
$$
with the {Newtonian action} $S_{\rm ID}[x]$ given by (Einstein's summation convention over repeated indices is always assumed)
\begin{equation}
S_{\rm ID}[x]=\int_{t_{ini}}^{t_{fin}}[\varphi+{1\over2}g_{ij}\,\dot{x}^{i}\dot{x}^{j}]\,dt,
\label{actmot}
\end{equation}
where $\varphi=\varphi(t,x^i)$ denotes the mental LSF--potential field, while the second term,
$$T={1\over2}g_{ij}\,\dot{x}^{i}\dot{x}^{j},$$ represents the physical (biodynamic) kinetic energy
generated by the {Riemannian inertial
metric tensor} $g_{ij}$ of the configuration biodynamic manifold $M_{\rm ID}$ (see Figure \ref{SpineSE(3)}). The corresponding Euler--Lagrangian equations give the
Newtonian equations of human movement
\begin{equation}
\frac{d}{dt}T_{\dot{x}^{i}}-T_{x^{i}}=F_i,
\label{Newton}
\end{equation}
where subscripts denote the partial derivatives and we have defined the covariant muscular forces $F_i=F_i(t,x^i,\dot{x}^{i})$ as negative gradients of the mental potential $\varphi(x^i)$,
\begin{equation}
F_i=-\varphi_{x^i}. \label{cforc1}
\end{equation}
Equation (\ref{Newton}) can be put into
the standard Lagrangian form as
\begin{equation}\frac{d}{dt}L_{\dot{x}^{i}}=L_{x^{i}},\qquad\text{with}\qquad
L=T-\varphi(x^i), \label{LgrEq}
\end{equation}
or (using the Legendre transform) into the forced, dissipative Hamiltonian form \cite{SIAM,GaneshSprSml}
\begin{equation}
\dot{x}^{i} = \partial _{p_{i}}H+\partial _{p_{i}}R, \qquad
\dot{p}_{i} = F_{i}-\partial _{x^{i}}H+\partial _{x^{i}}R, \label{ham1}
\end{equation}
where $p_{i}$ are the generalized momenta (canonically--conjugate to the coordinates $x^i$), $H=H(p,x)$ is the Hamiltonian (total energy function) and $R=R(p,x)$ is the general dissipative function.

The human motor system possesses many independently
controllable components that often allow for more than a
single movement pattern to be performed in order to
achieve a goal. Hence, the motor system is endowed with
a high level of adaptability to different tasks and also
environmental contexts \cite{Hong2}. The multiple SE(3)--dynamics applied to human musculo--skeletal system gives the fundamental law of biodynamics, which is the \emph{covariant force law}:
\begin{equation}
\text{Force co-vector field}=\text{Mass distribution}\times \text{%
Acceleration vector-field}, \label{covForce}
\end{equation}
which is formally written:
\begin{equation*}
F_{i}=g_{ij}a^{j},\qquad (i,j=1,...,n=\dim(M))
\end{equation*}
where $F_{i}$ are the covariant force/torque components, $g_{ij}$ is the inertial metric tensor of the configuration Riemannian manifold $M=\prod_{i}SE(3)^{i}$ ($g_{ij}$ defines the mass--distribution of the human body),
while $a^{j}$ are the contravariant components of the
linear and angular acceleration vector-field. (This fundamental biodynamic law states that contrary to common perception, acceleration and force are not quantities of
the same nature: while acceleration is a non-inertial vector-field, force is
an inertial co-vector-field. This apparently insignificant difference
becomes crucial in injury prediction/prevention, especially in its derivative form in which the `massless jerk' ($=\dot{a}$) is relatively benign, while the `massive jolt' ($=\dot{F}$) is deadly.) Both Lagrangian and (topologically equivalent) Hamiltonian development of the covariant force law is fully elaborated in \cite%
{GaneshSprSml,GaneshWSc,GaneshSprBig,GaneshADG}. This is consistent with the postulation that human action is guided primarily by natural law \cite{Kugler}.

On the micro--ID level, instead of each single trajectory defined by the
Newtonian equation of motion (\ref{Newton}), we have an
ensemble of fluctuating and crossing paths on the configuration manifold $M$ with weighted
probabilities (of the unit total sum). This ensemble of
micro--paths is defined by the simplest instance of our adaptive
path integral (\ref{pathInt2}), similar to the Feynman's original
{sum over histories},
\begin{equation}
\langle {\rm Physical~Action\,|\,Mental~Preparation}\rangle_M =\int_{\rm ID}\mathcal{D}[wx]\,
{\mathrm e}^{\mathrm i S[x]}, \label{Feynman}
\end{equation}
where $\mathcal{D}[wx]$ is the functional ID--measure on the
{space of all weighted paths}, and the exponential depends
on the action $S_{\rm ID}[x]$ given by (\ref{actmot}).

\subsection{Crowd behavioral--compositional dynamics ($\cal CD$)}

In this subsection we develop a generic crowd $\cal CD$, as a unique and globally smooth, causal and entropic phase--transition map (\ref{cd}),
in which agents (or, crowd's individual entities) can be both humans and robots. This crowd psycho--physical action takes place in a crowd smooth Riemannian $3n-$manifold $M$. Recall from Figure \ref{SpineSE(3)} that each individual segment of a human body moves in the Euclidean 3--space $\mathbb{R}^3$  according to its own constrained SE(3)--group. Similarly, each individual agent's trajectory, $x^i=x^i(t),~i=1,...n$, is governed by the Euclidean SE(2)--group of rigid body motions in the plane. (Recall
that a Lie group $SE(2)\equiv SO(2)\times \mathbb{R}$ is a set of all $%
3\times 3-$ matrices of the form:
\begin{equation*}
\left[
\begin{array}{ccc}
\cos \theta  & \sin \theta  & x \\
-\sin \theta  & \cos \theta  & y \\
0 & 0 & 1%
\end{array}%
\right] ,
\end{equation*}%
including both rigid translations (i.e., Cartesian $x,y-$coordinates) and
rotation matrix $\left[
\begin{array}{cc}
\cos \theta  & \sin \theta  \\
-\sin \theta  & \cos \theta
\end{array}%
\right] $ in Euclidean plane $\mathbb{R}^{2}$ (see \cite%
{GaneshSprBig,GaneshADG}).)
The crowd configuration manifold $M$ is defined as a union of Euclidean SE(2)--groups for all $n$ individual agents in the crowd, that is
crowd's
configuration $3n-$manifold is defined as a set
\begin{eqnarray}
M &=&\sum_{k=1}^{n}SE(2)^{k}\equiv \sum_{k=1}^{n}SO(2)^{k}\times \mathbb{R}%
^{k},  \label{crwdMan} \\
\text{coordinated by }\mathbf{x}^{k} &\mathbf{=}&\mathbf{\{}%
x^{k},y^{k},\theta ^{k}\},\ (\text{for }k=1,2,...,n).  \notag
\end{eqnarray}
In other words, the crowd configuration manifold $M$ is a \emph{dynamical planar graph} with individual agents' SE(2)--groups of motion in the vertices and time-dependent inter-agent distances $I_{ij} =\left[ x^{i}(t_{i})-x^{j}(t_{j})\right]$ as edges.

Similarly to the individual case, the crowd action functional includes mental cognitive potential and physical kinetic energy, formally given by (with $i,j=1,...,3n$):
\begin{eqnarray}
A[x^i,x^j;t_{i},t_{j}] &=&
\frac{1}{2}\int_{t_{i}}\int_{t_{j}}\,\delta
(I_{ij}^{2})\,\,\dot{x}^{i}(t_{i})\,\dot{x}^{j}(t_{j})\,\,dt_{i}dt_{j}
~+~{\frac{1}{2}}\int_{t}g_{ij}\,\dot{x}^{i}(t)\dot{x}^{j}(t)\,dt,
\label{Fey1} \\
\text{with} && I_{ij}^{2} =\left[ x^{i}(t_{i})-x^{j}(t_{j})\right] ^{2},
\qquad
\text{where \ \ }IN \leq t_{i},t_{j},t\leq OUT.\hspace{2cm}  \notag
\end{eqnarray}
The first term in (\ref{Fey1}) represents the mental potential for the interaction between any two agents $x^i$ and $x^i$ within the total crowd matrix $x^{ij}$. (Although, formally, this term contains cognitive velocities, it still
represents `potential energy' from the physical point of view.) It is defined as a
double integral over a delta function of the square of interval $I^{2}$
between two points on the paths in their individual cognitive LSFs. Interaction
occurs only when this LSF--distance between the two agents $x^i$ and $x^j$ vanishes. Note that the cognitive intentions of any two agents generally occur at different times $t_{i}$ and $t_{j}$
unless $t_{i}=t_{j},$ when cognitive synchronization occurs. This term effectively represents the \emph{crowd cognitive controller} (see \cite{IAY}).

The second term in (\ref{Fey1}) represents {kinetic energy of the physical
interaction of agents}. Namely, after the above cognitive synchronization is completed, the second term of physical kinetic energy is activated in the
common CD manifold, reducing it to just one of the agents' individual manifolds, which is equivalent to the center-of-mass segment in the human musculo-skeletal system. Therefore, from (\ref{Fey1}) we can derive a generic Euler--Lagrangian dynamics that is a composition of (\ref{LgrEq}), which also means that we have in place a generic Hamiltonian dynamics that is a amalgamate of (\ref{ham1}), as well as the crowd covariant force law (\ref{covForce}), the governing law of crowd biodynamics:
\begin{eqnarray}
&&\text{Crowd force co-vector field}=\text{Crowd mass distribution}\times \text{%
Crowd acceleration vector-field}, \notag\\
&&\text{formally:}~~~F_i=g_{ij}a^j,\qquad \text{where}~~g_{ij}~~\text{is the inertial metric tensor of crowd manifold~} M. \label{covForceCrowd}
\end{eqnarray}
The left-hand side of this equation defines forces acting on the crowd, while right-hand defines its mass distribution coupled to the crowd kinematics ($\cal CK$, described in the next subsection).

At the slave level, the adaptive path integral, representing
an $\infty-$dimensional neural network, corresponding to the psycho--physical crowd action (\ref{Fey1}), reads
\begin{equation}
\langle {\rm Physical~Action\,|\,Mental~Preparation}\rangle_{\mathrm{CD}} =\int_{\rm CD}\mathcal{D}[w,x,y]\, {\rm e}^{iA[x,y;t_{i},t_{j}]},  \label{pathInt}
\end{equation}
where the Lebesgue-type integration is performed over all continuous paths $%
x^{i}=x^{i}(t_{i})$ and $y^{j}=y^{j}(t_{j})$, while summation is performed
over all associated discrete Markov fluctuations and jumps. The symbolic differential in the path integral (\ref{pathInt})
represents an {adaptive path measure}, defined as the weighted product
\begin{equation}
\mathcal{D}[w,x,y]=\lim_{N\rightarrow \infty
}\prod_{s=1}^{N}w_{ij}^{s}dx^{i}dy^{j},  \qquad ({i,j=1,...,n}).  \label{prod}
\end{equation}
The quantum--field path integral (\ref{pathInt})--(\ref{prod}) defines the microstate $\cal CD-$level, an
ensemble of fluctuating and crossing paths on the crowd $3n-$manifold $M$.

The crowd manifold $M$ itself has quite a sophisticated
topological structure defined by its macrostate Euler--Lagrangian dynamics. As a Riemannian smooth $n-$manifold, $M$ gives rise to its fundamental
$n-${groupoid}, or $n-$category $\Pi _{n}(M)$ (see
\cite{GaneshSprBig,GaneshADG}). In $\Pi _{n}(M)$, 0--cells
are {points} in $M$; 1--cells are {paths} in
$M$ (i.e.,
parameterized smooth maps $f:[0,1]\rightarrow M$); 2--cells are {%
smooth homotopies} (denoted by $\simeq $) {of paths} relative
to endpoints (i.e., parameterized smooth maps $h:[0,1]\times
\lbrack 0,1]\rightarrow M$); 3--cells are {smooth
homotopies of homotopies} of paths in $M$ (i.e.,
parameterized smooth maps $j:[0,1]\times \lbrack 0,1]\times
\lbrack 0,1]\rightarrow M$). Categorical {composition}
is defined by {pasting} paths and homotopies. In this way,
the following {recursive homotopy dynamics} emerges on the crowd
$3n-$manifold $M$:

\label{ivncat}
\begin{eqnarray*}
&&\mathtt{0-cell:}\,\,x_{0}\,\node\,\,\,\qquad x_{0}\in M; \qquad
\text{in
the higher cells below: }t,s\in[0,1]; \\
&&\mathtt{1-cell:}\,\,x_{0}\,\node\cone{f}\node\,x_{1}\qquad
f:x_{0}\simeq
x_{1}\in M, \\
&&f:[0,1]\rightarrow M,\,f:x_{0}\mapsto
x_{1},\,x_{1}=f(x_{0}),\,f(0)=x_{0},\,f(1)=x_{1}; \\
&&\text{e.g., linear path: }f(t)=(1-t)\,x_{0}+t\,x_{1};\qquad \text{or} \\
&&\text{Euler--Lagrangian }f-\text{dynamics with endpoint conditions }%
(x_0,x_1): \\
&&\frac{d}{dt}f_{\dot{x}^{i}}=f_{x^{i}},\quad \text{with}\quad
x(0)=x_{0},\quad x(1)=x_{1},\quad (i=1,...,n); \\
&&\mathtt{2-cell:}\,\,x_{0}\,\node\ctwodbl{f}{g}{h}\node\,x_{1}\qquad
h:f\simeq g\in M, \\
&&h:[0,1]\times \lbrack 0,1]\rightarrow M,\,h:f\mapsto g,\,g=h(f(x_{0})), \\
&&h(x_{0},0)=f(x_{0}),\,h(x_{0},1)=g(x_{0}),\,h(0,t)=x_{0},\,h(1,t)=x_{1} \\
&&\text{e.g., linear homotopy: }h(x_{0},t)=(1-t)\,f(x_{0})+t\,g(x_{0});\qquad%
\text{or} \\
&&\text{homotopy between two Euler--Lagrangian
}(f,g)-\text{dynamics}
\\
&&\text{with the same endpoint conditions }(x_0,x_1): \\
&&\frac{d}{dt}f_{\dot{x}^{i}}=f_{x^{i}},\quad \text{and} \quad \frac{d}{dt}%
g_{\dot{x}^{i}}=g_{x^{i}}\quad\text{with}\quad x(0)=x_{0},\quad
x(1)=x_{1};
\\
&&\mathtt{3-cell:}\,\,x_{0}\,\node\cthreecelltrp{f}{g}{h}{i}{j}\node%
\,x_{1}\qquad j:h\simeq i\in M, \\
&&j:[0,1]\times \lbrack 0,1]\times \lbrack 0,1]\rightarrow
M,\,j:h\mapsto
i,\,i=j(h(f(x_{0}))) \\
&&j(x_{0},t,0)=h(f(x_{0})),\,j(x_{0},t,1)=i(f(x_{0})), \\
&&j(x_{0},0,s)=f(x_{0}),\,j(x_{0},1,s)=g(x_{0}), \\
&&j(0,t,s)=x_{0},\,j(1,t,s)=x_{1} \\
&&\text{e.g., linear composite homotopy: }j(x_{0},t,s)=(1-t)\,h(f(x_{0}))+t%
\,i(f(x_{0})); \\
&&\text{or, homotopy between two homotopies between above two Euler-} \\
&&\text{Lagrangian }(f,g)-\text{dynamics with the same endpoint conditions }%
(x_0,x_1).
\end{eqnarray*}

$~$

\subsection{Dissipative crowd kinematics ($\cal CK$)}

The crowd action (\ref{Fey1}) with its amalgamate Lagrangian dynamics (\ref{LgrEq}) and amalgamate Hamiltonian dynamics (\ref{ham1}), as well as the crowd force law (\ref{covForceCrowd})
define the macroscopic crowd dynamics, $\cal CD$. Suppose, for a moment, that $\cal CD$ is force--free and dissipation free, therefore conservative. Now, the basic characteristic of the conservative Lagrangian/Hamiltonian systems evolving in the phase space spanned by the system coordinates and their velocities/momenta, is that their \emph{flow} $\varphi _{t}^L$ (explained below) preserves the phase--space volume,
as proposed by the Liouville theorem, which is the well-known fact in
statistical mechanics. However, the preservation of the phase volume causes
structural instability of the conservative system, i.e., the
phase--space spreading effect by which small phase regions $R_{t}$ will
tend to get distorted from the initial one $R_{o}$ during the conservative
system evolution. This problem, governed by entropy growth ($\partial_t S>0$), is much more serious in higher dimensions
than in lower dimensions, since there are so many `directions' in which the
region can locally spread (see \cite{Penrose,GaneshSprBig}). This phenomenon is related to \emph{conservative Hamiltonian chaos} (see section \ref{entr} below).

However, this situation is not very frequent in case of `organized' human crowd. Its self-organization mechanisms are clearly much stronger than the conservative statistical mechanics effects, which we interpret in terms of Prigogine's dissipative structures (see Appendix). Formally, if dissipation of energy in a system is much stronger then its inertial characteristics, then instead of the second-order Newton--Lagrangian dynamic equations of motion, we are actually dealing with the first-order driftless (non-acceleration, non-inertial) kinematic equations of motion (see Appendix, eq. (\ref{eq:1-1})), which is related to \emph{dissipative chaos} \cite{Nicolis2}. Briefly, the dissipative crowd flow can be depicted like this: from the set of initial conditions for individual agents, the crowd evolves in time towards the set of the corresponding \emph{entangled attractors,}\footnote{Recall that quantum entanglement is a quantum mechanical phenomenon in which the quantum states of two or more objects are linked together so that one object can no longer be adequately described without full mention of its counterpart -- even though the individual objects may be spatially separated. This interconnection leads to correlations between observable physical properties of remote systems. The related phenomenon of wave-function collapse gives an impression that measurements performed on one system instantaneously influence the other systems entangled with the measured system, even when far apart.

Entanglement has many applications in quantum information theory. Mixed state entanglement can be viewed as a resource for quantum communication.
A common measure of entanglement is the entropy of a mixed quantum state (see, e.g. \cite{QuLeap}). Since a mixed quantum state $\rho$ is a probability distribution over a quantum ensemble, this leads naturally to the definition of the \emph{von Neumann entropy}, ~
$S(\rho) = -  \hbox{Tr} \left( \rho \log_2 {\rho} \right),$~
which is obviously similar to the classical \emph{Shannon entropy} for probability distributions $(p_1, \cdots, p_n)$, defined as
~$S(p_1, \cdots, p_n) = - \sum_i p_i \log_2 p_i.$~
As in statistical mechanics, one can say that the more uncertainty (number of microstates) the system should possess, the larger is its entropy. Entropy gives a tool which can be used to quantify entanglement. If the overall system is pure, the entropy of one subsystem can be used to measure its degree of entanglement with the other subsystems.

The most popular issue in a research on dissipative quantum
brain modelling has been
\textit{quantum entanglement} between the \textit{brain} and its
\textit{environment} \cite{MM,PV}, where the
brain--environment system has an entangled `memory' state,
identified with the ground (vacuum) state $| 0>_{\mathcal{N}}$,
that cannot be factorized into two single--mode
states. (In the Vitiello--Pessa dissipative quantum brain
model \cite{MM,PV}, the evolution of the $\mathcal{N}$--coded
memory system was represented as a trajectory of given initial
condition running over time--dependent states
$|0(t)>_{\mathcal{N}}$, each one minimizing the free energy
functional.) Similar to this microscopic brain--environment
entanglement, we propose a kind of \textit{macroscopic
entanglement} between the operating modes of the crowd psycho--physical controller and its biodynamics, which can be considered as a `long--range
correlation'.

Applied externally to the dimension of the crowd $3n-$manifold $M$, entanglement effectively reduces the number of active degrees of freedom in (\ref{crwdMan}).} ~which are mutually separated by fractal (non-integer dimension) separatrices.

In this subsection we elaborate on the dissipative crowd kinematics ($\cal CK$), which is self--controlled and dominates the $\cal CD$ if the crowd's inertial forces are much weaker then the the crowd's dissipation of energy, presented here in the form of nonlinear velocity controllers.

Recall that the essential concept
in dynamical systems theory is the notion of a \textit{vector--field} (that
we will denote by a boldface symbol), which assigns a tangent vector to each
point $p$ in the manifold in case. In particular, $\mathbf{v}$ is a gradient
vector--field if it equals the gradient of some scalar function. A \textit{%
flow--line} of a vector--field $\mathbf{v}$ is a path $\mathbf{\gamma }(t)$
satisfying the vector ODE, ~
$
\mathbf{\dot{\gamma}}(t)=\mathbf{v}(\mathbf{\gamma }(t)),~
$
that is, $\mathbf{v}$ yields the velocity field of the path $\mathbf{\gamma }%
(t)$. The set of all flow lines of a vector--field $\mathbf{v}$ comprises
its flow ~$\varphi _{t}$~ that is (technically, see e.g., \cite%
{GaneshSprBig,GaneshADG}) a one--parameter Lie group of diffeomorphisms
(smooth bijective functions) generated by a vector-field $\mathbf{v}$ on $M$%
, such that
\begin{equation*}
\varphi _{t}\circ \varphi _{s}=\varphi _{t+s},\qquad \varphi _{0}=\text{%
identity},\qquad\text{which gives:}\quad \gamma(t) = \varphi_t(\gamma(0)).
\end{equation*}
Analytically, a vector-field $\mathbf{v}$ is defined as a set of autonomous
ODEs. Its solution gives the flow $\varphi _{t}$, consisting of integral
curves (or, flow lines) $\mathbf{\gamma }(t)$ of the vector--field, such
that all the vectors from the vector-field are tangent to integral curves at
different representative points $p\in M$. In this way, through every
representative point $p\in M$ passes both a curve from the flow and its
tangent vector from the vector-field.
Geometrically, vector-field is defined as a cross-section of the tangent
bundle $TM$ of the manifold $M$.

In general, given an $n$D frame $\{\partial _{i}\}\equiv \{\partial
/\partial x^{i}\}$ on a smooth $n-$manifold $M$ (that is, a basis of tangent
vectors in a local coordinate chart $x^{i}=(x^{1},...,x^{n})\subset M$), we
can define any vector-field $\mathbf{v}$ on $M$ by its components $%
v^{i}=v^{i}(t)$ as
\begin{equation*}
\mathbf{v}=v^{i}\partial _{i}=v^{i}\frac{\partial }{\partial x^{i}}=v^{1}%
\frac{\partial }{\partial x^{1}}+...+v^{n}\frac{\partial }{\partial x^{n}}.
\end{equation*}
Thus, a vector-field $\mathbf{v}\in \mathcal{X}(M)$ (where $\mathcal{X}(M)$
is the set of all smooth vector-fields on $M$) is actually a differential
operator that can be used to differentiate any smooth scalar function $%
f=f(x^{1},...,x^{n})$ on $M$, as a \emph{directional derivative} of $f$ in
the direction of $\mathbf{v}.$ This is denoted simply $\mathbf{v}f$, such
that
\begin{equation*}
\mathbf{v}f=v^{i}\partial _{i}f=v^{i}\frac{\partial f}{\partial x^{i}}=v^{1}%
\frac{\partial f}{\partial x^{1}}+...+v^{n}\frac{\partial f}{\partial x^{n}}.
\end{equation*}

In particular, if $\mathbf{v}=\dot{\gamma}(t)$ is a velocity vector-field of
a space curve $\gamma (t)=(x^{1}(t),...,x^{n}(t)),$ defined by its
components $v^{i}=\dot{x}^{i}(t),$ directional derivative of $f(x^{i})$ in
the direction of $\mathbf{v}$ becomes
\begin{equation*}
\mathbf{v}f=\dot{x}^{i}\partial _{i}f=\frac{dx^{i}}{dt}\frac{\partial f}{%
\partial x^{i}}=\frac{df}{dt}=\dot{f},
\end{equation*}
which is a rate-of-change of $f$ along the curve $\gamma (t)$ at a point $%
x^{i}(t).$

Given two vector-fields, $\mathbf{u}=u^{i}\partial _{i},\mathbf{v}%
=v^{i}\partial _{i}\in \mathcal{X}(M)$, their Lie bracket (or, commutator)
is another vector-field $[\mathbf{u},\mathbf{v}]$ $\in \mathcal{X}(M),$
defined by
\begin{equation*}
\lbrack \mathbf{u},\mathbf{v}]=\mathbf{uv}-\mathbf{vu}=u^{i}\partial
_{i}\,v^{j}\partial _{j}-v^{j}\partial _{j}\,u^{i}\partial _{i},
\end{equation*}
which, applied to any smooth function $f$ on $M,$ gives
\begin{equation*}
\lbrack \mathbf{u},\mathbf{v}](f)=\mathbf{u}\left( \mathbf{v}(f)\right) -%
\mathbf{v}\left( \mathbf{u}(f)\right) .
\end{equation*}

The Lie bracket measures the failure of `mixed directional derivatives' to
commute. Clearly, mixed partial derivatives \textit{do} commute, ~$\lbrack \partial _{i},\partial _{j}]=0$, while in general it is \emph{not} the case, ~$\lbrack \mathbf{u},\mathbf{v}]\neq 0$. In addition, suppose that $\mathbf{u}$ generates the flow $\varphi _{t}$ and $\mathbf{v}$
generates the flow $\varphi _{s}$. Then, for any smooth function $f$ on $M,$
we have at any point $p$ on $M,$%
\begin{equation*}
\lbrack \mathbf{u},\mathbf{v}](f)(p)=\frac{\partial ^{2}}{\partial t\partial
s}\left( f(\varphi _{s}(\varphi _{t}(p))\right) -f(\varphi _{t}(\varphi
_{s}(p))),
\end{equation*}%
which means that in $f(\varphi _{s}(\varphi _{t}(p))$ we are starting at $p$%
, flowing along $\mathbf{v}$ a little bit, then along $\mathbf{u}$ a little
bit, and then evaluating $f$, while in $f(\varphi _{t}(\varphi _{s}(p))$ we
are flowing first along $\mathbf{u}$ and then $\mathbf{v}$. Therefore, the
Lie bracket infinitesimally measures how these flows fail to commute.

The Lie bracket satisfies the following three properties (for any three
vector-fields $\mathbf{u,v,w}\in M$ and two constants $a,b$ -- thus forming a Lie algebra on the crowd manifold $M$):

(i)~~ $[\mathbf{u},\mathbf{v}]=-[\mathbf{v},\mathbf{u}]-$ skew-symmetry;

(ii)~ $[\mathbf{u},a\mathbf{v}+b\mathbf{w}]=a[\mathbf{u},\mathbf{v}]+b[%
\mathbf{u},\mathbf{w}]-$ bilinearity; ~and

(iii) $[\mathbf{u},[\mathbf{v},\mathbf{w}]]+[\mathbf{v},[\mathbf{w},\mathbf{u%
}]]+[\mathbf{w},[\mathbf{u},\mathbf{v}]]-$ Jacobi identity.\newline
A new set of vector-fields on $M$ can be generated by repeated Lie brackets
of $\mathbf{u,v,w}\in M$.

The Lie bracket is a standard tool in geometric nonlinear control theory
(see, e.g. \cite{GaneshSprBig,GaneshADG}). Its action on vector-fields can
be best visualized using the popular car parking example, in which the
driver has two different vector--field transformations at his disposal. They
can turn the steering wheel, or they can drive the car forward or backward.
Here, we specify the state of a car by four coordinates: the $(x,y)$
coordinates of the center of the rear axle, the direction $\theta $ of the
car, and the angle $\phi $ between the front wheels and the direction of the
car. $l$ is the constant length of the car. Therefore, the 4D configuration
manifold of a car is a set $M\equiv SO(2)\times \mathbb{R}^{2},$ coordinated
by $\mathbf{x=\{}x,y,\theta ,\phi \}$, which is slightly more complicated
than the individual crowd agent's 3D configuration manifold $SE(2)\equiv
SO(2)\times \mathbb{R},$ coordinated by $\mathbf{x=\{}x,y,\theta \}$. The
driftless car kinematics can be defined as a vector ODE:
\begin{equation}
\mathbf{\,\dot{x}}=\mathbf{u}(\mathbf{x})\,c_{1}+\mathbf{v}(\mathbf{x}%
)\,c_{2},  \label{Car2}
\end{equation}%
with two vector--fields, $\mathbf{u},\mathbf{v}\in \mathcal{X}(M)$, and two
scalar control inputs, $c_{1}$ and $c_{2}$. The infinitesimal car--parking transformations will be the following vector--fields
\begin{eqnarray*}
\mathbf{u}(\mathbf{x}) &\equiv &\text{\textsc{drive}}=\cos \theta \frac{%
\partial }{\partial x}+\sin \theta \frac{\partial }{\partial y}+\frac{\tan
\phi }{l}\frac{\partial }{\partial \theta }\equiv \left(
\begin{array}{c}
\cos \theta \\
\sin \theta \\
\frac{1}{l}\tan \phi \\
0%
\end{array}%
\right) , \\
\text{and}\qquad \mathbf{v}(\mathbf{x}) &\equiv &\text{\textsc{steer}}=\frac{%
\partial }{\partial \phi }\equiv \left(
\begin{array}{c}
0 \\
0 \\
0 \\
1%
\end{array}%
\right) .
\end{eqnarray*}%
The car kinematics (\ref{Car2}) therefore expands into a matrix ODE:
\begin{equation*}
\left(
\begin{array}{c}
\dot{x} \\
\dot{y} \\
\dot{\theta} \\
\dot{\phi}%
\end{array}%
\right) =\text{\textsc{drive}}\cdot c_{1\text{ }}+\text{\textsc{steer}}\cdot
c_{2\text{ }}\equiv \left(
\begin{array}{c}
\cos \theta \\
\sin \theta \\
\frac{1}{l}\tan \phi \\
0%
\end{array}%
\right) \cdot c_{1\text{ }}+\left(
\begin{array}{c}
0 \\
0 \\
0 \\
1%
\end{array}%
\right) \cdot c_{2\text{ }}.
\end{equation*}

However, \textsc{steer}\ and \textsc{drive}\ do not commute (otherwise we
could do all your steering at home before driving of on a trip). Their
combination is given by the Lie bracket
\begin{equation*}
\lbrack \mathbf{v},\mathbf{u}]\equiv \lbrack \text{\textsc{steer}},\text{%
\textsc{drive}}]=\frac{1}{l\cos ^{2}\phi }\frac{\partial }{\partial \theta }%
\equiv \text{\textsc{wriggle}}.
\end{equation*}%
The operation $[\mathbf{v},\mathbf{u}]\equiv $ \textsc{wriggle }$\equiv
\lbrack $\textsc{steer}$,$\textsc{drive}$]$ is the infinitesimal version of
the sequence of transformations: steer, drive, steer back, and drive back,
i.e.,
\begin{equation*}
\{\text{\textsc{steer}},\text{\textsc{drive}},\text{\textsc{steer}}^{-1},%
\text{\textsc{drive}}^{-1}\}.
\end{equation*}%
Now, \textsc{wriggle}\ can get us out of some parking spaces, but not tight
ones: we may not have enough room to \textsc{wriggle} out. The usual tight
parking space restricts the \textsc{drive}\ transformation, but not \textsc{%
steer}. A truly tight parking space restricts \textsc{steer}\ as well by
putting your front wheels against the curb.

Fortunately, there is still another commutator available:
\begin{eqnarray*}
\lbrack \mathbf{u},[\mathbf{v},\mathbf{u}]] &\equiv &[\text{\textsc{drive}},[%
\text{\textsc{steer}},\text{\textsc{drive}}]]=[[\mathbf{u},\mathbf{v}],%
\mathbf{u}]\equiv \\
\lbrack \text{\textsc{drive}},\text{\textsc{wriggle}}] &=&\frac{1}{l\cos
^{2}\phi }\left( \sin \theta \frac{\partial }{\partial x}-\cos \theta \frac{%
\partial }{\partial y}\right) \equiv \text{\textsc{slide}}.
\end{eqnarray*}%
The operation $[[\mathbf{u},\mathbf{v}],\mathbf{u}]\equiv $ \textsc{slide }$%
\equiv \lbrack $\textsc{drive}$,$\textsc{wriggle}$]$ is a displacement at
right angles to the car, and can get us out of any parking place. We just
need to remember to steer, drive, steer back, drive some more, steer, drive
back, steer back, and drive back:
\begin{equation*}
\{\text{\textsc{steer}},\text{\textsc{drive}},\text{\textsc{steer}}^{-1},%
\text{\textsc{drive}},\text{\textsc{steer}},\text{\textsc{drive}}^{-1},\text{%
\textsc{steer}}^{-1},\text{\textsc{drive}}^{-1}\}.
\end{equation*}%
We have to reverse steer in the middle of the parking place. This is not
intuitive, and no doubt is part of a common problem with parallel parking.

Thus, from only two controls, $c_{1}$ and $c_{2}$, we can form the
vector--fields \textsc{drive }$\equiv \mathbf{u}$, \textsc{steer }$\equiv
\mathbf{v}$, \textsc{wriggle }$\equiv $ $[\mathbf{v},\mathbf{u}],$ and
\textsc{slide }$\equiv \lbrack \lbrack \mathbf{u},\mathbf{v}],\mathbf{u}]$,
allowing us to move anywhere in the car configuration manifold $M\equiv
SO(2)\times \mathbb{R}^{2}$. All above computations are straightforward in $Mathematica^{TM}$\footnote{The above computations could instead be done in other available packages, such as Maple, by suitably translating the provided example code.} if we define the following three symbolic functions:

1. Jacobian matrix: ~~JacMat[v\_List, x\_List] := Outer[D, v, x];

2. Lie bracket: ~~LieBrc[u\_List, v\_List, x\_List] := JacMat[v, x] . u - JacMat[u, x] . v;

3. Repeated Lie bracket: ~~Adj[u\_List, v\_List, x\_List, k\_] := \\
$~$\hspace{5.6cm} If[k == 0, v, LieBrc[u, Adj[u, v, x, k - 1], x] ];

In case of the human crowd, we have a slightly simpler, but multiplied problem, i.e., superposition of $n$ individual agents' motions. So, we can define the dissipative crowd kinematics as a system of $n$ vector ODEs:
\begin{equation}
\,\mathbf{\dot{x}}^{k}=\mathbf{u}^{k}(\mathbf{x})\,c_{1}^{k}+\mathbf{v}^{k}(%
\mathbf{x})\,c_{2}^{k},\qquad\text{where}  \label{CrwdKin1}
\end{equation}
\begin{eqnarray*}
\mathbf{u}^{k}(\mathbf{x}) &\equiv &\text{\textsc{drive}}^{k}=\cos
^{k}\theta \frac{\partial }{\partial x^{k}}+\sin ^{k}\theta \frac{\partial }{%
\partial y^{k}}\equiv \left(
\begin{array}{c}
\cos ^{k}\theta  \\
\sin ^{k}\theta  \\
0%
\end{array}%
\right) ,\qquad \text{and} \\
\mathbf{v}^{k}(\mathbf{x}) &\equiv &\text{\textsc{steer}}^{k}=\frac{\partial
}{\partial \theta ^{k}}\equiv \left(
\begin{array}{c}
0 \\
0 \\
1%
\end{array}%
\right) ,\qquad \text{while \ }c_{1}^{k}\ \text{\ and \ }c_{2}^{k}\text{ \
are crowd controls.}
\end{eqnarray*}%
Thus, the crowd kinematics (\ref{CrwdKin1}) expands into the matrix ODE:
\begin{equation}
\left(
\begin{array}{c}
\dot{x} \\
\dot{y} \\
\dot{\theta}%
\end{array}%
\right) =\text{\textsc{drive}}^{k}\cdot c_{1\text{ }}^{k}+\text{\textsc{steer%
}}^{k}\cdot c_{2\text{ }}^{k}\equiv \left(
\begin{array}{c}
\cos ^{k}\theta  \\
\sin ^{k}\theta  \\
0%
\end{array}%
\right) \cdot c_{1\text{ }}^{k}+\left(
\begin{array}{c}
0 \\
0 \\
1%
\end{array}%
\right) \cdot c_{2\text{ }}^{k}.  \label{CrwdKin2}
\end{equation}

The dissipative crowd kinematics (\ref{CrwdKin1})--(\ref{CrwdKin2}) obeys the set of $n$%
-tuple integral rules of motion that are similar (though slightly simpler) to
the above rules of the car kinematics, including the following derived
vector-fields:

\textsc{wriggle}$^{k}$\textsc{\ }$\equiv \lbrack $\textsc{steer}$^{k},$%
\textsc{drive}$^{k}]\text{\textsc{\ }}\equiv \lbrack \mathbf{v}^{k},\mathbf{u%
}^{k}]$ \  and  \
\textsc{slide}$^{k}$\textsc{\ }$\equiv \lbrack $\textsc{drive}$^{k},$\textsc{%
wriggle}$^{k}]\equiv \lbrack \lbrack \mathbf{u}^{k},\mathbf{v}^{k}],\mathbf{u%
}^{k}].$

Thus, controlled by the two vector controls $c_{1}^{k}$ and $c_{2}^{k},$ the
crowd can form the vector--fields: \textsc{drive }$\equiv \mathbf{u}^{k}$,
\textsc{steer }$\equiv \mathbf{v}^{k}$, \textsc{wriggle }$\equiv $ $[\mathbf{%
v}^{k},\mathbf{u}^{k}],$ and \textsc{slide }$\equiv \lbrack \lbrack \mathbf{u%
}^{k},\mathbf{v}^{k}],\mathbf{u}^{k}]$, allowing it to move anywhere within
its configuration manifold $M$ given by (\ref{crwdMan}). Solution of the dissipative crowd kinematics (\ref{CrwdKin1})--(\ref{CrwdKin2}) defines the dissipative crowd flow, $\phi_t^{K}$.

Now, the general $\cal CD$--$\cal CK$ crowd behavior can be defined as a amalgamate flow (psycho--physical Lagrangian flow, $\phi_t^{L}$, plus dissipative kinematic flow, $\phi_t^{K}$) on the crowd manifold $M$ defined by (\ref{crwdMan}), \begin{equation*}C_t=\phi_t^{L}+\phi_t^{K}:t\mapsto (M(t),g(t)),\end{equation*} which is
a one-parameter family of homeomorphic (topologically equivalent) Riemannian manifolds\footnote{Proper differentiation of vector and tensor fields on a smooth Riemannian
manifold (like the crowd $3n-$manifold $M$) is performed using the \textit{%
Levi--Civita covariant derivative} (see, e.g., \cite{GaneshSprBig,GaneshADG}%
). Formally, let $M$ be a Riemannian $N-$manifold with the tangent bundle $TM
$ and a local coordinate system $\{x^{i}\}_{i=1}^{N}$ defined in an open set
$U\subset M$. The covariant derivative operator, $\nabla _{X}:C^{\infty
}(TM)\rightarrow C^{\infty }(TM)$, is the unique linear map such that for
any vector-fields $X,Y,Z,$ constant $c$, and scalar function $f$ the
following properties are valid:
\[
\nabla _{X+cY}=\nabla _{X}+c\nabla _{Y},\qquad \nabla _{X}(Y+fZ)=\nabla
_{X}Y+(Xf)Z+f\nabla _{X}Z,\qquad \nabla _{X}Y-\nabla _{Y}X=[X,Y],
\]%
where $[X,Y]$ is the Lie bracket of $X$ and $Y$. In local coordinates, the
metric $g$ is defined for any orthonormal basis $(\partial _{i}=\partial
/\partial x^{i})$ in $U\subset M$ by \ $g_{ij}=g(\partial _{i},\partial
_{j})=\delta _{ij},$ $\ \partial _{k}g_{ij}=0.$ Then the affine \textit{%
Levi--Civita connection} is defined on $M$ by
\[
\nabla _{\partial _{i}}\partial _{j}=\Gamma _{ij}^{k}\partial _{k},~~\text{ \
\ where \ }~~\Gamma _{ij}^{k}=\frac{1}{2}g^{kl}\left( \partial
_{i}g_{jl}+\partial _{j}g_{il}-\partial _{l}g_{ij}\right) \text{ \ are the
Christoffel symbols}.
\]

Now, using the covariant derivative operator $\nabla _{X}$ we can define the
\textit{Riemann curvature} $(3,1)-$tensor $\mathfrak{Rm}$ by
\[
\mathfrak{Rm}(X,Y)Z=\nabla _{X}\nabla _{Y}Z-\nabla _{Y}\nabla _{X}Z-\nabla
_{\lbrack X,Y]}Z,
\]%
which measures the curvature of the manifold by expressing how
noncommutative covariant differentiation is. The $(3,1)-$components $%
R_{ijk}^{l}$ of $\mathfrak{Rm}$ are defined in $U\subset M$ by
\[
\mathfrak{Rm}\left( \partial _{i},\partial _{j}\right) \partial
_{k}=R_{ijk}^{l}\partial _{l},~~\text{ \ \ or \ }~~~R_{ijk}^{l}=\partial
_{i}\Gamma _{jk}^{l}-\partial _{j}\Gamma _{ik}^{l}+\Gamma _{jk}^{m}\Gamma
_{im}^{l}-\Gamma _{ik}^{m}\Gamma _{jm}^{l}.
\]%
Also, the Riemann $(4,0)-$tensor $R_{ijkl}=g_{lm}R_{ijk}^{m}$ is defined as
the $g-$based inner product on $M$,
\[
R_{ijkl}=\left\langle \mathfrak{Rm}\left( \partial _{i},\partial _{j}\right)
\partial _{k},\partial _{l}\right\rangle .
\]

The first and second Bianchi identities for the Riemann $(4,0)-$tensor $%
R_{ijkl}$ hold,
\[
R_{ijkl}+R_{jkil}+R_{kijl}=0,\qquad \nabla _{i}R_{jklm}+\nabla
_{j}R_{kilm}+\nabla _{k}R_{ijlm}=0,
\]%
while the twice contracted second Bianchi identity reads: \ $2\nabla
_{j}R_{ij}=\nabla _{i}R.$

The $(0,2)$ \textit{Ricci tensor} $\mathfrak{Rc}$ is the trace of the
Riemann $(3,1)-$tensor $\mathfrak{Rm}$,
\[
\mathfrak{Rc}(Y,Z)+\mathrm{tr}(X\rightarrow \mathfrak{Rm}(X,Y)Z),~~\text{ \ \
so that \ }~~\mathfrak{Rc}(X,Y)=g(\mathfrak{Rm}(\partial _{i},X)\partial
_{i},Y),
\]%
Its components $R_{jk}=\mathfrak{Rc}\left( \partial _{j},\partial
_{k}\right) $ are given in $U\subset M$ by the contraction
\[
R_{jk}=R_{ijk}^{i},~\quad \text{or \ }~~R_{jk}=\partial _{i}\Gamma
_{jk}^{i}-\partial _{k}\Gamma _{ji}^{i}+\Gamma _{mi}^{i}\Gamma
_{jk}^{m}-\Gamma _{mk}^{i}\Gamma _{ji}^{m}.
\]

Finally, the scalar curvature $R$ is the trace of the Ricci tensor $%
\mathfrak{Rc}$, given in $U\subset M$ by:~ $R=g^{ij}R_{ij}. $} $(M,g=g_{ij})$, parameterized by a `time' parameter $t$. That is, $C_{t}$ can
be used for describing smooth deformations of the crowd manifold $M$ over time. The manifold family $(M(t),g(t))$ at time $t$ determines the manifold family $(M(t+dt),g(t+dt))
$ at an infinitesimal time $t+dt$ into the future, according to some presecribed geometric flow, like the celebrated \emph{Ricci
flow} \cite{Ham82,4-manifold,Harnack,surface} (that was an instrument for a proof of a 100--year old Poincar\'e conjecture),
\begin{equation}
\partial _{t}g_{ij}(t)=-2R_{ij}(t),  \label{RF}
\end{equation}%
where $R_{ij}$ is the Ricci curvature tensor (see Appendix) of the crowd manifold $M$ and $\partial _{t}g(t)$ is defined as
\begin{equation}
\partial _{t}g(t)\equiv \frac{d}{dt}g(t):=\lim_{dt\rightarrow 0}\frac{%
g(t+dt)-g(t)}{dt}. \label{RFlim}
\end{equation}%

\subsection{Aggregate behavioral--compositional dynamics ($\mathcal{AD}$)} \label{aggreg}

To formally develop the meso-level aggregate behavioral--compositional dynamics ($\mathcal{AD}$), we start with the crowd path integral (\ref{pathInt}), which can
be redefined if we Wick--rotate
the time variable $t$ to imaginary values, $t\mapsto \tau={\mathrm
i} t$, thereby transforming the Lorentzian path integral in real time into the Euclidean path integral in imaginary time. Furthermore, if we rectify the time axis back to the real line, we get the adaptive SFT--partition function as our proposed $\mathcal{AD}$--model:
\begin{equation}
\langle {\rm Physical~Action\,|\,Mental~Preparation}\rangle_{\mathrm{AD}} =\int_{\rm CD}\mathcal{D}[w,x,y]\, {\rm e}^{-A[x,y;t_{i},t_{j}]}. \label{sft}
\end{equation}

The adaptive $\mathcal{AD}$--transition amplitude $\langle {\rm Physical~Action\,|\,Mental~Preparation}\rangle_{\mathrm{AD}}$ as defined by the SFT--partition function (\ref{sft}) is a general model of a\textit{Markov stochastic process}. Recall that Markov process is a random
process characterized by a \emph{lack of memory}, i.e., the
statistical properties of the immediate future are uniquely
determined by the present, regardless of the past (see, e.g. \cite{Gardiner,GaneshSprBig}).
The $N-$dimensional Markov process can be defined by the Ito
stochastic differential equation,
\begin{eqnarray}
dx_{i}(t) &=&A_{i}[x^{i}(t),t]dt+B_{ij}[x^{i}(t),t]\,dW^{j}(t), \\
x^{i}(0) &=&x_{i0},\qquad (i,j=1,\dots ,N) \label{Ito1}
\end{eqnarray}%
or corresponding \emph{Ito stochastic integral equation}
\begin{eqnarray}
x^{i}(t)=x^{i}(0)+\int_{0}^{t}ds\,A_{i}[x^{i}(s),s]+\int_{0}^{t}dW^{j}(s)%
\,B_{ij}[x^{i}(s),s], \label{Ito2}
\end{eqnarray}%
in which $x^{i}(t)$ is the variable of interest, the vector
$A_{i}[x(t),t]$ denotes deterministic \emph{drift}, the matrix
$B_{ij}[x(t),t]$ represents
continuous stochastic \emph{diffusion fluctuations}, and $W^{j}(t)$ is an $%
N-$ variable \textit{Wiener process} (i.e., generalized Brownian
motion \cite{Gardiner}) and $$dW^{j}(t)=W^{j}(t+dt)-W^{j}(t).$$

The two Ito equations (\ref{Ito1})--(\ref{Ito2}) are equivalent to the general \textit{Chapman--Kolmogorov probability equation} (see equation (\ref{CK}) below). There are
three well--known special cases of the
Chapman--Kolmogorov equation (see \cite{Gardiner}):
\begin{enumerate}
    \item When both $B_{ij}[x(t),t]$ and $W(t)$ are zero, i.e., in the case of
pure deterministic motion, it reduces to the \textit{Liouville
equation}
\begin{eqnarray*}
\partial _{t}P(x',t'|x'',t'')=-\sum_{i}\frac{\partial }
{\partial x^{i}}\left\{ A_{i}[x(t),t]\,P(x'%
,t'|x'',t'')\right\} .
\end{eqnarray*}
    \item When only $W(t)$ is zero, it reduces to the
    \textit{Fokker--Planck equation}
\begin{eqnarray*}
\partial _{t}P(x',t'|x'',t'')&=&
-\sum_{i}\frac{\partial }{\partial x^{i}}\left\{ A_{i}[x(t),t]\,P(x'%
,t'|x'',t'')\right\} \\
&+&\frac{1}{2}\sum_{ij}\frac{\partial ^{2}}{\partial x^{i}\partial
x^{j}}\left\{ B_{ij}[x(t),t]\,P(x',t'|x'',t'')\right\} .
\end{eqnarray*}
    \item When both $A_{i}[x(t),t]$ and $B_{ij}[x(t),t]$ are zero,
i.e., the state--space consists of integers only, it reduces to
the \textit{Master equation} of discontinuous jumps
\begin{eqnarray*}
\partial _{t}P(x',t'|x'',t'') =
\int dx\,W(x'|x'',t)\,P(x',t'|x'',t'')-\int dx\,W(x''|x',t)\,P(%
x',t'|x'',t'').
\end{eqnarray*}
\end{enumerate}

The \textit{Markov assumption} can now be formulated in terms of
the conditional probabilities $P(x^{i},t_{i})$: if the times
$t_{i}$ increase from right to left, the conditional probability
is determined entirely by the knowledge of the most recent
condition. Markov process is generated by a set of conditional
probabilities whose probability--density $P=P(x',t'|x'',t'')$
evolution obeys the general \textit{Chapman--Kolmogorov
integro--differential equation}
\begin{eqnarray*}
\partial _{t}P &=&-\sum_{i}\frac{\partial }{\partial x^{i}}\left\{
A_{i}[x(t),t]\,P\right\}
~+~\frac{1}{2}\sum_{ij}\frac{\partial ^{2}}{\partial x^{i}\partial x^{j}}%
\left\{ B_{ij}[x(t),t]\,P\right\}\\
&+&\int dx\left\{ W(x^{\prime
}|x^{\prime \prime },t)\,P-W(x^{\prime \prime }|x^{\prime
},t)\,P\right\} \label{CK}
\end{eqnarray*}
including \emph{deterministic drift}, \emph{diffusion
fluctuations} and \emph{discontinuous jumps} (given respectively
in the first, second and third terms on the r.h.s.). This general Chapman--Kolmogorov integro-differential
equation (\ref{CK}), with its conditional probability density evolution,
$P=P(x',t'|x'',t'')$, is represented by our SFT--partition function (\ref{sft}).

Furthermore, discretization of the adaptive SFT--partition function (\ref{sft}) gives the standard
\emph{partition function} (see Appendix)
\begin{equation}
Z=\sum_j{\mathrm e}^{-w_jE^j/T}, \label{partition}
\end{equation}
where $E^j$ is the motion energy eigenvalue (reflecting each
possible motivational energetic state), $T$ is the
temperature--like environmental control parameter, and the sum
runs over all ID energy eigenstates (labelled by the index
$j$). From (\ref{partition}), we can calculate the \emph{transition entropy}, as $S = k_B\ln Z$ (see the next section).

\section{Entropy, chaos and phase transitions in the crowd manifold} \label{entr}

Recall that nonequilibrium phase transitions \cite%
{Haken1,Haken2,Haken3,Haken4,HakenBrain} are phenomena which bring about {qualitative}
physical changes at the macroscopic level in presence of the same
microscopic forces acting among the constituents of a system. In this
section we extend the $\cal CD$ formalism to incorporate both algorithmic and geometrical entropy as well as dynamical chaos \cite%
{TacaNODY,StrAttr,Complexity} between the entropy--growing phase of Mental Preparation and the entropy--conserving phase of Physical Action, together with the associated topological phase transitions.

\subsection{Algorithmic entropy}

The Boltzmann and Shannon (hence also Gibbs entropy, which is Shannon
entropy scaled by $k\ln2$, where $k$ is the Bolzmann constant) entropy
definitions involve the notion of \emph{ensembles}. Membership of
microscopic states in ensembles defines the probability density function
that underpins the entropy function; the result is that the entropy
of a definite and completely known microscopic state is precisely
zero. Bolzmann entropy defines the probabilistic model of the system
by effectively discarding part of the information about the system,
while the Shannon entropy is concerned with measuring the ignorance
of the observer -- the amount of missing information -- about the system.

Zurek proposed a new physical entropy measure that can be applied
to individual microscopic system states and does not use the ensemble
structure. This is based on the notion of a fixed individually random
object provided by Algorithmic Information Theory and Kolmogorov Complexity:
put simply, the randomness $K(x)$ of a binary string $x$ is the
length in terms of number of bits of the smallest program $p$ on
a universal computer that can produce $x$.

While this is the basic idea, there are some important technical details
involved with this definition. The randomness definition uses the
prefix complexity $K(.)$ rather than the older Kolmogorov complexity
measure $C(.)$: the prefix complexity $K(x|y)$ of $x$ given $y$
is the Kolmogorov complexity $C_{\phi_{u}}(x|y)=\min\left\{ p\:|\: x=\phi_{u}(\left\langle y,p\right\rangle )\right\} $
(with the convention that $C_{\phi_{u}}(x|y)=\infty$ if there is
no such $p$) that is taken with respect to a reference universal
partial recursive function $\phi_{u}$ that is a universal prefix
function. Then the prefix complexity $K(x)$ of $x$ is just $K(x|\varepsilon)$
where $\varepsilon$ is the empty string. A partial recursive prefix
function $\phi:M\rightarrow\mathbb{N}$ is a partial recursive
function such that if $\phi(p)<\infty$ and $\phi(q)<\infty$ then
$p$ is not a proper prefix of $q$: that is, we restrict the complexity
definition to a set of strings (which are descriptions of effective
procedures) such that none is a proper prefix of any other. In this
way, all effective procedure descriptions are \emph{self-delimiting}:
the total length of the description is given within the description
itself. A universal prefix function $\phi_{u}$is a prefix function
such that $\forall n\in\mathbb{N\;}\phi_{u}(\left\langle y,\left\langle n,p\right\rangle \right\rangle )=\phi_{n}(\left\langle y,p\right\rangle )$,
where $\phi_{n}$ is numbered $n$ according to some Godel numbering
of the partial recursive functions; that is, a universal prefix function
is a partial recursive function that simulates any partial recursive
function. Here, $\left\langle x,y\right\rangle $ stands for a total
recusive one-one mapping from $\mathbb{N}$$\times\mathbb{N}$ into
$\mathbb{N}$, $\left\langle x_{1},x_{2},\ldots,x_{n}\right\rangle =\left\langle x_{1},\left\langle x_{2},\ldots,x_{n}\right\rangle \right\rangle $,
$\mathbb{N}$ is the set of natural numbers, and $M=\left\{ 0,1\right\} ^{*}$is
the set of all binary strings.

This notion of entropy circumvents the use of probability to give
a concept of entropy that can be applied to a fully specified macroscopic
state: the algorithmic randomness of the state is the length of the
shortest possible effective description of it. To illustrate, suppose
for the moment that the set of microscopic states is countably infinite,
with each state identified with some natural number. It is known that
the discrete version of the Gibbs entropy (and hence of Shannon's
entropy) and the algorithmic entropy are asymptotically consistent
under mild assumptions. Consider a system with a countably infinite
set of microscopic states $X$ supporting a probability density function
$P(.)$ so that $P(x)$ is the probability that the system is in microscopic
state $x\in X$. Then the Gibbs entropy is $S_{G}(P)=-(k\ln2)\underset{x\in X}{\sum}P(x)\log P(x)$
(which is Shannon's information-theoretic entropy $H(P)$ scaled by
$k\ln2$). Supposing that $P(.)$ is recursive, then $S_{G}(P)=(k\ln2)\underset{x\in X}{\sum}P(x)K(x)+C$,
where $C_{\phi}$ is a constant depending only on the choice of the
reference universal prefix function $\phi$. Hence, as a measure of
entropy, the function $K(.)$ manifests the same kind of behavior
as Shannon's and Gibbs entropy measures.

Zurek's proposal was of a new physical entropy measure that includes
contributions from both the randomness of a state and ignorance about
it. Assume now that we have determined the macroscopic parameters
of the system, and encode this as a string - which can always be converted
into an equivalent binary string, which is just a natural number under
a standard encoding. It is standard to denote the binary string and
its corresponding natural number interchangeably; here let $x$ be
the encoded macroscopic parameters. Zurek's definition of \emph{algorithmic
entropy} of the macroscopic state is then $K(x)+H_{x}$, where $H_{x}=S_{B}(x)/(k\ln2)$,
where $S_{B}(x)$ is the Bolzmann entropy of the system constrained
by $x$ and $k$ is Bolzmann's constant; the physical version of the
algorithmic entropy is therefore defined as $S_{A}(x)=(k\ln2)(K(x)+H_{x})$.
Here $H_{x}$represents the level of ignorance about the microscopic
state, given the parameter set $x$; it can decrease towards zero
as knowledge about the state of the system increases, at which point
the algorithmic entropy reduces to the Bolzmann entropy.

\subsection{Ricci flow and Perelman entropy--action on the crowd manifold}

Recall that the inertial metric crowd flow, $C_{t}:t\mapsto (M(t),g(t))$ on the crowd $3n-$manifold (\ref{crwdMan}) is
a one-parameter family of homeomorphic Riemannian manifolds $(M,g)$, evolving by the Ricci flow (\ref{RF})--(\ref{RFlim}).

Now, given a smooth scalar function $u:M \rightarrow \mathbb{R}$ on the Riemannian crowd $3n-$manifold $M $, its Laplacian operator $\Delta $ is
locally defined as
\begin{equation*}
\Delta u=g^{ij}\nabla _{i}\nabla _{j}u,
\end{equation*}%
where $\nabla _{i}$ is the covariant derivative (or, Levi--Civita
connection, see Appendix). We say that a smooth
function $u:M \times \lbrack 0,T)\rightarrow \mathbb{R},$ where $T\in
(0,\infty ],$ is a solution to the heat equation (see Appendix, eq. (\ref{t81})) on $M $ if
\begin{equation}
\partial _{t}u=\Delta u.  \label{heat1}
\end{equation}%
One of the most important properties satisfied by the heat equation is the {%
maximum principle}, which says that for any smooth solution to the heat
equation, whatever point-wise bounds hold at $t=0$ also hold for $t>0$ \cite%
{CaoChow}. This property exhibits the smoothing behavior of the heat
diffusion (\ref{heat1}) on $M $.

Closely related to the heat diffusion (\ref{heat1}) is the (the Fields medal winning) Perelman entropy--action functional, which is on a $3n-$manifold $M $ with a Riemannian
metric $g_{ij}$ and a (temperature-like) scalar function $f$ given by \cite%
{Perel1}
\begin{equation}
\mathcal{E}=\int_{M }(R+|\nabla f|^{2}){\mathrm{e}}^{-f}d\mu   \label{F}
\end{equation}%
where $R$ is the scalar Riemann curvature on $M$, while $d\mu $ is the volume $3n-$form on $M$, defined as
\begin{equation}
d\mu =\sqrt{\det (g_{ij})}\,dx^{1}\wedge dx^{2}\wedge ...\wedge dx^{3n}.
\label{dmu}
\end{equation}

During the {Ricci flow} (\ref{RF})--(\ref{RFlim}) on the crowd manifold (\ref{crwdMan}), that is, during the inertial metric crowd flow, $C_{t}:t\mapsto (M(t),g(t))$, the Perelman entropy functional (\ref{F}) evolves as
\begin{equation}
\partial _{t}\mathcal{E}=2\int |R_{ij}+\nabla _{i}\nabla _{j}f|^{2}{\mathrm{e%
}^{-f}}d\mu {.}  \label{dF}
\end{equation}

Now, the \emph{crowd breathers} are solitonic crowd behaviors, which could be given by localized
periodic solutions of some nonlinear soliton PDEs, including the exactly
solvable sine--Gordon equation and the focusing nonlinear Schr\"{o}dinger
equation. In particular, the time--dependent crowd inertial metric $g_{ij}(t)$, evolving by the Ricci flow $g(t)$ given by (\ref{RF})--(\ref{RFlim}) on the crowd $3n-$manifold $M $ is the \emph{Ricci crowd
breather}, if for some $t_{1}<t_{2}$ and $\alpha >0$ the metrics $\alpha
g_{ij}(t_{1})$ and $g_{ij}(t_{2})$ differ only by a diffeomorphism; the
cases $\alpha =1,\alpha <1,\alpha >1$ correspond to steady, shrinking and
expanding crowd breathers, respectively. Trivial crowd breathers, for which
the metrics $g_{ij}(t_{1})$ and $g_{ij}(t_{2})$ on $M $ differ only by
diffeomorphism and scaling for each pair of $t_{1}$ and $t_{2}$, are the \emph{crowd
Ricci solitons}. Thus, if we consider the Ricci flow (\ref{RF})--(\ref{RFlim}) as a biodynamical system on
the space of Riemannian metrics modulo diffeomorphism and scaling, then
crowd breathers and solitons correspond to periodic orbits and fixed points
respectively. At each time the Ricci soliton metric satisfies on $M $
an equation of the form \cite{Perel1}
\begin{equation*}
R_{ij}+cg_{ij}+\nabla _{i}b_{j}+\nabla _{j}b_{i}=0,
\end{equation*}%
where $c$ is a number and $b_{i}$ is a 1--form; in particular, when $b_{i}=%
\frac{1}{2}\nabla _{i}a$ for some function $a$ on $M ,$ we get a
gradient Ricci soliton.

Define $\lambda (g_{ij})=\inf \mathcal{E}(g_{ij},f),$ where infimum is taken
over all smooth $f,$ satisfying
\begin{equation}
\int_{M }{\mathrm{e}^{-f}}d\mu =1.  \label{eDm}
\end{equation}%
$\lambda (g_{ij})$ is the lowest eigenvalue of the operator $-4\Delta +R.$
Then the entropy evolution formula (\ref{dF}) implies that $\lambda
(g_{ij}(t))$ is non-decreasing in $t,$ and moreover, if $\lambda
(t_{1})=\lambda (t_{2}),$ then for $t\in \lbrack t_{1},t_{2}]$ we have $%
R_{ij}+\nabla _{i}\nabla _{j}f=0$ for $f$ which minimizes $\mathcal{E}$ on $%
M $ \cite{Perel1}. Therefore, a steady breather on $M $ is
necessarily a steady soliton.

If we define the conjugate {heat operator} on $M $ as
\begin{equation*}
\Box ^{\ast }=-\partial /\partial t-\Delta +R
\end{equation*}%
then we have the {conjugate heat equation}:
~$
\Box ^{\ast }u=0.
$~

The entropy functional (\ref{F}) is nondecreasing under the coupled {%
Ricci--diffusion flow} on $M $ \cite{IvRicciSiam}
\begin{equation}
\partial _{t}g_{ij} =-2R_{ij}, \qquad
\partial _{t}u =-\Delta u+\frac{R}{2}u-\frac{|\nabla u|^{2}}{u},
\label{conHeat}
\end{equation}%
where the second equation ensures
~$
\int_{M }u^{2}d\mu =1,
$~
to be preserved by the Ricci flow $g(t)$ on $M $. If we define $\ u=%
\mathrm{e}^{-\frac{f}{2}}$, then (\ref{conHeat}) is equivalent to $f-$%
evolution equation on $M $ (the nonlinear backward heat equation),
\begin{equation*}
\partial _{t}f=-\Delta f+|\nabla f|^{2}-R,
\end{equation*}%
which instead preserves (\ref{eDm}). The coupled Ricci--diffusion flow (\ref%
{conHeat}) is the most general biodynamic model of the crowd reaction--diffusion
processes on $M$. In a recent study \cite{DifJennings} this general model has
been implemented for modelling a generic perception--action cycle with
applications to robot navigation in the form of a dynamical grid.

Perelman's functional $\mathcal{E}$ is analogous to negative thermodynamic
entropy \cite{Perel1}. Recall (see Appendix) that thermodynamic {partition function} for a
generic canonical ensemble at temperature $\beta ^{-1}$ is given by
\begin{equation}
Z=\int \mathrm{e}^{{-\beta E}}d\omega (E),  \label{Z}
\end{equation}%
where $\omega (E)$ is a `density measure', which does not depend on $\beta .$
From it, the {average energy} is given by
~$
\left\langle E\right\rangle =-\partial _{\beta }\ln Z,
$
the {entropy }is
~$
S=\beta \left\langle E\right\rangle +\ln Z,
$
and the {fluctuation }is
~$
\sigma =\left\langle (E-\left\langle E\right\rangle )^{2}\right\rangle
=\partial _{\beta ^{2}}\ln Z.
$

If we now fix a closed $3n-$manifold $M $ with a probability measure $m$
and a metric $g_{ij}(\tau )$ that depends on the temperature $\tau $, then
according to equation
\begin{equation*}
\partial _{\tau }g_{ij}=2(R_{ij}+\nabla _{i}\nabla _{j}f),
\end{equation*}%
the partition function (\ref{Z}) is given by
\begin{equation}
\ln Z=\int (-f+\frac{n}{2})\, dm.  \label{lnZ}
\end{equation}%
From (\ref{lnZ}) we get (see \cite{Perel1})
\begin{eqnarray*}
&&\left\langle E\right\rangle  =-\tau ^{2}\int_{M}(R+|\nabla f|^{2}-\frac{n}{%
2\tau })\, dm,\qquad S=-\int_{M}(\tau (R+|\nabla f|^{2})+f-n)\, dm, \\
&&\sigma  =2\tau ^{4}\int_{M}|R_{ij}+\nabla _{i}\nabla _{j}f-\frac{1}{2\tau }%
g_{ij}|^{2}\, dm,\qquad\text{where~~}\, dm=u\,dV,~~ u=(4\pi \tau )^{-\frac{n}{2}}\mathrm{e}^{-f}.
\end{eqnarray*}

From the above formulas, we see that the fluctuation $\sigma $ is
nonnegative; it vanishes only on a gradient shrinking soliton. $\left\langle
E\right\rangle $ is nonnegative as well, whenever the flow exists for all
sufficiently small $\tau >0$. Furthermore, if the heat function $u$: (a)
tends to a $\delta -$function as $\tau \rightarrow 0,$ or (b) is a limit of
a sequence of partial heat functions $u_{i},$ such that each $u_{i}$ tends
to a $\delta -$function as $\tau \rightarrow \tau _{i}>0,$ and $\tau
_{i}\rightarrow 0,$ then the entropy $S$ is also nonnegative. In case (a),
all the quantities $\left\langle E\right\rangle ,S,\sigma $ tend to zero as $%
\tau \rightarrow 0,$ while in case (b), which may be interesting if $%
g_{ij}(\tau )$ becomes singular at $\tau =0,$ the entropy $S$ may tend to a
positive limit.

\subsection{Chaotic inter-phase in crowd dynamics induced by its Riemannian geometry change}

Recall that $\cal CD$ transition map (\ref{cd}) is defined by the chaotic
crowd phase--transition amplitude
\begin{equation*}
\left\langle \overset{\partial _{t}S=0}{\mathrm{PHYS.~ACTION}}\right\vert
CHAOS\left\vert \overset{\partial _{t}S>0}{\mathrm{MENTAL~PREP.}}%
\right\rangle :=\int_M \mathcal{D}[x]\,\mathrm{e}^{iA[x]},
\end{equation*}
where we expect the inter-phase chaotic behavior (see \cite{IAY}). To show that this chaotic inter-phase is caused by the change in Riemannian geometry of the crowd $3n-$manifold $M$, we will first simplify the $\cal CD$ action functional (\ref{Fey1}) as
\begin{equation}
A[x]={\frac{1}{2}}\int_{t_{ini}}^{t_{fin}}[g_{ij}\,\dot{x}^{i}\dot{x}%
^{j}-V(x,\dot{x})]\,dt,  \label{locAct}
\end{equation}%
with the associated standard Hamiltonian, corresponding to the amalgamate version of (\ref{ham1}),
\begin{equation}
H(p,x)=\sum_{i=1}^{N}\frac{1}{2}p_{i}^{2}+V(x,\dot{x}),  \label{Ham}
\end{equation}%
where $p_i$ are the SE(2)--momenta, canonically conjugate to the individual agents' SE(2)--coordinates $x^i,~(i=1,...,3n)$. Biodynamics of systems with action (\ref{locAct}) and
Hamiltonian (\ref{Ham}) are given by the set of \emph{geodesic equations}
\cite{GaneshSprBig,GaneshADG}
\begin{equation}
\frac{d^{2}x^{i}}{ds^{2}}+\Gamma _{jk}^{i}\frac{dx^{j}}{ds}\frac{dx^{k}}{ds}%
=0,  \label{geod-mot}
\end{equation}%
where $\Gamma _{jk}^{i}$ are the Christoffel symbols of the affine
Levi--Civita connection of the Riemannian $\cal CD$ manifold $M$ (see Appendix).
In this geometrical framework, the instability of the trajectories is the
instability of the geodesics, and it is completely determined by the
curvature properties of the $\cal CD$ manifold $M$ according to
the {Jacobi equation} of geodesic deviation \cite{GaneshSprBig,GaneshADG}
\begin{equation}
\frac{D^{2}J^{i}}{ds^{2}}+R_{~jkm}^{i}\frac{dx^{j}}{ds}J^{k}\frac{dx^{m}}{ds}%
=0,  \label{eqJ}
\end{equation}%
whose solution $J$, usually called {Jacobi variation field}, locally
measures the distance between nearby geodesics; $D/ds$ stands for the {%
covariant derivative} along a geodesic and $R_{~jkm}^{i}$ are the components
of the {Riemann curvature tensor} of the $\cal CD$ manifold $M$.

The relevant part of the Jacobi equation (\ref{eqJ}) is given by the {%
tangent dynamics equation} \cite{pre96,CCC97}
\begin{equation}
\ddot{J}^{\,i}+R_{~0k0}^{i}J^{k}=0,\qquad (i,k=1,\dots ,3n),
\label{eqdintang}
\end{equation}%
where the only non-vanishing components of the curvature tensor of the
$\cal CD$ manifold $M$ are
\begin{equation}
R_{~0k0}^{i}=\partial ^{2}V/\partial x^{i}\partial x^{k}. \label{rok}
\end{equation}

The tangent dynamics equation (\ref{eqdintang}) can be used to define {%
Lyapunov exponents} in dynamical systems given by the Riemannian action (\ref%
{locAct}) and Hamiltonian (\ref{Ham}), using the formula \cite{physrep}
\begin{equation}
\lambda _{1}=\lim_{t\rightarrow \infty }1/2t\log (M
_{i=1}^{N}[J_{i}^{2}(t)+J_{i}^{2}(t)]/M
_{i=1}^{N}[J_{i}^{2}(0)+J_{i}^{2}(0)]).  \label{Lyap1}
\end{equation}%
Lyapunov exponents measure the {strength of dynamical chaos in the crowd
psycho--physical behavior}. The sum of positive Lyapunov exponents defines the \emph{Kolmogorov--Sinai entropy} (see Appendix).

\subsection{Crowd nonequilibrium phase transitions induced by manifold topology change}

Now, to relate these results to topological phase transitions within the
$\cal CD$ manifold $M$ given by (\ref{crwdMan}), recall that any two high--dimensional
manifolds $M _{v}$ and $M _{v^{\prime }}$ have the same topology
if they can be continuously and differentiably deformed into one another,
that is if they are diffeomorphic. Thus by {topology change} the `loss of
diffeomorphicity' is meant \cite{Pet07}. In this respect, the so--called {%
topological theorem} \cite{FP04} says that non--analyticity is the `shadow'
of a more fundamental phenomenon occurring in the system's configuration
manifold (in our case the $\cal CD$ manifold): a topology change within the
family of equipotential hypersurfaces
\begin{equation*}
M _{v}=\{(x^{1},\dots ,x^{3n})\in \mathbb{R}^{3n}|\ V(x^{1},\dots
,x^{3n})=v\},
\end{equation*}%
where $V$ and $x^{i}$ are the microscopic interaction potential and
coordinates respectively. This topological approach to PTs stems from the
numerical study of the dynamical counterpart of phase transitions, and
precisely from the observation of discontinuous or cuspy patterns displayed
by the largest Lyapunov exponent $\lambda _{1}$ at the {transition energy}
\cite{physrep}. Lyapunov exponents cannot be measured in laboratory
experiments, at variance with thermodynamic observables, thus, being genuine
dynamical observables they are only be estimated in numerical simulations of
the microscopic dynamics. If there are critical points of $V$ in
configuration space, that is points $x_{c}=[{\overline{x}}_{1},\dots ,{%
\overline{x}}_{3n}]$ such that $\left. \nabla V(x)\right\vert _{x=x_{c}}=0$,
according to the {Morse Lemma} \cite{hirsch}, in the neighborhood of any {%
critical point} $x_{c}$ there always exists a coordinate system {$x$}$(t)=[${%
$x$}$^{1}(t),...,${$x$}$^{3n}(t)]$ for which \cite{physrep}
\begin{equation}
V({x})=V(x_{c})-{x}_{1}^{2}-\dots -{x}_{k}^{2}+{x}_{k+1}^{2}+\dots +{x}%
_{3n}^{2},  \label{morsechart}
\end{equation}%
where $k$ is the {index of the critical point}, i.e., the number of negative
eigenvalues of the Hessian of the potential energy $V$. In the neighborhood
of a critical point of the $\cal CD$--manifold $M $, equation (\ref{morsechart})
yields the simplified form of (\ref{rok}),
~$
\partial ^{2}V/\partial x^{i}\partial x^{j}=\pm \delta _{ij},
$
giving $j$ unstable directions that contribute to the exponential
growth of the norm of the tangent vector $J$.

This means that the strength of dynamical chaos within the $\cal CD$--manifold $M $, measured by the largest Lyapunov exponent $\lambda
_{1}$ given by (\ref{Lyap1}), is affected by the existence of critical
points $x_{c}$ of the potential energy $V(x)$. However, as $V(x)$ is bounded
below, it is a good {Morse function}, with no vanishing eigenvalues of its
Hessian matrix. According to {Morse theory} \cite{hirsch}, the existence of
critical points of $V$ is associated with topology changes of the
hypersurfaces $\{M _{v}\}_{v\in \mathbb{R}}$.
The topology change of the $\{M _{v}\}_{v\in
\mathbb{R}}$ at some $v_{c}$ is a {necessary} condition for a phase
transition to take place at the corresponding energy value \cite{FP04}. The topology
changes implied here are those described within the framework of Morse
theory through `attachment of handles' \cite{hirsch} to the $\cal CD$--manifold $%
M$.

In our path--integral language this means that suitable
topology changes of equipotential submanifolds of the $\cal CD$--manifold $M $ can entail thermodynamic--like phase transitions
\cite{Haken1,Haken2,Haken3}, according to the general formula:
\begin{equation*}
\langle \mathrm{phase~out}\,|\,\mathrm{phase~in}\rangle\,:=
\int_{\rm top-ch}\mathcal{D}[w\Phi]\, {\rm e}^{iS[\Phi]}.
\end{equation*}
The statistical behavior of the crowd biodynamics system with the action functional (\ref{locAct}) and
the Hamiltonian (\ref{Ham}) is encompassed, in the canonical
ensemble, by its {partition function}, given by the Hamiltonian path
integral \cite{GaneshADG}%
\begin{equation}
Z_{3n}=\int_{\rm top-ch}\mathcal{D}[p]\mathcal{D}[x]\exp \{ \mathrm{i}\int_{t}^{t{%
^{\prime }}}[p_i\,\dot{x}^i-H(p,x)]\,d\tau \} ,  \label{PI1}
\end{equation}%
where we have used the shorthand notation
\begin{eqnarray*}
\int_{\rm top-ch}\mathcal{D}[p]\mathcal{D}[x]\equiv \int \prod_{\tau }\frac{dx(\tau )dp(\tau )%
}{2\pi }.
\end{eqnarray*}%
The path integral (\ref{PI1}) can be calculated as the
partition function \cite{FPS00},
\begin{eqnarray}
&&Z_{3n}(\beta )=\int \prod_{i=1}^{3n}dp_{i}\,dx^{i}\mathrm{e}^{-\beta
H(p,x)}=\left( \frac{\pi }{\beta }\right) ^{\frac{3n}{2}}\!\!\int
\prod_{i=1}^{3n}dx^{i}\mathrm{e}^{-\beta V(x)}  \notag \\
&&=\left( \frac{\pi }{\beta }\right) ^{\frac{3n}{2}}\int_{0}^{\infty }dv\,%
\mathrm{e}^{-\beta v}\int_{M _{v}}\frac{d\sigma }{\Vert \nabla V\Vert } , \label{zeta}
\end{eqnarray}%
where the last term is written using the so--called {co--area formula} \cite%
{federer}, and $v$ labels the equipotential hypersurfaces $M _{v}$ of
the $\cal CD$ manifold $M $,
\begin{equation*}
M _{v}=\{(x^{1},\dots ,x^{3n})\in \mathbb{R}^{3n}|V(x^{1},\dots
,x^{3n})=v\}.
\end{equation*}%
Equation (\ref{zeta}) shows that the relevant statistical information is
contained in the canonical configurational partition function
\begin{equation*}
Z_{3n}^{C}=\int \prod dx^{i}\,V(x)\,\mathrm{e}^{-\beta V(x)}.
\end{equation*}%
Note that $Z_{3n}^{C}$ is decomposed, in the last term of (\ref{zeta}), into
an infinite summation of geometric integrals,
\begin{equation*}
\int_{M _{v}}d\sigma /\Vert \nabla V\Vert ,
\end{equation*}%
defined on the $\{M _{v}\}_{v\in \mathbb{R}}$. Once the microscopic
interaction potential $V(x)$ is given, the configuration space of the system
is automatically foliated into the family $\{M _{v}\}_{v\in \mathbb{R}}$
of these equipotential hypersurfaces. Now, from standard statistical
mechanical arguments we know that, at any given value of the inverse
temperature $\beta $, the larger the number $3n$, the closer to $M
_{v}\equiv M _{u_{\beta }}$ are the microstates that significantly
contribute to the averages, computed through $Z_{3n}(\beta )$, of
thermodynamic observables. The hypersurface $M _{u_{\beta }}$ is the
one associated with
\begin{equation*}
u_{\beta }=(Z_{3n}^{C})^{-1}\int \prod dx^{i}V(x)\,\mathrm{e}^{-\beta V(x)},
\end{equation*}%
the average potential energy computed at a given $\beta $. Thus, at any $%
\beta $, if $3n$ is very large the effective support of the canonical measure
shrinks very close to a single $M _{v}=M _{u_{\beta }}$. Hence,
the basic origin of a phase transition lies in a suitable topology change of
the $\{M _{v}\}$, occurring at some $v_{c}$ \cite{FPS00}. This topology
change induces the singular behavior of the thermodynamic observables at a
phase transition. It is conjectured that the counterpart of a phase
transition is a breaking of diffeomorphicity among the surfaces $M _{v}$%
, it is appropriate to choose a {diffeomorphism invariant} to probe if and
how the topology of the $M _{v}$ changes as a function of $v$.
Fortunately, such a topological invariant exists, the {Euler characteristic}
of the crowd manifold $M $, defined by \cite{GaneshSprBig,GaneshADG}
\begin{equation}
\chi (M )=\sum_{k=0}^{3n}(-1)^{k}b_{k}(M ),  \label{chi}
\end{equation}%
where the {Betti numbers} $b_{k}(M )$ are diffeomorphism invariants ($b_{k}$ are the dimensions of the de Rham's cohomology
groups $H^{k}(M ;\mathbb{R})$; therefore the $b_{k}$ are
integers). This homological formula can be simplified by the use of the {%
Gauss--Bonnet theorem}, that relates $\chi (M )$ with the total {%
Gauss--Kronecker curvature} $K_{G}$ of the $\cal CD$--manifold $M $ given by \cite{GaneshADG,Complexity}
$$
\chi (M ) =\int_{M }K_{G}\, d\mu ,\qquad \text{where  $d\mu$  is given by (\ref{dmu}). }$$

\section{Conclusion}

Our understanding of crowd dynamics is presently limited in important ways; in particular, the lack of a geometrically \emph{predictive} theory of crowd behavior restricts the ability for authorities to intervene appropriately, or even to recognize when such intervention is needed. This is not merely an idle theoretical investigation: given increasing population sizes and thus increasing opportunity for the formation of large congregations of people, death and injury due to trampling and crushing -- even within crowds that have not formed under common malicious intent -- is a growing concern among police, military and emergency services. This paper represents a contribution towards the understanding of crowd behavior for the purpose of better informing decision--makers about the dangers and likely consequences of different intervention strategies in particular circumstances.

In this paper, we have proposed an entropic geometrical model of crowd dynamics, with dissipative kinematics, that operates across macro--, micro-- and meso--levels. This proposition is motivated by the need to explain the dynamics of crowds across these levels simultaneously: we contend that only by doing this can we expect to adequately characterize the geometrical properties of crowds with respect to regimes of behavior and the changes of state that mark the boundaries between such regimes.

In pursuing this idea, we have set aside traditional assumptions with respect to the separation of mind and body. Furthermore, we have attempted to transcend the long--running debate between contagion and convergence theories of crowd behavior with our multi-layered approach: rather than representing a reduction of the whole into parts or the emergence of the whole from the parts, our approach is build on the supposition that the direction of logical implication can and does flow in both directions simultaneously. We refer to this third alternative, which effectively unifies the other two, as \emph{behavioral composition.}

The most natural statistical descriptor is crowd entropy, which satisfies the extended second thermodynamics law applicable to open systems comprised of many components. Similarities between the configuration manifolds of individual (micro--level) and crowds (macro--level) motivate our claim that goal--directed movement operates under entropy conservation, while natural crowd dynamics operates under monotonically increasing entropy functions. Of particular interest is what happens between these distinct topological phases: the phase transition is marked by chaotic movement.

We contend that backdrop gives us a basis on which we can build a geometrically predictive model--theory of crowd psycho--physical behavior. This contrasts with previous approaches, which are explanatory only (explanation that is really narrative in nature). We propose an entropy formulation of crowd dynamics as a three step process involving individual and collective psycho-dynamics, and -- crucially -- non-equilibrium phase transitions whereby the forces operating at the microscopic level result in geometrical change at the macroscopic level. Here we have incorporated both geometrical and algorithmic notions of entropy as well as chaos in studying the topological phase transition between the entropy conservation of physical action and the entropy increase of mental preparation.

\section{Appendix}

\subsection{Extended second law of thermodynamics}

According to Boltzmann's interpretation of the second law of thermodynamics,
there exists a function of the state variables, usually chosen to be the
\textit{physical entropy} $S$ of the system that varies monotonically during the
approach to the unique final state of thermodynamic equilibrium:
\begin{equation}
\partial _{t}S\geq 0\qquad (\text{for any isolated system}).
\label{2law_isolated}
\end{equation}%
It is usually interpreted as a \emph{tendency to increased disorder}, i.e.,
an irreversible trend to maximum disorder. The above interpretation of
entropy and a second law is fairly obvious for systems of \emph{weakly
interacting particles}, to which the arguments developed by Boltzmann
referred.

However, according to Prigogine \cite{Nicolis}, the above interpretation of
entropy and a second law is fairly obvious \textit{only} for systems of
\emph{weakly interacting particles}, to which the arguments developed by
Boltzmann referred. On the other hand, for strongly interacting systems like
the crowd, the above interpretation does not apply in a straightforward
manner since, we know that for such systems there exists the possibility of
evolving to more ordered states through the mechanism of \textit{phase
transitions}.

Let us now turn to nonisolated systems (like a human crowd), which exchange
energy/matter with the environment. The entropy variation will now be the
sum of two terms. One, entropy flux, $d_{e}S$, is due to these exchanges;
the other, entropy production, $d_{i}S$, is due \ to the phenomena going on
within the system. Thus the entropy variation is
\begin{equation}
\partial _{t}S=\frac{d_{i}S}{dt}+\frac{d_{e}S}{dt}.  \label{entropy_tot}
\end{equation}%
For an isolated system $d_{e}S=0,$ and (\ref{entropy_tot}) together with (%
\ref{2law_isolated}) reduces to $dS=d_{i}S\geq 0$, the usual statement of
the second law. But even if the system is nonisolated, $d_{i}S$ will
describe those (irreversible) processes that would still go on even in the
absence of the flux term $d_{e}S$. We thus require the following extended
form of the second law:
\begin{equation}
\partial _{t}S\geq 0\qquad (\text{for any nonisolated system}).
\label{2law_nonisolated}
\end{equation}%
As long as $d_{i}S$ is strictly positive, irreversible processes will go on
continuously within the system.\footnote{%
Among the most common irreversible processes contributing to $d_{i}S$ are
chemical reactions, heat conduction, diffusion, viscous dissipation, and
relaxation phenomena in electrically or magnetically polarized systems. For
each of these phenomena two factors can be defined: an appropriate internal
\textit{flux}, $J_{i},$ denoting essentially its rate, and a driving \textit{%
force}, $X_{i},$ related to the maintenance of the nonequilibrium
constraint. A most remarkable feature is that $d_{i}S$ becomes a \emph{%
bilinear form} of $J_{i}$ and $X_{i}$. The following table summarizes the
fluxes and forces associated with some commonly observed irreversible
phenomena (see \cite{Nicolis,GaneshWSc})
\par
\begin{center}
$\left[
\begin{tabular}{llll}
\textbf{Phenomenon} & \textbf{Flux} & \textbf{Force} & \textbf{Rank} \\
Heat conduction & Heat flux, $\mathbf{J}_{th}$ & $grad(1/T)$ & Vector \\
Diffusion & Mass flux, $\mathbf{J}_{d}$ & $-[grad(\mu /T)-\mathbf{F}]$ &
Vector \\
Viscous flow & Pressure tensor, $\mathbf{P}$ & $(1/T)$ $grad\,\mathbf{v}$ &
Tensor \\
Chemical reaction & Rate of reaction, $\omega $ & Affinity of reaction &
Scalar%
\end{tabular}%
\ \right] $%
\end{center}
\par
In general, the fluxes $J_{k}$ are very complicated functions of the forces $%
X_{i}$. A particularly simple situation arises when their relation is
linear, then we have the celebrated \textit{Onsager relations},
\begin{equation}
J_{i}= L_{ik}X_{k},  \qquad (i,k=1,...,n) \label{Onsager}
\end{equation}%
in which $L_{ik}$ denote the set of \emph{phenomenological coefficients}.
This is what happens near equilibrium where they are also symmetric, $%
L_{ik}=L_{ki}$. Note, however, that certain states far from equilibrium can
still be characterized by a linear dependence of the form of (\ref{Onsager})
that occurs either accidentally or because of the presence of special types
of regulatory processes.} ~~Thus, $d_{i}S>0$ is equivalent to the condition of
dissipativity as time irreversibility. If, on the other hand, $d_{i}S$
reduces to zero, the process will be reversible and will merely join
neighboring states of equilibrium through a slow variation of the flux term $%
d_{e}S$.

From a computational perspective, we have a related \emph{algorithmic entropy}. Suppose we have a universal machine capable of simulating any effective procedure (i.e., a universal machine that can compute any computable function). There are several models to choose from, classically we would use a Universal Turing Machine but for technical reasons we are more interested in Lambda--type Calculi or Combinatory Logics. Let us describe the system of interest through some encoding as a combinatorial structure (classically this would be a binary string, but again I prefer for technical reasons Normal Forms with respect to alpha/beta/eta, weak, strong reduction, which are basically the Lambda--type Calculi and Combinatory Logic notions roughly akin to a ``computational" step). In other words, we have states of our system now represented as sentences in some language. The entropy is simply the minimum effective procedure against our computational model that generates the description of the system state. This is a universal and absolute notion of compression of our data -- the entropy is the strongest compression over all possible compression schemes, in effect. Now here is the `magic': this minimum is absolute in the sense that it does not vary (except by a constant) with respect to our reference choice of machine.

\subsection{Thermodynamic partition function}

Recall that the partition function $Z$ is a
quantity that encodes the statistical properties of a system in
thermodynamic equilibrium. It is a function of temperature and other
parameters, such as the volume enclosing a gas. Other thermodynamic
variables of the system, such as the total energy, free energy, entropy, and
pressure, can be expressed in terms of the partition function or its
derivatives.

A canonical ensemble is a statistical ensemble representing a probability
distribution of microscopic states of the system. Its probability
distribution is characterized by the proportion $p_{i}$ of members of the
ensemble which exhibit a measurable macroscopic state $i$, where the
proportion of microscopic states for each macroscopic state $i$ is given by
the Boltzmann distribution,
\begin{equation*}
p_{i}=\tfrac{1}{Z}\mathrm{e}^{-E_{i}/(kT)}=\mathrm{e}^{-(E_{i}-A)/(kT)},
\end{equation*}%
where $E_{i}$ is the energy of state $i$. It can be shown that this is the
distribution which is most likely, if each system in the ensemble can
exchange energy with a heat bath, or alternatively with a large number of
similar systems. In other words, it is the distribution which has \emph{%
maximum entropy} for a given average energy $<E_{i}>$.

The partition function of a \emph{canonical ensemble}~is defined as a sum ~~$%
Z(\beta )=\sum_{j}\mathrm{e}^{-\beta E_{j}},$~ where $\beta =1/(k_{B}T)$ is
the `inverse temperature', where $T$ is an ordinary temperature and $k_{B}$
is the Boltzmann's constant. However, as the position $x^{i}$ and momentum $%
p_{i}$ variables of an $i$th particle in a system can vary continuously, the
set of microstates is actually uncountable. In this case, some form of
\textit{coarse--graining} procedure must be carried out, which essentially
amounts to treating two mechanical states as the same microstate if the
differences in their position and momentum variables are `small enough'. The
partition function then takes the form of an integral. For instance, the
partition function of a gas consisting of $N$ molecules is proportional to
the $6N-$dimensional phase--space integral,
\begin{equation*}
Z(\beta )\sim \int_{\mathbb{R}^{6N}}\,d^{3}p_{i}\,d^{3}x^{i}\exp [-\beta
H(p_{i},x^{i})],
\end{equation*}%
where $H=H(p_{i},x^{i}),$ ($i=1,...,N$) \ is the classical Hamiltonian
(total energy) function.

More generally, the so--called \textit{configuration integral}, as used in
probability theory, information science and dynamical systems, is an
abstraction of the above definition of a partition function in statistical
mechanics. It is a special case of a normalizing constant in probability
theory, for the Boltzmann distribution. The partition function occurs in
many problems of probability theory because, in situations where there is a
natural symmetry, its associated probability measure, the \emph{Gibbs measure%
} (see below), which generalizes the notion of the canonical ensemble, has
the \emph{Markov property}.

Given a set of random variables $X_{i}$ taking on values $x^{i}$, and purely
potential Hamiltonian function $H(x^{i}),$ ($i=1,...,N$), the partition
function is defined as
\begin{equation}
Z(\beta )=\sum_{x^{i}}\exp \left[ -\beta H(x^{i})\right] . \label{partFun1}
\end{equation}%
The function $H$ is understood to be a real-valued function on the space of
states $\{X_{1},X_{2},\cdots \}$ while $\beta $ is a real-valued free
parameter (conventionally, the inverse temperature). The sum over the $x^{i}$
is understood to be a sum over all possible values that the random variable $%
X_{i}$ may take. Thus, the sum is to be replaced by an integral when the $%
X_{i}$ are continuous, rather than discrete. Thus, one writes
\begin{equation*}
Z(\beta )=\int dx^{i}\exp \left[ -\beta H(x^{i})\right] ,
\end{equation*}%
for the case of continuously-varying random variables $X_{i}$.

The Gibbs measure of a random variable $X_{i}$ having the value $x^{i}$ is
defined as the probability density function
\begin{equation*}
P(X_{i}=x^{i})=\frac{1}{Z(\beta )}\exp \left[ -\beta E(x^{i})\right] =\frac{%
\exp \left[ -\beta H(x^{i})\right] }{\sum_{x^{i}}\exp \left[ -\beta H(x^{i})%
\right] }.
\end{equation*}%
where $E(x^{i})=H(x^{i})$ is the energy of the configuration $x^{i}$. This
probability, which is now properly normalized so that $0\leq P(x^{i})\leq 1,$
can be interpreted as a likelihood that a specific configuration of values $%
x^{i},$ ($i=1,2,...N$) occurs in the system. $P(x^{i})$\ is also closely
related to $\Omega ,$\ the probability of a \emph{random partial recursive
function halting}.

As such, the partition function $Z(\beta )$\ can be understood to provide
the Gibbs measure on the space of states, which is the unique statistical
distribution that maximizes the entropy for a fixed expectation value of the
energy,
\begin{equation*}
\langle H\rangle =-\frac{\partial \log (Z(\beta ))}{\partial \beta }.
\end{equation*}%
The associated entropy is given by

\begin{equation*}
S=-\sum_{x^{i}}P(x^{i})\ln P(x^{i})=\beta \langle H\rangle +\log Z(\beta ),
\end{equation*}%
representing `ignorance' + `randomness'.

The principle of maximum entropy related to the expectation value of the
energy $\langle H\rangle ,$ is a postulate about a universal feature of any
probability assignment on a given set of propositions (events, hypotheses,
indices, etc.). Let some testable information about a probability
distribution function be given. Consider the set of all trial probability
distributions which encode this information. Then the probability
distribution which maximizes the information entropy is the true probability
distribution, with respect to the testable information prescribed.

Applied to the crowd dynamics, the Boltzman's theorem of \textit{%
equipartition of energy} states that the expectation value of the energy $%
\langle H\rangle $ is uniformly spread among all degrees-of-freedom of the
crowd (that is, across the whole crowd manifold $M$).

\subsection{Free energy, Landau's phase transitions and Haken's synergetics}

All thermodynamic--like properties of a multi-component system like a human
(or robot) crowd may be expressed in terms of its \textit{free energy
potential,} $\mathcal{F}=-k_B T\ln Z(\beta),$ and its partial derivatives. In particular, the
physical entropy $S$ of the crowd is defined as the (negative) first partial
derivative of the free energy $\mathcal{F}$ with respect to the control
parameter temperature $T$, i.e.,
~$
S=-\partial_T \mathcal{F},
$
while the \textit{specific heat capacity} $C$ is the second derivative,
~$
C=T\partial_TS.
$

A \textit{phase} of the crowd denotes a set of its states that have
relatively uniform psycho--physical properties. A \textit{crowd phase
transition }represents the its transformation from one phase to another (see
e.g., \cite{GaneshWSc,Complexity}). In general, the crowd phase transitions are
divided into two categories:

\begin{itemize}
\item The \textit{first--order phase transitions}, or, \textit{discontinuous
phase transitions}, are those that involve a latent heat $C$. During such a
transition, a crowd either absorbs or releases a fixed (and typically large)
amount of energy. Because energy cannot be instantaneously transferred
between the system and its environment, first--order crowd transitions are
associated with \emph{mixed--phase regimes} in which some parts of the crowd
have completed the transition and others have not. This forms a turbulent
spatioi-temporal chaotic interphase,  difficult to study, because its
dynamics can be violent and hard to control.

\item The \textit{second--order phase transitions} are the \textit{%
continuous phase transitions}, in the entropy $S$ is continuous$,$without
any latent heat $C$. They are purely entropic crowd transitions, which are
at the focus of the present study.
\end{itemize}

In Landau's theory od phase transitions (see \cite{GaneshWSc,Complexity}), the
probability density function $P$ is exponentially related to the free energy
potential $\mathcal{F}$, i.e.,
~$
P\approx \mathrm{e}^{-\mathcal{F}(T)},
$
if $\mathcal{F}$ is considered as a function of some order parameter $o$.
Thus, the most probable order parameter is determined by the requirement $%
\mathcal{F}=\mathrm{min}$. Therefore, the most natural order parameter for
the crowd dynamics would be its entropy $S$.

The following table gives the analogy between various systems in thermal
equilibrium and the corresponding nonequilibrium systems analyzed in Haken's
synergetics \cite{Haken1,Haken2,Haken3}:
\begin{center}
\begin{tabular}{l|l}
\hline
\textbf{System in thermal equilibrium} & \textbf{Nonequilibrium system} \\ \hline
Free energy potential $\mathcal{F}$ & Generalized potential $V$ \\ \hline
Order parameters $o_{i}$ & Order parameters $o_{i}$ \\ \hline
$\dot{o}_{i}=-\frac{\partial \mathcal{F}}{\partial o_{i}}$ & $\dot{o}_{i}=-%
\frac{\partial V}{\partial o_{i}}$ \\ \hline
Temperature $T$ & Control input $u$ \\ \hline
Entropy $S$ & System output $y$ \\ \hline
Specific Heat $c$ & System efficiency $e$ \\ \hline
\end{tabular}
\end{center}\bigbreak

In particular, in case of human biodynamics \cite{GaneshWSc,Complexity},
natural control inputs $u_{i}$ are muscular forces and torques, $F_{i}$, natural system
outputs $y_{i}$ are joint coordinates $q^{i}$ and momenta $p_{i}$, while the
system efficiencies $e_{i}$ represent the changes of coordinates and momenta
with changes of corresponding muscular torques for the $i$th active human
joint,
~~
$e_{i}^{q}=\frac{\partial q^{i}}{\partial F_{i}},\quad e_{i}^{p}=\frac{%
\partial p_{i}}{\partial F_{i}}.
$

\subsection{Heat equation, Dirichlet action and gradient flow on a Riemannian manifold}

The heat equation
\begin{equation}
\dot{u}=\Delta u,  \label{t81}
\end{equation}%
on a compact Riemannian manifold $M$ with static metric ($\partial _{t}g=0$), where $u:[0,T]\times M\rightarrow \mathbb{R}$ is a
scalar field, can be interpreted as the gradient flow for the \textit{%
Dirichlet action}
\begin{equation}
E(u):=\frac{1}{2}\int_{M}|\nabla u|_{g}^{2}\ d\mu ,  \label{t82}
\end{equation}%
using the inner product, $\langle u_{1},u_{2}\rangle _{\mu
}:=\int_{M}u_{1}u_{2}\ d\mu ,$ associated to the volume measure $d\mu $.
This can be proved if we evolve $u$ in time at some arbitrary rate $u$, an
application of integration by parts formula,
\begin{equation*}
\int_{M}u\nabla _{\alpha }X^{\alpha }\ d\mu =-\int_{M}(\nabla _{\alpha
}u)X^{\alpha }\ d\mu
\end{equation*}%
(where $\mathrm{div}(X):=\nabla _{\alpha }X^{\alpha }$ is the divergence of
the vector-field $X^{\alpha }$, which validates the Stokes theorem, $\int_{M}%
\mathrm{div}(X)\ d\mu =0),$ gives
\begin{equation}
\partial _{t}E(u)=-\int_{M}(\Delta u)\dot{u}\ d\mu =\langle -\Delta u,\dot{u}%
\rangle _{\mu },  \label{t83}
\end{equation}%
from which we see that (\ref{t81}) is indeed the gradient flow for (\ref{t83}%
) with respect to the inner product. In particular, if $u$ solves the heat
equation (\ref{t81}), we see that the Dirichlet energy is decreasing in
time,
\begin{equation}
\partial _{t}E(u)=-\int_{M}|\Delta u|^{2}\ d\mu .  \label{t84}
\end{equation}%
Thus we see that by representing the parabolic PDE (\ref{t81}) as a gradient
flow, we automatically gain a controlled quantity of the evolution, namely
the energy functional that is generating the gradient flow. This
representation also strongly suggests that solutions of (\ref{t81}) should
eventually converge to stationary points of the Dirichlet energy (\ref{t82}%
), which by (\ref{t83}) are harmonic functions (i.e., the functions $u$ with
$\Delta u=0$). As an application of the gradient flow interpretation, we can
assert that the only periodic (or, \textquotedblleft
breather\textquotedblright ) solutions to the heat equation (\ref{t81}) are
the harmonic functions (which must be constant if the manifold $M$ is
compact). Indeed, if a solution $u$ was periodic, then the monotone
functional $E$ must be constant, which by (\ref{t84}) implies that $u$ is
harmonic as claimed.

\subsection{Lyapunov exponents and Kolmogorov--Sinai entropy}

A branch of nonlinear dynamics has been developed with the aim of
formalizing and quantitatively characterizing the general sensitivity to
initial conditions. The \textit{largest Lyapunov exponent} $\lambda ,$
together with the related \emph{Kaplan--Yorke dimension} $d_{KY}$ and the
\textit{Kolmogorov--Sinai entropy} $h_{KS}$ are the three indicators for
measuring the \textit{rate of error growth} produced by a dynamical system
\cite{ER85,StrAttr,TacaNODY}.

The characteristic Lyapunov exponents are somehow an extension of the linear
stability analysis to the case of aperiodic motions. Roughly speaking, they
measure the typical rate of exponential divergence of nearby trajectories.
In this sense they give information on the rate of growth of a very small
error on the initial state of a system \cite{BLV01,PredictCompl}.

Consider an $n$D dynamical system given by the set of ODEs of the form
\begin{equation}
\dot{x}=f(x),  \label{eq:1-1}
\end{equation}
where $x=(x_{1},\dots ,x_{n})\in \mathbb{R}^{n}$ and $f:\mathbb{R}%
^{n}\rightarrow \mathbb{R}^{n}$. Recall that since the r.h.s of equation (%
\ref{eq:1-1}) does not depend on $t$ explicitly, the system is called \emph{%
autonomous}. We assume that ${f}$ is smooth enough that the evolution is
well--defined for time intervals of arbitrary extension, and that the motion
occurs in a bounded region $R$ of the system phase space $M$. We intend to
study the separation between two trajectories in $M$, $x(t)$ and $x^{\prime
}(t)$, starting from two close initial conditions, $x(0)$ and $x^{\prime
}(0)=x(0)+\delta x(0)$ in $R_0 \subset M$, respectively.

As long as the difference between the trajectories, $\delta x(t)=x^{\prime
}(t)-x(t)$, remains infinitesimal, it can be regarded as a vector, $z(t)$,
in the tangent space $T_{x}M$ of $M$. The time evolution of $z(t)$ is given
by the linearized differential equations:
\begin{equation*}
\dot{z}_{i}(t)=\,\left. \frac{\partial f_{i}}{\partial x_{j}}\right\vert
_{x(t)}\,z_{j}(t).
\end{equation*}%
Under rather general hypothesis, Oseledets \cite{Oseledets} proved that for
almost all initial conditions $x(0)\in R$, there exists an orthonormal basis
$\{e_{i}\}$ in the tangent space $T_{x}M$ such that, for large times,
\begin{equation}
z(t)=c_{i}e_{i}\exp (\lambda _{i}t),  \label{eq:1-5}
\end{equation}%
where the coefficients $\{c_{i}\}$ depend on $z(0)$. The exponents $\lambda
_{1}\geq \lambda _{2}\geq \cdots \geq \lambda _{d}$ are called \textit{%
characteristic Lyapunov exponents}. If the dynamical system has an ergodic
invariant measure on $M$, the spectrum of LEs $\{\lambda _{i}\}$ does not
depend on the initial conditions, except for a set of measure zero with
respect to the natural invariant measure.

Equation~(\ref{eq:1-5}) describes how an $n$D spherical region $%
R=S^{n}\subset M$, with radius $\epsilon $ centered in $x(0)$, deforms, with
time, into an ellipsoid of semi--axes $\epsilon _{i}(t)=\epsilon \exp
(\lambda _{i}t)$, directed along the $e_{i}$ vectors. Furthermore, for a
generic small perturbation $\delta x(0)$, the distance between the reference
and the perturbed trajectory behaves as
\begin{equation*}
|\delta x(t)|\sim |\delta x(0)|\,\exp (\lambda _{1}t)\,\left[ 1+O\left( \exp
{-(\lambda _{1}-\lambda _{2})t}\right) \right] .
\end{equation*}%
If $\lambda _{1}>0$ we have a rapid (exponential) amplification of an error
on the initial condition. In such a case, the system is chaotic and,
unpredictable on the long times. Indeed, if the initial error amounts to $%
\delta _{0}=|\delta x(0)|$, and we purpose to predict the states of the
system with a certain tolerance $\Delta $, then the prediction is reliable
just up to a \textit{predictability time} given by
\begin{equation*}
T_{p}\sim \frac{1}{\lambda _{1}}\ln \left( {\frac{\Delta }{\delta _{0}}}%
\right) .
\end{equation*}%
This equation shows that $T_{p}$ is basically determined by the \textit{%
positive leading Lyapunov exponent}, since its dependence on $\delta _{0}$
and $\Delta $ is logarithmically weak. Because of its preeminent role, $%
\lambda _{1}$ is often referred as $\mathbf{`}$the leading positive Lyapunov
exponent', and denoted by $\lambda $.

Therefore, Lyapunov exponents are average rates of expansion or contraction
along the principal axes. For the $i$th principal axis, the corresponding
Lyapunov exponent is defined as
\begin{equation}
\lambda _{i}=\lim_{t\rightarrow \infty }\{(1/t)\ln [L_{i}(t)/L_{i}(0)]\},
\label{LyapExpon}
\end{equation}%
where $L_{i}(t)$ is the radius of the ellipsoid along the $i$th principal
axis at time $t$.

An initial volume $V_{0}$ of the phase--space region $R_0$ evolves on
average as
\begin{equation}
V(t)=V_{0}e^{(\lambda _{1}+\lambda _{2}+\cdots +\lambda _{2n})t},
\label{vol}
\end{equation}
and therefore the rate of change of $V(t)$ is simply
\begin{equation*}
\dot{V}(t)=\sum_{i=1}^{2n}\lambda _{i}V(t).
\end{equation*}

In the case of a 2D phase area $A$, evolving as $A(t)=A_{0}e^{(\lambda
_{1}+\lambda _{2})t}$, a \textit{Lyapunov dimension} $d_{L}$ is defined as
\begin{equation*}
d_{L}=\lim_{\epsilon \rightarrow 0}\left[ \frac{d(\ln (N(\epsilon )))}{d(\ln
(1/\epsilon ))}\right] ,
\end{equation*}%
where $N(\epsilon )$ is the number of squares with sides of length $\epsilon
$ required to cover $A(t)$, and $d$ represents an ordinary \textit{capacity
dimension},
\begin{equation*}
d_{c}=\lim_{\epsilon \rightarrow 0}\left( \frac{\ln N}{\ln (1/\epsilon )}%
\right) .
\end{equation*}

Lyapunov dimension can be extended to the case of $n$D phase--space by means
of the \textit{Kaplan--Yorke dimension} \cite{Kaplan,Yorke,Ott90}) as
\begin{equation*}
d_{KY}=j+\frac{\lambda _{1}+\lambda _{2}+\cdots +\lambda _{j}}{|\lambda
_{j+1}|},
\end{equation*}%
where the $\lambda _{i}$ are ordered ($\lambda _{1}$ being the largest) and $%
j$ is the index of the smallest nonnegative Lyapunov exponent.

On the other hand, a state, initially determined with an error $\delta x(0)$%
, after a time enough larger than $1/\lambda $, may be found almost
everywhere in the region of motion $R\in M$. In this respect, the \textit{%
Kolmogorov--Sinai} (KS) \emph{entropy}, $h_{KS}$, supplies a more refined
information. The error on the initial state is due to the maximal resolution
we use for observing the system. For simplicity, let us assume the same
resolution $\epsilon $ for each degree of freedom. We build a partition of
the phase space $M$ with cells of volume $\epsilon ^{d}$, so that the state
of the system at $t=t_{0}$ is found in a region $R_{0}$ of volume $%
V_{0}=\epsilon ^{d}$ around $x(t_{0})$. Now we consider the trajectories
starting from $V_{0}$ at $t_{0}$ and sampled at discrete times $%
t_{j}=j\,\tau $ ($j=1,2,3,\dots ,t$). Since we are considering motions that
evolve in a bounded region $R\subset M$, all the trajectories visit a finite
number of different cells, each one identified by a symbol. In this way a
unique sequence of symbols $\{s(0),s(1),s(2),\dots \}$ is associated with a
given trajectory $x(t)$. In a chaotic system, although each evolution $x(t)$
is univocally determined by $x(t_{0})$, a great number of different symbolic
sequences originates by the same initial cell, because of the divergence of
nearby trajectories. The total number of the admissible symbolic sequences, $%
\widetilde{N}(\epsilon ,t)$, increases exponentially with a rate given by
the topological entropy
\begin{equation*}
h_{T}=\lim_{\epsilon \rightarrow 0}\lim_{t\rightarrow \infty }{\frac{1}{t}}%
\ln {\widetilde{N}}(\epsilon ,t)\,.
\end{equation*}

However, if we consider only the number of sequences $N_{eff}(\epsilon
,t)\leq \widetilde{N}(\epsilon ,t)$ which appear with very high probability
in the long time limit -- those that can be numerically or experimentally
detected and that are associated with the natural measure -- we arrive at a
more physical quantity called the Kolmogorov--Sinai (or metric) entropy, which is the key entropy notion in ergodic theory \cite%
{ER85}:
\begin{equation}
h_{KS}=\lim_{\epsilon \rightarrow 0}\lim_{t\rightarrow \infty }{\frac{1}{t}}%
\ln N_{eff}(\epsilon ,t)\leq h_{T}.  \label{metentro}
\end{equation}%
$h_{KS}$ quantifies the long time exponential rate of growth of the number
of the effective coarse-grained trajectories of a system. This suggests a
link with information theory where the Shannon entropy measures the mean
asymptotic growth of the number of the typical sequences -- the ensemble of
which has probability almost one -- emitted by a source.

We may wonder what is the number of cells where, at a time $t>t_{0}$, the
points that evolved from $R_{0}$ can be found, i.e., we wish to know how big
is the coarse--grained volume $V(\epsilon ,t)$, occupied by the states
evolved from the volume $V_{0}$ of the region $R_{0}$, if the minimum volume
we can observe is $V_{min}=\epsilon ^{d}$. As stated above (\ref{vol}), we
have
\begin{equation*}
V(t)\sim V_{0}\exp (t\sum_{i=1}^{d}\lambda _{i}).
\end{equation*}

However, this is true only in the limit $\epsilon \rightarrow 0$. In this
(unrealistic) limit, $V(t)=V_{0}$ for a conservative system (where $%
\sum_{i=1}^{d}\lambda _{i}=0$) and $V(t)<V_{0}$ for a dissipative system
(where $\sum_{i=1}^{d}\lambda _{i}<0$). As a consequence of limited
resolution power, in the evolution of the volume $V_{0}=\epsilon ^{d}$ the
effect of the contracting directions (associated with the negative Lyapunov
exponents) is completely lost. We can experience only the effect of the
expanding directions, associated with the positive Lyapunov exponents. As a
consequence, in the typical case, the coarse grained volume behaves as
\begin{equation*}
V(\epsilon ,t)\sim V_{0}\,\mathrm{e}^{(\sum_{\lambda _{i}>0}\lambda
_{i})\,t},
\end{equation*}%
when $V_{0}$ is small enough. Since $N_{eff}(\epsilon ,t)\propto V(\epsilon
,t)/V_{0}$, one has: ~$
h_{KS}=\sum_{\lambda _{i}>0}\lambda _{i}.
$
This argument can be made more rigorous with a proper mathematical
definition of the metric entropy. In this case one derives the Pesin
relation \cite{Pesin,ER85}: ~
$
h_{KS}\leq \sum_{\lambda _{i}>0}\lambda _{i}.
$
Because of its relation with the Lyapunov exponents, or by the definition (%
\ref{metentro}), it is clear that also $h_{KS}$ is a fine-grained and
global characterization of a dynamical system.

The metric entropy is an invariant characteristic quantity of a dynamical
system, i.e., given two systems with invariant measures, their KS--entropies
exist and they are equal iff the systems are isomorphic \cite{bill65}.

Finally, the \emph{topological entropy} on the manifold $M$ equals the supremum of the Kolmogorov-Sinai entropies, $$h(u)=\sup\{h_{KS}(u)=h_{\mu}(u):\mu\in P_u(M)\},$$ where $u:M\to M$ is a continuous map on $M$, and $\mu$ ranges over all $u-$invariant (Borel) probability measures on $M$. Dynamical systems of positive topological entropy are often considered topologically chaotic.

\bigbreak\bigbreak

\end{document}